%% file: final_version.tex
\documentclass[journal]{IEEEtran}
\usepackage{cite}
\usepackage{setspace}
\usepackage{amsmath,amssymb,amsthm}
\usepackage{graphicx}
\usepackage{caption}
\usepackage{subcaption}
\usepackage{float}
\usepackage{bm}

\usepackage{url}
\usepackage{algorithm}
\usepackage{algorithmicx}
\usepackage{algpseudocode}
\usepackage{filecontents}

\usepackage{indentfirst}
\usepackage{color}
\usepackage{booktabs}

\usepackage{epstopdf}
\usepackage{grffile}
\usepackage{multirow}
\usepackage{diagbox}
\DeclareMathOperator*{\argmin}{argmin} % no space, limits underneath in displays

\newtheorem{theorem}{Theorem}
\newtheorem{lemma}{Lemma}
\newcommand{\x}		{\mathbf{x}}	
	
\newcommand{\y}		{\mathbf{y}}
\newcommand{\A}		{\mathbf{A}}
\newcommand{\z}	 {\mathbf{z}}
\newcommand{\W}	 {\mathbf{W}}
\renewcommand{\P}	 {\mathbf{P}}
\newcommand{\omg}	 {\mathbf{\Omega}}

\newcommand{\Z}		{\mathbf{Z}}

\newcommand{\D}	 {\mathbf{D}}

\newcommand{\s}	 {\mathbf{s}}
\newcommand{\g}	 {\mathbf{g}}

\newcommand{\ze}	 {\bm{\zeta}}

\newcommand{\R}	     {\mathsf{R}}

\newcommand{\DIFadd}{\textcolor{black}}

\newcommand{\diag}  {\mathsf{diag}}

\title{SPULTRA: Low-Dose CT Image Reconstruction with Joint Statistical and Learned Image Models}

\author{Siqi Ye, Saiprasad Ravishankar, \textit{Member, IEEE}, Yong Long*, \textit{Member, IEEE}, \\ and Jeffrey A. Fessler, \textit{Fellow, IEEE}
	\thanks{ 
		
		This work was supported in part by the SJTU-UM Collaborative Research Program, NSFC (61501292), NIH grant U01 EB018753, ONR grant N00014-15-1-2141, DARPA Young Faculty Award D14AP00086, and ARO MURI grants W911NF-11-1-0391 and 2015-05174-05. \textit{Asterisk indicates the corresponding author.} 
		
		S. Ye and Y. Long are with the University of Michigan - Shanghai Jiao Tong University Joint Institute, Shanghai Jiao Tong University, Shanghai 200240, China (email: yesiqi@sjtu.edu.cn, yong.long@sjtu.edu.cn).
		
		\DIFadd{S. Ravishankar is with the Department of Computational Mathematics, Science and Engineering, and the Department of Biomedical Engineering, Michigan State University, East Lansing, MI, 48824 USA (email: {ravisha3@msu.edu}).}
		
		J. A. Fessler is with the Department of Electrical Engineering and Computer Science, University of Michigan, Ann Arbor, MI, 48109 USA (email: fessler@umich.edu).
		
		Copyright (c) 2019 IEEE. Personal use of this material is permitted.  Permission from IEEE must be obtained for all other uses, in any current or future media, including reprinting/republishing this material for advertising or promotional purposes, creating new collective works, for resale or redistribution to servers or lists, or reuse of any copyrighted component of this work in other works.
}}
\begin{document}
	\maketitle
	\begin{abstract}
		Low-dose CT image reconstruction has been a popular research topic in recent years. A typical reconstruction method based on post-log measurements is called penalized weighted-least squares (PWLS). Due to the underlying limitations of the post-log statistical model, the PWLS reconstruction quality is often degraded in low-dose scans. This paper investigates a shifted-Poisson (SP) model based likelihood function that uses the pre-log raw measurements that better represents the measurement statistics, together with a data-driven regularizer exploiting a Union of Learned TRAnsforms (SPULTRA).
		Both the SP induced data-fidelity term and the regularizer in the proposed framework are nonconvex. \DIFadd{The proposed SPULTRA algorithm uses} quadratic surrogate functions for the SP induced data-fidelity term. \DIFadd{Each iteration involves} a quadratic subproblem for updating the image, and a sparse coding and clustering subproblem that has a closed-form solution.
		The SPULTRA algorithm has a similar computational cost per iteration as its recent counterpart PWLS-ULTRA that uses post-log measurements, and it provides \DIFadd{better} image reconstruction quality than PWLS-ULTRA, especially in low-dose scans.
	\end{abstract}

	\begin{IEEEkeywords}
	Inverse problems, sparse representation, transform learning, shifted-Poisson model, nonconvex optimization, efficient algorithms, machine learning. 
	\end{IEEEkeywords}
%	\IEEEpeerreviewmaketitle

\vspace{-0.1in}
\input{introduction_r2}

\section{Problem Formulation for SPULTRA}
\label{sec:formulation}
The goal in LDCT image reconstruction is to estimate the linear attenuation coefficients $\x \in \mathbb{R}^{N_p}$ from CT measurements $\y \in \mathbb{R}^{N_d}$. We propose to obtain the reconstructed image by solving a SP model-based penalized-likelihood problem:
\begin{equation}\label{eq:P0}
\hat{\x} = \arg\min\DIFadd{_{\x \in\mathcal{X}}\ G (\x), \DIFadd{~~G}(\x) =  \mathsf{L}(\x) + \R(\x),
} \tag{P0}
\end{equation}
\DIFadd{where $\mathcal{X}=\{\x| 0 \leq \x_j \leq x_{\mathrm{max}}\}$, $x_{\mathrm{max}}$ is a large constant.} The objective function \DIFadd{$ G(\x) $} is composed of a negative log-likelihood function $\mathsf{L}(\x)$ based on the SP model for the measurements, and a penalty term $\R(\x)$ that is based on the \DIFadd{ULTRA model \cite{ravishankar:16:tci, pwls-ultra2018}}. 
The SP model can be described as \DIFadd{${Y_i\sim \text{Poisson}\{{I_0e^{-f_i([\A\x]_i)}} + \sigma^2\}}$}, where $Y_i$ is the shifted quantity of the $i$th measurement for $i = 1, \ldots, N_d$, $\sigma^2$ is the variance of the electronic noise, $I_0$ is the incident photon count per ray from the source, \DIFadd{${f_i(\cdot) }$ models the beam-hardening effect,} and $\A \in \mathbb{R}^{N_d \times N_p}$ is the CT system matrix. Denoting $l_i(\x) \triangleq [\A\x]_i$ (or $l_i$ in short), the data-fidelity term $\mathsf{L}(\x)$ can be written as
\vspace{-0.02in}
\begin{equation}\label{eq:L}	
\begin{aligned}
\mathsf{L}(\x) &= \sum_{i = 1}^{N_d}h_i(l_i(\x)),
\end{aligned} 
\vspace{-0.1in}
\end{equation}	
where
\begin{equation}
h_i(l_i) \triangleq (I_0e \DIFadd{^{-f_i(l_i)} } + \sigma^2) - Y_i\log(I_0e \DIFadd{^{-f_i(l_i)} } + \sigma^2)  .
\end{equation}
\DIFadd{The beam-hardening model $f_i(\cdot)$ is usually approximated as a polynomial \cite{pre-post-log}. For simplicity, we use a second order polynomial, i.e., ${f_i(l_i) = s_{1_i} l_i + s_{2_i} l_i^2}$, where $s_{1_i}$ and $s_{2_i}$ are coefficients of the polynomial for the $i$th measurement.
}

The ULTRA regularizer $\R(\x)$ has the following form \cite{pwls-ultra2018}:
\begin{equation}\label{eq:Rx}
\begin{aligned}
\R(\x) \triangleq  &\min_{\{\z_j, C_k\}}  \beta  \sum_{k=1}^{K}  \bigg\{  \sum_{j\in C_k} \tau_j \{\|\omg_k \P_j \x - \z_{j}\|^2_2 + \gamma_c^2\|\z_{j}\|_0 \}\bigg\} \\
&\quad\text{s.t.}\ \{C_k\} \in \bm{\mathcal{G}},
\end{aligned}
%\vspace{-0.05in}
\end{equation}	 
where $\bm{\mathcal{G}}$ denotes the set consisting of all possible partitions of $\{1, 2, \ldots,N_p\}$ into $K$ disjoint subsets, $K$ is the number of \DIFadd{classes} and $C_k$ denotes the set of indices of patches belonging to the $k$th \DIFadd{class}. The operator $\P_j \in \mathbb{R}^{v \times N_p}$ is the patch extraction operator that extracts the $j$th patch of $v$ voxels for \DIFadd{$j = 1, \ldots, \tilde{N}$, from $\x$, where $\tilde{N}$ is the number of extracted patches.} The learned transform corresponding to the $k$th \DIFadd{class} $\omg_k \in \mathbb{R}^{v \times v}$ maps the patches to the transform domain. Vector $\z_j~\in~\mathbb{R}^v$ denotes the sparse approximation of the transformed $j$th patch, with the parameter $\gamma_c^{2}$ ($\gamma_c>0$) controlling its sparsity level. We use the $\ell_0$ ``norm" (that counts the number of nonzero elements in $\z_j$) to enforce sparsity. The patch-based weight $\tau_j$ is defined as \DIFadd{$\|\P_j\bm{\kappa}\|_1/v$\cite{pwls-ultra2018, chun:17:svx}}, where $\bm{\kappa} \in \mathbb{R}^{N_p}$ is defined to help encourage resolution uniformity as $\kappa_j \triangleq \sqrt{\sum_{i=1}^{N_d}a_{ij} \DIFadd{\tilde{w}_i}\sum_{i=1}^{N_d}a_{ij}}$ \cite[eq(39)]{kappa}, with $a_{ij}$ \DIFadd{denoting} the entries of $\A$, and \DIFadd{$\tilde{w}_i$ is approximated as ${\tilde{w}_i = \dot{f_i}(\tilde{l}_i)^2 \frac{y_i^2}{y_i + \sigma^2}}$ \cite[eq(10)]{pre-post-log}, where $\tilde{l}_i$ is the beam-hardening corrected, post-log sinogram data. }To balance the data-fidelity term and the regularizer in the formulation, $\R(\x)$ is scaled by a positive parameter $\beta$.

\vspace{-0.1in}
\input{Algorithm_r2}

\vspace{-0.05in}
\section{Convergence Analysis}
\label{sec:convergenceAnalysis}
\DIFadd{The objective function \eqref{eq:P0} of SPULTRA is highly nonconvex due to the nonconvexity of the data-fidelity term and the regularizer. The proposed algorithm efficiently optimizes it by using surrogate functions and alternating minimization. This section provides a convergence analysis for the general optimization approach. While a recent work~\cite{ravishankar:16:tci} analyzed the convergence of a related optimization method, it did not involve the use of surrogate functions and involved adaptive learning of transforms.}
	
\DIFadd{In the proposed method, the sparse coding and clustering step is solved exactly. For the image update step, where the cost function is quadratic as in \eqref{eq:Phi1}, many approaches may be used to optimize it, e.g., \cite{nien:16:rla, OGM2016, Momentum}. Our convergence proof in the supplement assumes for simplicity that the image update step is solved exactly. }
	
\DIFadd{The convergence result uses the following notation. We use $\Z$ for the sparse code matrix concatenated by column vectors $\z_{j}$,  and use a vector ${\Gamma \in \mathbb{R}^{\tilde{N}}}$, whose elements represent the classes indices for the patches, i.e., $\Gamma_j \in \{1, \cdots, K\}$. 
	For an initial $(\x^0, \Z^0, \Gamma^0)$, we let $\{\x^n, \Z^n, \Gamma^n\}$ denote the sequence of iterates generated by alternating algorithm. The objective function in \eqref{eq:P0} is denoted as $G(\x,\Z,\Gamma)$ and includes the constraint on $\x$ as an added barrier penalty (which takes the value $+\infty$ when the constraint is violated and is zero otherwise). The convergence result is as follows.}
	\vspace{-0.1in}
	\begin{theorem}
\DIFadd{\DIFadd{Assume the image update step is solved exactly}. For an initial $(\x^0, \Z^0, \Gamma^0)$, iterative sequence $\{\x^n, \Z^n, \Gamma^n\}$ generated by the SPULTRA algorithm is bounded, and the corresponding objective sequence $\{G(\x^n, \Z^n, \Gamma^n)\}$ decreases monotonically and converges to ${G^* \triangleq G^*(\x^0, \Z^0, \Gamma^0)}$. Moreover, all the accumulation points of the iterate sequence are equivalent and achieve the same value $G^*$ of the objective. Each accumulation point $(\x^*,\Z^*,\Gamma^*)$ also satisfies the following partial optimality conditions:
	\begin{equation}\label{eq:critical-point}
	\begin{aligned}
	&\mathbf{0} \in \partial_{\x}G(\x,\Z^*,\Gamma^*)|_{\x = \x^*},\\
	&(\Z^*,\Gamma^*) \in \arg\min_{\Z,\Gamma} G(\x^*,\Z,\Gamma),
	\end{aligned}
	\tag{14}
	\end{equation}
	where $\partial_{\x}$ denotes the sub-differential operator for the function $G$ with respect to $\x$ \cite{rockafellar2009variational,mordukhovich2006variational,2015BCS-ST}.
	Finally,  ${\|\x^{n+1} - \x^n\|_2 \to 0}$ as $n\to \infty$.}
	\end{theorem}
	
\DIFadd{The above theorem implies that for each initial $(\x^0, \Z^0, \Gamma^0)$, the objective sequence converges (although the limit may depend on initialization) and the iterate sequence in the optimization framework converges to an equivalence class of accumulation points (i.e., all accumulation points have the same objective value $G^*$) that are also partial optimizers satisfying~\eqref{eq:critical-point}. Moreover, the image sequence satisfies $\|\x^{n+1} - \x^n\|_2 \to 0 $.}
	
\DIFadd{When $K = 1$, \eqref{eq:critical-point} readily implies that the iterate sequence in the algorithm converges to an equivalence class of critical points~\cite{rockafellar2009variational} (that are generalized stationary points) of the nonconvex cost $G(\x, \Z, \Gamma)$.
}

\DIFadd{A detailed proof is included in the supplement\footnote{Supplementary material is available in the supplementary materials / multimedia tab.}.
}
\vspace{-0.05in}
\input{Experiments_r2}
\vspace{-0.05in}

%\clearpage
\input{conclusions_r1}
\vspace{-0.15in}
%\clearpage
\bibliographystyle{IEEEbib}
\bibliography{spultra-submit}

\clearpage

%%%%%%%% supplement %%%%%%%%
%% Title 
{\twocolumn[
	\begin{center}
		\Huge SPULTRA: Low-Dose CT Image Reconstruction with Joint Statistical and Learned Image Models \\-- Supplementary Materials
		\vspace{0.4in}
	\end{center}]}
\input{multimedia}

\end{document}

%% file: introduction_r2.tex
\section{Introduction} 
Recent years have witnessed the growing deployment of X-ray computed tomography (CT) in medical \DIFadd{applications}. Simultaneously there has been great concern to reduce the potential risks caused by exposure to X-ray radiation. Strategies for reducing the X-ray radiation in CT include reducing the photon intensity at the X-ray source, i.e., low-dose CT (LDCT), or lowering the number of projection views obtained by the CT machine, i.e., sparse-view CT. \DIFadd{In the case where the X-ray radiation is extremely low, the CT image may not be suitable for medical diagnosis, but it is still quite helpful for non-diagnostic applications such as attenuation correction for PET/CT imaging~\cite{kinahan1998attenuation, ACforPET2011, ACforPET2015} and virtual CT colonoscopy screening~\cite{wang2008virtual}. Reconstructing CT images }with reduced radiation is challenging, \DIFadd{and} many reconstruction methods have been proposed for this setting. Model-based iterative reconstruction (MBIR) is widely used~\cite{fessler2000statistical} \DIFadd{among these approaches}. Based on maximum a posteriori (MAP) estimation, MBIR approaches form a cost function that incorporates the statistical model for the acquired measurements and the prior knowledge (model) of the images. This section first reviews some of the statistical models for CT measurements along with recent works on extracting prior knowledge about images for LDCT image reconstruction, and then presents our contributions.
\vspace{-0.12in}
\subsection{Background}
Accurate statistical modeling of the measurements in CT scanners is challenging, especially in low-dose imaging, when the electronic noise in the data acquisition system (DAS) becomes significant \cite{streakArtifact,whiting2002signal,elbakri2003efficient,yu2012development, ma2012variance,ding:16:mmp,survey2013-ct}.
Approximations of the measurement statistics can be categorized into \textit{post-log} and \textit{pre-log} models \cite{pre-post-log}, which are detailed next. 

The post-log models work on data obtained from the logarithmic transformation of the raw measurements, which is often assumed Gaussian distributed. Since the logarithmic transformation approximately linearizes the raw measurements, methods based on post-log data can readily exploit various optimization approaches and regularization designs with efficiency and convergence guarantees for this reconstruction problem \cite{beister2012iterative, thibault:07:atd, Momentum}. The post-log methods however have a major drawback: the raw measurements may contain non-positive values on which the logarithmic transformation cannot be taken (or near-zero positive measurements whose logarithm can be very negative), particularly when the electronic noise becomes significant as compared to the photon statistical noise in low-dose cases. 

There are many pre-correction approaches to deal with \DIFadd{non-positive} raw measurements \DIFadd{for post-log methods}. %the non-positive values. 
Examples of such approaches include using a statistical weight of zero for \DIFadd{such} measurements \cite{Polyenergetic02}, replacing the non-positive measurements with a small positive value~\cite{PWLS06} and filtering neighboring measurements~\cite{streakArtifact}.
Thibault et al.~\cite{recurFilter} proposed a recursive filter which preserves the local mean to pre-process noisy measurements, but still used a non-linear function to map all noisy measurements to strictly positive values. \DIFadd{Chang et al. \cite{chang2016:sino-correct} applied the local linear minimum mean-square error (LLMMSE) filter to pre-process the raw measurements, but the LLMMSE filter does not guarantee positivity in its output sinograms and introduces correlations among neighbouring channels. This correlation violates the assumption of independence of sinogram data on which MAP reconstruction formulations rely. Chang et al. \cite{chang2016:sino-correct} also proposed a pointwise Bayesian restoration (PBR) approach, which better preserves the independence of sinogram data while reducing bias for photon-starved CT data.}
When pre-processing a large percentage of non-positive values for LDCT measurements, these pre-correction methods \DIFadd{may still} introduce bias in the reconstructed image and can degrade image quality~\cite{pre-post-log, recurFilter}. The logarithmic transformation itself causes a positive bias in the line integrals from which the image is reconstructed~\cite{pre-post-log, fessler1995hybrid}. 
A typical method for reconstructing images from the post-log data is penalized weighted least squares (PWLS) \cite{PWLS06} that optimizes an objective consisting of a weighted least squares data fidelity term and a regularization penalty. \DIFadd{However, the} pre-correction process and non-linear logarithmic operation create challenges in estimating the statistical weights for the PWLS methods \DIFadd{\cite{recurFilter,hayes2019unbiased}}.

Contrary to the post-log methods, the pre-log methods work directly with the raw measurements. A robust statistical model for the pre-log raw CT measurements is the shifted-Poisson (SP) model. This model shifts the measurements by the variance of the electric readout noise. The shifted measurement has its variance equal to its mean, so that it could be approximated to be Poisson distributed. Since the shifted-Poisson model is a better approximation for CT measurement statistics compared to the Gaussian model \cite{pre-post-log}, \DIFadd{\cite{PL-smooth, PL-restore, ElecNoiseModeling,tilley2017penalized}}, and no pre-correction of the data is needed for most LDCT dose levels \cite{pre-post-log}, this paper uses this SP model for LDCT image reconstruction.

There has been growing interest in improving CT image reconstruction by extracting prior knowledge from previous patient scans.  
Many methods have been proposed in this regard, such as prior image constrained compressed sensing methods (PICCS) \cite{PICCS2008, ncpiccs2011, piccs-app2012}, or the previous normal-dose scan induced nonlocal means method \cite{ndiNLM, zhang2017applications}. 
More recently, inspired by the success of learning-based methods in image processing and computer vision, researchers have incorporated data-driven approaches along with statistical models for LDCT image reconstruction. 
One such approach proposed by Xu et al. \cite{xu:12:ldx} combined dictionary learning techniques with the PWLS method for LDCT image reconstruction. 
The dictionary they used was either pre-learned from a training image set (consisting of 2D images) and fixed during reconstruction, or adaptively learned while reconstructing the image. The 2D dictionary model for image patches was later extended to a 2.5D dictionary (where different dictionaries were trained from 2D image patches extracted from axial, sagittal, and coronal planes of 3D data) \cite{2.5D}, and then to a 3D dictionary trained from 3D image patches \cite{3D-dict-18}.
These dictionary learning and reconstruction methods are typically computationally expensive, because they involve repeatedly optimizing \DIFadd{NP-hard problems~\cite{NPhard02}} for estimating the sparse coefficients of patches.
%are  when estimating the sparse coefficients of training signals. 
The learning of sparsifying transforms (ST) was proposed in recent works \cite{STlearning13, STlearning15} as a generalized analysis dictionary learning method, where the sparse coefficients are estimated directly by simple and efficient thresholding.
Pre-learned square sparsifying transforms have been recently incorporated into 2D LDCT image reconstruction with both post-log Gaussian statistics~\DIFadd{\cite{pwls-ultra2018}} and pre-log SP measurement models \cite{ye:17:asm}.
\DIFadd{Especially, Zheng et al.~}\cite{pwls-ultra2018} showed promise for PWLS with a union of pre-learned sparsifying transforms \cite{wen:14:sos} regularization that generalizes the square sparsifying transform approach.

In addition to the dictionary learning-based approaches, some works have incorporated neural networks in CT image reconstruction. Adler and {\"O}ktem proposed a learned primal-dual reconstruction method \cite{adler2018primaldual}, that uses convolutional neural networks (CNNs) to learn parameterized proximal operators. This method was applied to relatively simple 2D phantoms. 
Wu et. al \cite{wu2017KSAE} proposed a k-sparse autoencoder (KSAE) based regularizer for LDCT image reconstruction, where they trained three independent KSAEs from axial, sagittal and coronal slices for 3D reconstruction via artificial neural networks. Chen et al. \cite{chen2018learn} proposed to unfold the classical iterative reconstruction procedure into a CNN-based recurrent residual network so that the original fixed regularizers and the balancing parameters within the iterative scheme can vary for each layer. The reconstruction with this network was only performed slice by slice. \DIFadd{He et al. proposed a parameterized plug-and-play alternating direction method (3pADMM) for PWLS model based low-dose CT image reconstruction \cite{3pADMM}. By regarding the ADMM optimization steps as network modules, this method can optimize the 3p prior and the related parameters simultaneously.} These methods are fully supervised learning methods requiring large datasets consisting of both undersampled images or measurements and the corresponding high-quality images. Some post-processing approaches involving neural networks such as a U-Net or a residual net also improve CT image quality \cite{han2016deep,WavResNet18}, but such post-processing methods usually construct an image-to-image mapping without fully incorporating the physics of the imaging process. \DIFadd{Additionally, the generalization of supervised learning methods may be limited} in the sense that the trained model may only work well on the data that is similar to the training set.
\vspace{-0.1in}
\subsection{Contributions}
Considering the robustness and accuracy offered by the SP statistics, %statistics, 
and inspired by the data-driven image modeling methods not requiring paired training data or previous registered normal-dose images, here we propose a new LDCT image reconstruction method named SPULTRA that combines robust SP measurement modeling with a union of learned sparsifying transforms (ULTRA) based regularizer. 
Since the SP model leads to a nonconvex data-fidelity term, we design a series of quadratic surrogate functions for this term in our optimization.
For each surrogate function combined with the ULTRA regularizer (a majorizer of the SPULTRA objective), we optimize it by alternating between an \emph{image update step} and a \emph{sparse coding and clustering step}. \DIFadd{The proposed SPULTRA scheme is proved to converge to the critical points of the overall nonconvex problem. In the experiments, we compare SPULTRA with the recent PWLS-ULTRA scheme~\cite{pwls-ultra2018} under different incident photon intensity levels for 3D XCAT phantom simulations. The }results demonstrate that the proposed method avoids bias in image regions caused by the PWLS-ULTRA method, especially for low  X-ray doses. 
At the same time, SPULTRA achieves better image reconstruction quality than PWLS-ULTRA given the same number of iterations, or alternatively, SPULTRA achieves a desired image reconstruction quality much faster than the competing PWLS-ULTRA scheme, especially for low X-ray doses. \DIFadd{We verify the bias avoidance property of SPULTRA on a synthesized 3D clinical chest scan, and an ultra low-dose 2D shoulder phantom scan simulated from standard-dose raw measurements that also involve beam-hardening effects. We compared SPULTRA with a recent deep-learning based denoising framework~\cite{WavResNet18} on the 2D data demonstrating the better reconstruction and generalization ability of SPULTRA.}

This paper significantly extends our previous conference work~\cite{ye:17:asm} by incorporating the ULTRA regularizer and proposing a faster optimization procedure \DIFadd{with convergence guarantees. We performed extensive numerical evaluations compared to the 2D LDCT XCAT phantom results in~\cite{ye:17:asm}.}

\vspace{-0.1in}
\subsection{Organization}
The rest of this paper is organized as follows. Section~\ref{sec:formulation} presents the proposed problem formulation for low-dose CT image reconstruction. Section~\ref{sec:algorithm} briefly reviews the ULTRA learning method and describes the proposed SPULTRA image reconstruction algorithm. Section~\DIFadd{\ref{sec:convergenceAnalysis} discusses the convergence properties of the SPULTRA methodology. Section~}\ref{sec:experiments} presents detailed experimental results and comparisons. Section~\ref{sec:conclusion} presents conclusions.

%% file: Algorithm_r2.tex
\vspace{-0.1in}
\section{Algorithm}\label{sec:algorithm}%\vspace{-0.01in}
The proposed SPULTRA algorithm is based on a pre-learned union of sparsifying transforms. The process of learning such a union of transforms from a dataset of image patches has been detailed in~\cite{pwls-ultra2018}. The learning problem in \cite{pwls-ultra2018} simultaneously groups the training patches into $K$ \DIFadd{classes }and learns a transform in each \DIFadd{class }along with the sparse coefficients (in the transform domain) of the patches. This learning is accomplished by an alternating algorithm (see~\cite{pwls-ultra2018}). 
This section focuses on describing the algorithm in the reconstruction stage for SPULTRA, i.e., for \eqref{eq:P0}.

The data-fidelity term $\mathsf{L}(\x)$ in \eqref{eq:P0} is nonconvex when the electronic noise variance $\sigma^2$ is nonzero. It is challenging to directly optimize such a logarithmic nonconvex function. We propose to iteratively design quadratic surrogate functions for this data-fidelity term $\mathsf{L}(\x)$. In each iteration, we optimize the surrogate function that is a quadratic data-fidelity term together with the ULTRA regularizer using alternating minimization that alternates between an image update step and a sparse coding and clustering step that has closed-form solution~\cite{wen:14:sos}. We use the relaxed OS-LALM algorithm for the image update step~\cite{nien:16:rla}. We perform only one alternation between the two steps for each designed surrogate function, which saves runtime and works well in practice.
\vspace{-0.15in}
\subsection{Surrogate function design} \vspace{-0.03in}
We design a series of quadratic surrogate functions for $\mathsf{L}(\x)$ as follows:\\
\begin{equation}\label{eq:surrogate}
\begin{aligned}
\phi(\x;\x^n) &= \mathsf{L}(\x^n) + \bm{d}_h(l^n)\A(\x - \x^n)\\
&\qquad+\frac{1}{2}(\x - \x^n)^T\A^T\W^n\A(\x - \x^n),			
\end{aligned}
\end{equation}
where $(\cdot)^n$ denotes values at the $n$th iteration and ${\bm{d}_h(l^n) \in \mathbb{R}^{N_d}}$ is a row vector capturing the gradient information and is defined as $\bm{d}_h(l^n) \triangleq [\dot{h_i}(l_i^n)]_{i = 1}^{N_d}$. The curvatures of the $n$th updated parabola (surrogate) are described by $\W^n\triangleq \text{diag}\{c_i(l^n_i)\}$. In this paper, we use the optimum curvatures \cite{TMI99} that are defined as follows:
\begin{equation} \label{eq:OptCuv}	
c_i(l^n_i) =  \begin{cases}
\big[2\frac{h_i(0) - h_i(l_i^n) + (l_i^n)\dot{h_i}(l_i^n)}{(l_i^n)^2}\big]_+, &l_i^n > 0\\
\big[\ddot{h_i}(0)\big]_+, &l_i^n = 0, 
\end{cases}
\end{equation}
where $\ddot{h}_i$ is the second-order derivative, and operator $[\cdot]_+$ sets the non-positive values to zero. In practice, we replace negative values with a small positive number so that the diagonal matrix $\W^n$ is invertible. 
Due to numerical precision, \eqref{eq:OptCuv} might become extremely large when $l_i^n$ is nonzero but small. To avoid this problem, we use an upper bound of the maximum second derivative $\big[\ddot{h_i}(0)\big]_+$ for the curvature $c_i(l^n_i)$ when $l_i^n > 0$ \cite{TMI99}.

By ignoring the terms irrelevant to $\x$ in \eqref{eq:surrogate}, we get the following equivalent form of $\phi(\x;\x^n)$:
\begin{equation}\label{deduce2}
\phi(\x;\x^n) \equiv \frac{1}{2}||\tilde{\y}^n - \A\x||_{\W^n}^2,
\end{equation}	
where ``$\equiv$'' means equal to within irrelevant constants of $\x$, and $\tilde{\y}^n \triangleq \A\x^n -\big(\W^n\big)^{-1}[\bm{d}_h(l^n)]^T$. The overall surrogate function at the $n$th iteration for the penalized-likelihood objective function \DIFadd{$G(\x)$} in \eqref{eq:P0} is then
\begin{equation}\label{eq:surr_all}
\Phi(\x;\x^n) = \frac{1}{2}||\tilde{\y}^n - \A\x||_{\W^n}^2 + \R(\x),\ \DIFadd{\text{s.t.}\ \x \in \mathcal{X}.}
\end{equation}
We \DIFadd{descend} the surrogate function $\Phi(\x;\x^n)$ in \eqref{eq:surr_all} by alternating \DIFadd{once} between an image update step, and a sparse coding and clustering step.
\vspace{-0.2in}
\subsection{Image Update Step}\vspace{-0.05in}
In the image update step, we update the image $\x$ with fixed sparse codes $\{\z_j\}$ and \DIFadd{class} assignments $\{C_k\}$. The relevant part of the majorizer for this step is 
\begin{equation}\label{eq:Phi1}
 \begin{aligned}
 \Phi_1(\x;\x^n) = \phi(\x;\x^n)  + \beta \sum_{k=1}^{K} \sum_{j\in C_k} \tau_j \|\omg_k \P_j \x - \z_j\|^2_2
 \end{aligned}
 \end{equation}
\DIFadd{Although we have a box} constraint on $\x$\DIFadd{, i.e., $\x \in \mathcal{X}$, in practice, the upper bound $x_{\mathrm{max}}$ can be set high such that it will not be active. We applied the }relaxed OS-LALM algorithm~\cite{nien:16:rla} \DIFadd{to minimize }\eqref{eq:Phi1} \DIFadd{with the constraint. This algorithm is } shown in Algorithm \ref{alg: spultra-alg} (steps 7-10). The OS-LALM method uses majorizing matrices. In particular, the matrix $\A^T\W^n\A$ is majorized by \DIFadd{$\D_{\A} \triangleq \text{diag}\{\A^T\W^n \A \mathbf{1}\}$, where $\mathbf{1}$ denotes a vector of ones.} Denoting the regularization term in \eqref{eq:Phi1} as $\R_2(\x)$, its gradient is 
\begin{equation}\label{eq:grad_r2}
\nabla \R_2(\x) = 2\beta\sum_{k=1}^{K} \sum_{j\in C_k} \tau_j \P_j^T \omg_{k}^T (\omg_k \P_j \x - \z_j ).
\vspace{-0.05in}
\end{equation}
The Hessian of $\R_2(\x)$ is majorized by the following diagonal matrix:\\
\begin{equation}\label{eq: D_R}
%\begin{aligned}
\D_{\R} \triangleq 2\beta \bigg\{\max_k \|\omg_{k}^T\omg_{k}\|_2\bigg\}\sum_{k=1}^{K} \sum_{j\in C_k}\tau_j \P^T_j \P_j .
%\end{aligned}
\end{equation}
The (over-)relaxation parameter $\alpha \in [1,2)$ and the parameter $\rho_t >0$ decreases with iterations $t$ in OS-LALM according to the following equation \cite{nien:16:rla}:
\begin{equation}\label{eq:rho}		
\rho_t(\alpha)= \begin{cases}
1,      &  t=0\\ 
\frac{\pi}{\alpha(t+1)}\sqrt{1-\big(\frac{\pi}{2\alpha(t+1)}\big)^2},     &  \text{otherwise.}
\end{cases}
\vspace{-0.1in}
\end{equation}	
\vspace{-0.1in}
\subsection{Sparse Coding and Clustering Step}
Here, with $\x$ fixed, we jointly update the sparse codes and the \DIFadd{class }memberships of patches. The relevant part of the cost function for the sparse coding and clustering step is 
\begin{equation}\label{eq:codes-clusters} 
\min_{\{\z_j, C_k\}} \sum_{k=1}^{K}  \bigg\{  \sum_{j\in C_k} \tau_j \{\|\omg_k\P_j \x - \z_{j}\|^2_2 + \gamma_c^2\|\z_j\|_0 \}\bigg\} .
\end{equation}
\DIFadd{Problem \eqref{eq:codes-clusters} is separable in terms of the patches, so each patch is clustered and sparse coded independently in parallel. The optimal sparse code $\z_j$ in \eqref{eq:codes-clusters} is obtained by hard-thresholding, i.e., ${\z_j= \mathit{H}_{\gamma_c}(\omg_{k_j}\P_j\x)}$, where ${\mathit{H}_{\gamma_c}(\cdot)}$ represents a vector hard-thresholding operator that zeros out elements whose magnitudes are smaller than $\gamma_c$, and leaves other entries unchanged. Then the optimized class $\hat{k}_j$ for the $j$th patch is computed as follows~\cite{pwls-ultra2018}:}
\begin{equation}
\label{eq:clustering}
 \small{\hat{k}_j = \argmin_{1\leq k \leq K} || \omg_{k} \P_j \x - \mathit{H}_{\gamma_c}(\omg_{k}\P_j\x)||^2_2 + \gamma_c^2\|\mathit{H}_{\gamma_c}(\omg_{k}\P_j\x)\|_0. }
\end{equation} 
\DIFadd{We compute the cost values on the right hand side of \eqref{eq:clustering} for each $k = 1, \cdots, K$, and determine the $\hat{k}_j \in \{1, \cdots, K\}$ that gives the minimal cost value, i.e., patch $\P_j \x$ is grouped with the transform that provides the smallest value of the cost in \eqref{eq:clustering}. Then, the corresponding optimal sparse code is ${\hat{\z}_j = \mathit{H}_{\gamma_c}(\omg_{\hat{k}_j}\P_j\x)}$.}

\textbf{Algorithm} \ref{alg: spultra-alg} illustrates the proposed optimization algorithm for Problem \eqref{eq:P0}.
\begin{algorithm}[!htp]  
	\caption{SPULTRA Algorithm}
	\label{alg: spultra-alg}
	\begin{algorithmic}[1]
		\Require~~\\
		initial image $\hat{\x}^{0}$; $\alpha = 1.999$ ; $\rho_0 = 1$; \\
		pre-computed $\D_\R $ according to \eqref{eq: D_R};\\
		number of outer iterations $N$, number of inner iterations $P$, and number of ordered-subsets $M$.	
		\Ensure reconstructed image $\hat{\x}^N$. 
		\For {$n=0,1,2,\cdots, N-1$}	
		\State \textbf{(1) Image Update}: Fix $\hat{\z}_j^{n}$ and $\hat{C}_k^n$; \\
		\textit{Initializations: 
		}\begin{enumerate}
			\item $\x^{(0)} =  \hat{\x}^{n}$,
			\item Determine $c_i(l^n_i)$ according to \eqref{eq:OptCuv},
			\item $\W^n = \diag\{c_i(l^n_i)\}$,
			\item $\D_{\A} \triangleq \diag\{\DIFadd{\A^T\W^n\A\mathbf{1}}\}$,			 				  		 
			\item \DIFadd{$\bm{d}_h(l^n) = [I_0e^{-f_i(l^n_i)}\dot{f_i}(l_i^n)(\frac{Y_i}{I_0e^{-\DIFadd{f_i(l_i^n)}} + \sigma^2} - 1)]_{i = 1}^{N_d}$,	 
			}\item $\tilde{\y}^n = \A\x^{(0)} -{(\W^n)}^{-1}[\bm{d}_h(l^n)]^T$,
			\item $\ze^{(0)} = \g^{(0)} = M\A_M^T\W^n_M(\A_M\x^{(0)}-\tilde{\y}_M^n) $,
			\item $\bm{\eta}^{(0)} = \D_\A \x^{(0)} - \ze^{(0)}$, 
			\item compute $\nabla \R_2(\x)$ according to \eqref{eq:grad_r2}.
		\end{enumerate}
		\For {$p=0,1,2,3,\cdots,P-1$}	
		\For {$m=0,1,2,3,\cdots,M-1$}	
		\begin{equation*}\label{eq:rlalm}
		\begin{aligned}
		&t ~= ~pM + ~m;\\
		&\hspace{-0.1in}\left\{		
		\begin{aligned}
		\s^{(t+1)} &= \rho_t(\D_\A \x^{(t)} -\bm{\eta}^{(t)}) + (1-\rho_t)\g^{(t)} \\
		\x^{(t+1)} &= [\x^{(t)} - (\rho_t\D_\A+\D_\R)^{-1}(\s^{(t+1)} +\nabla \R_2(\x^{(t)}))]_\mathcal{C} \\
		\ze^{(t+1)}& \triangleq M\A_m^T\W^n_m(\A_m\x^{(t+1)}-\tilde{\y}_m^n)   \\
		\g^{(t+1)} &= \frac{\rho_t}{\rho_t+1}(\alpha \ze^{(t+1)} + (1-\alpha)\g^{(t)}) +  \frac{1}{\rho_t+1}\g^{(t)}\\
		\bm{\eta}^{(t+1)} &= \alpha(\D_{\A} \x^{(t+1)} -\ze^{(t+1)}) + (1-\alpha)\bm{\eta}^{(t)}	
		\end{aligned}
		\right.\\	
		&\text{Decrease $\rho_t$ according to \eqref{eq:rho}};
		\end{aligned}
		\end{equation*}  
		\EndFor	
		\EndFor				  	
		\State  $\hat{\x}^{n+1} = \x^{(t+1)}$;
		\State \textbf{(2) Sparse Coding and Clustering}: Fix $\hat{\x}^{n+1}$, compute class assignments $\hat{k}_j^{n+1}$ using \eqref{eq:clustering}, and sparse codes  $\hat{\z}_{j}^{n+1} =  H_{\DIFadd{\gamma_c}}(\omg_{\hat{k}_j^{n+1}} \P_j \hat{\x}^{n+1}),\   \forall \ j$. 
		\EndFor		
	\end{algorithmic} 
\end{algorithm}	

\vspace{-0.13in}
\subsection{Computational Cost} \label{sec:computation analysis} \vspace{-0.02in}
The SPULTRA algorithm has a similar structure in each iteration as the recent PWLS-ULTRA \cite{pwls-ultra2018}, except for several initializations in the image update step. Since forward and backward projections are used to compute $\D_\A$ and $\tilde{\y}^n$ during initialization, the image update step of SPULTRA is slightly slower than PWLS-ULTRA. In our experiments, we observed that the initializations took around $20 \%$ of the runtime in each outer iteration. However, in practice, especially for low doses, SPULTRA reconstructs images better than PWLS-ULTRA for a given number of outer iterations. Or alternatively, SPULTRA takes much fewer outer iterations (and runtime) to achieve the same image reconstruction quality as PWLS-ULTRA. These results are detailed in Sec. \ref{sec:experiments}.

%% file: Experiments_r2.tex
\section{Experimental Results}\label{sec:experiments} \vspace{-0.02in}
Here we present numerical experiments demonstrating the behavior of SPULTRA. We evaluated the proposed SPULTRA method on the \DIFadd{3D XCAT phantom \cite{segars:08:rcs} and synthesized clinical data at multiply X-ray doses, as well as an ultra low-dose 2D shoulder phantom scan simulated from real raw data,} and compared its performance with that of the state-of-the-art PWLS-ULTRA \cite{pwls-ultra2018}. We computed the root mean square error (RMSE) and structural similarity index (SSIM) \cite{xu:12:ldx,ssim2} of \DIFadd{XCAT} images reconstructed by various methods in a region of interest (ROI). The RMSE is defined as $\sqrt{\sum_{i\in \mathrm{ROI}}(\hat{x}_i - x_i^*)^2/N_{p, ROI}}$, where $N_{p,ROI}$ is the number of pixels in the ROI, $\hat{\x}$ is the reconstructed image, and $\x^*$ is the ground-truth image. We also compared to PWLS reconstruction with an edge-preserving regularizer (PWLS-EP) ${\R(\x) = \sum_{j  =1}^{N_p} \sum_{k\in N_{j}}\kappa_{j} \kappa_{k} \varphi(x_j - x_k)}$, where $N_j$ represents the neighborhood of the $j$th pixel, $\kappa_j$ and $\kappa_k$ are elements of $\bm{\kappa}$ that encourages resolution uniformity \cite{kappa}. \DIFadd{The potential function for 3D reconstruction was ${\varphi (t) = \delta^2(|t/\delta| - \log(1+|t/\delta|))}$ with\footnote{``HU" used in this paper is the shifted Hounsfield unit, where air is 0 HU and water is 1000 HU.} ${\delta = 10\text{ HU}}$, and that for 2D shoulder phantom simulations was
	${\varphi (t) = \delta^2(\sqrt{1+|t/\delta|^2}-1)}$ with ${\delta = 100 \text{ HU}}$. The results obtained by PWLS-EP were taken as initial images for other methods we compared with in this section. }	

The SPULTRA method shifts uncorrected pre-log data by the variance of electronic noise. Such un-preprocessed pre-log data and the variance of the electronic noise on a CT scanner are proprietary to CT vendors, especially for LDCT. In our experiments \DIFadd{of XCAT phantom simulations and the synthesized clinical data}, we generated pre-log data $\hat{\y}$ from the XCAT phantom as well as from a clinical image $\tilde{\x}$ reconstructed by the PWLS-ULTRA method as follows:
\begin{equation}
\hat{\y}_i = \text{Poisson} \{I_0 e^{-[\A\tilde{\x}]_i}\} + \mathcal{N}\{0, \sigma^2\},
\end{equation}
where $\mathcal{N}\{\mu, \sigma^2\}$ denotes a Gaussian distribution with mean $\mu$ and variance $\sigma^2$. \DIFadd{We refer to the image $\tilde{\x}$ used for generating the synthesized clinical data as the \textit{``true'' clinical image}.	
	We also simulated an ultra low-dose scan from raw (pre-log) measurements of a standard-dose scan of a 2D shoulder phantom as:
	\vspace{-0.1in}
	\begin{equation}\label{eq:shoulder_gen}
	\hat{\y}_i = \text{Poisson} \{\frac{1}{\alpha}\y_{i_{s}}\} + \mathcal{N}\{0, \sigma^2\},
	\end{equation}
	where $\alpha$ is a scale factor we used to lower the dose from standard-dose measurements, and $\y_{i_{s}}$ denotes the raw standard-dose measurements. 
	We set $\sigma = 5$ for all the simulations, as suggested in prior works~\cite{pre-post-log, pwls-ultra2018}.
	We implemented the system model $\A$ via the separable footprint projector methods~\cite{long2010SF}.} \DIFadd{MATLAB code to reproduce the results in this work \DIFadd{is} released at \url{http://web.eecs.umich.edu/~fessler/}. Some additional results are included in the supplement. }

\vspace{-0.11in}
\subsection{XCAT phantom results} 
\subsubsection{Framework}
We pre-learned a union of 15 square transforms from $8 \times 8 \times 8$ overlapping patches extracted from a $420\times 420\times 54$ XCAT phantom with a patch stride $2 \times 2 \times 2$. These transforms were initialized during training \cite{pwls-ultra2018} with 3D DCT, and the clusters were initialized randomly. 
\begin{figure}[!htp]\vspace{-0.1in}
	\centering
	\includegraphics[width = 0.4\textwidth]{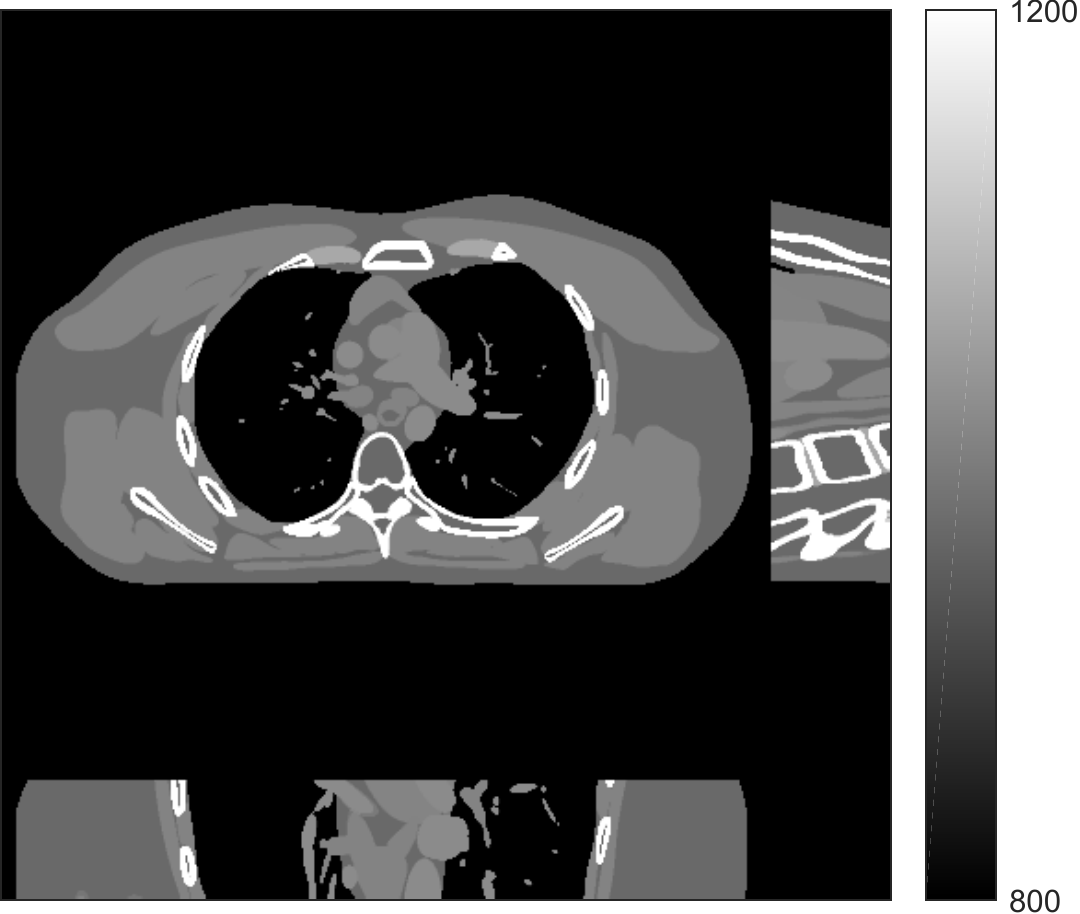}
	\caption{\DIFadd{Reconstruction targeted ROI of the true XCAT phantom displayed with central slices along the axial, sagittal and coronal directions. The display window is [800, 1200] HU.}}
	\label{fig:3d-trueimg}
	\vspace{-0.15in}
\end{figure}
We simulated 3D axial cone-beam scans using a $840\times 840\times 96$ XCAT phantom with $\Delta_x = \Delta_y = 0.4883$ mm and $\Delta_z = 0.625$ mm. We generated sinograms of size $888\times 64 \times 984$ using GE LightSpeed cone-beam geometry corresponding to a mono-energetic source with $I_0 = \DIFadd{1\times}10^4$, $5\times 10^3$, $3\times 10^3$, and ${2\times 10^3}$ incident photons per ray and no scatter, respectively. Tab.~\ref{tab:non-pos-perc} shows percentages of non-positive measurements under different dose levels. We set these non-positive measurements to $1\times 10^{-5}$ for generating the post-log sinogram that PWLS-based methods rely on \cite{pre-post-log}. We reconstructed the 3D volume with a size of $420\times 420\times 96$ at a coarser resolution of $\Delta_x = \Delta_y = 0.9766$ mm and $\Delta_z = 0.625$ mm. The patch size during reconstruction was $8\times 8 \times 8$ and the stride was $3\times 3 \times 3$. For evaluating reconstruction performance, \DIFadd{we chose an ROI that was composed of the central 64 out of 96 axial slices, and refer to it as the \textit{reconstruction targeted ROI}}. Fig.~\ref{fig:3d-trueimg} shows the central slices of the true XCAT phantom \DIFadd{inside this ROI }along three directions. In the reconstruction stage of PWLS-ULTRA and SPULTRA, we used 4 iterations for the image update step, i.e., $P = 4$, for a good trade-off between algorithms' convergence and computational costs. We used $12$ ordered subsets, i.e., $M =12$, to speed up the algorithm. The initial image for the ULTRA methods was reconstructed by PWLS-EP, whose regularization parameter was set empirically to ensure good reconstruction quality as $\beta_{ep} = 2^{13}$ for all the experimented dose cases. \DIFadd{We used an analytical filtered back-projection (FBP) method FDK \cite{feldkamp1984practical} to initialize PWLS-EP. The FDK images of XCAT phantom for all the dose levels are shown in the supplement.}
Due to the fact that SPULTRA has a similar cost function as PWLS-ULTRA in each outer iteration, we used the same parameter settings for both methods: $\beta = 4\times 10^4$ and $\gamma_c = 4\times 10^{-4}$, which we observed worked well for all the dose levels we tested.

\begin{table}[!htp]
	\centering
	\caption{Percentages of non-positive measurements under different dose levels for XCAT phantom simulations.}
	\label{tab:non-pos-perc}
	\begin{tabular}{c c c c c}
		\toprule
		{$I_0$} & $1\times 10^4$  & $5\times 10^3$  & $3\times 10^3$ & $2\times 10^3$ \\ 
		\midrule
		\begin{tabular}[c]{@{}l@{}}Non-positive\\ Percentage (\%)\end{tabular} &0.06  & 0.20 &0.48  & 0.96 \\ 
		\bottomrule
	\end{tabular} 
	\vspace{-0.1in}
\end{table}

\subsubsection{Behavior of the learned ULTRA Models}
The learned union of transforms contributes to the clustering and sparsification of image patches. To illustrate the behavior of the learned transforms, we selected 3 of the 15 transforms that capture important structures/features of the reconstructed image \DIFadd{(with $I_0 = 1\times 10^4$)} in their \DIFadd{classes}. 
\begin{figure*}[!tbhp]
	\centering
	\includegraphics[width=1\textwidth]{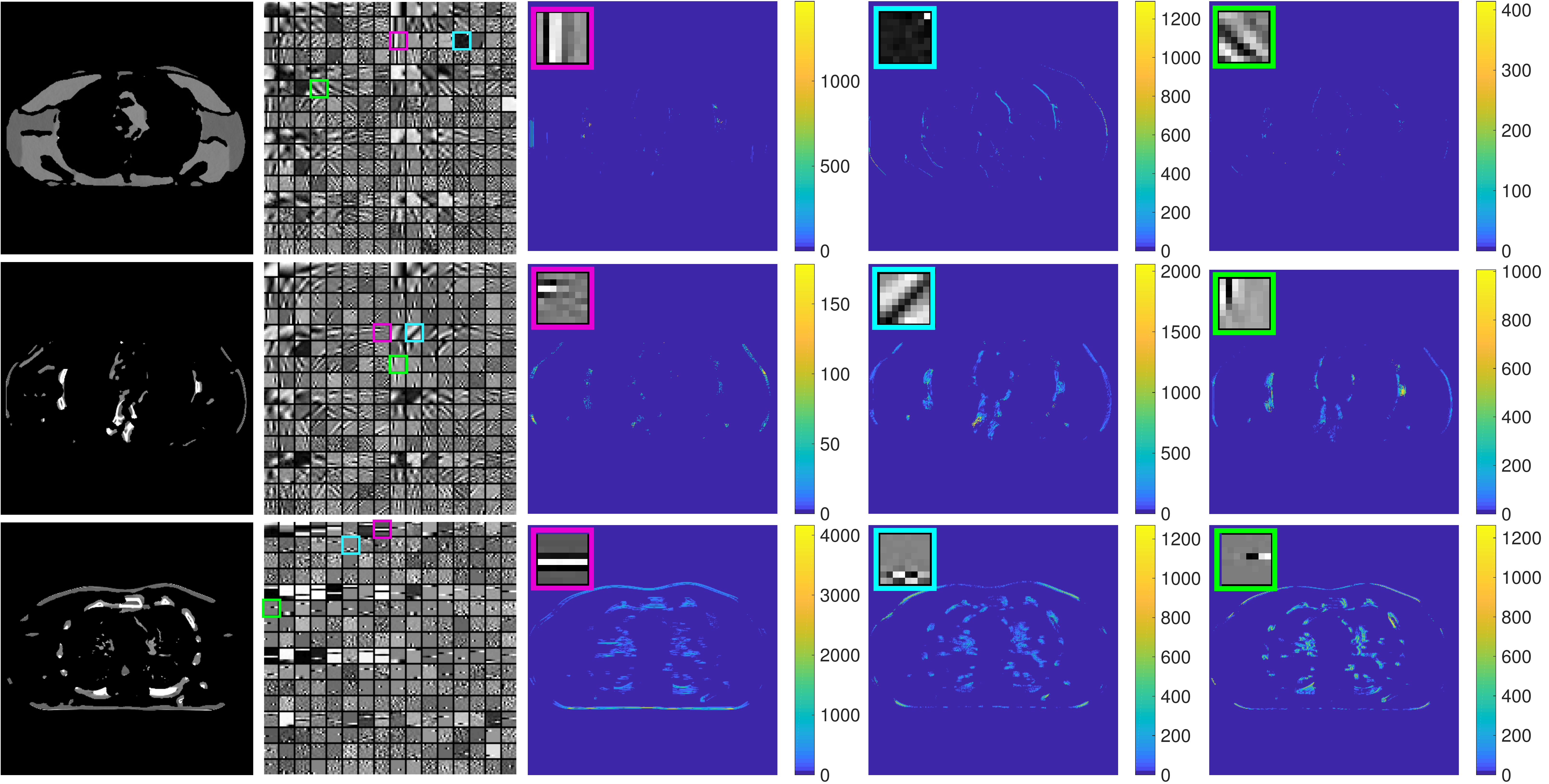}
	\caption{\DIFadd{The three rows correspond to the 1st, 13th, and 14th classes respectively. The first column displays three voxel-level clustered images of the central axial slice. Each of them is formed by image patches lie in the corresponding class. The second column displays part of the transforms for the corresponding classes. The third, fourth and fifth columns show the central axial slice of the sparse coefficient maps obtained by applying specific filters (shown in the top left corner) to patches belonging to the corresponding classes.}
		The patch stride for plotting these figures was $1\times 1\times 1$.}
	\label{fig:sparsecode}
\end{figure*}
\begin{table*}[ht]
	\centering
	\caption{\DIFadd{RMSE (HU) and SSIM of the reconstruction targeted ROI at various dose levels ($I_0$) using the PWLS-EP, PWLS-ULTRA, and SPULTRA methods for the XCAT phantom simulations.}} 
	%	\vspace{-0.1in}
	\label{tab:numerical}
	\begin{subtable}{0.45\textwidth}
		\centering
		\caption{RMSE (HU)}	
		\begin{tabular}{l c c c}
			\toprule
			$\quad I_0$	& PWLS-EP  & PWLS-ULTRA  & SPULTRA  \\ \midrule
			{$1\times 10^4$ } & 45.3 &29.1  & \textbf{28.9}  \\   \midrule
			{$5\times 10^3$ } &47.1  &33.3  & \textbf{32.8}  \\  \midrule
			{$3\times 10^3$ } &  49.7 & 37.7  &\textbf{36.4}   \\ \midrule
			{$2\times 10^3$} & 53.5  & 43.2 & \textbf{39.9}  \\    \bottomrule
		\end{tabular}
	\end{subtable}
	\hspace{0.2in}
	\begin{subtable}{0.45\textwidth}
		\centering
		\caption{SSIM}	
		\begin{tabular}{l c c c}
			\toprule
			$\quad I_0$	& PWLS-EP  & PWLS-ULTRA  & SPULTRA  \\ \midrule
			{$1\times 10^4$ } & 0.941	& 0.974  & \textbf{0.974}   \\   \midrule
			{$5\times 10^3$ } & 0.937 & 0.969 & \textbf{0.970}  \\  \midrule
			{$3\times 10^3$ } &0.927  &0.961   & \textbf{0.963}    \\ \midrule
			{$2\times 10^3$} &0.911   & 0.948 &  \textbf{0.956}  \\   \bottomrule
		\end{tabular}
	\end{subtable}
\end{table*}
Fig.~\ref{fig:sparsecode} (first column) shows three \DIFadd{voxel}-level \DIFadd{classes }(\DIFadd{voxels} are clustered by majority vote among patches overlapping them) for the reconstructed central axial slice. The top image only contains soft tissues, whereas the middle image shows some edges and bones in the vertical direction, and the bottom image captures some high-contrast structures. % along the horizontal direction. 
Fig.~\ref{fig:sparsecode} (second column) shows the transforms for the corresponding classes. Each learned transform has 512 $8\times 8 \times 8$ filters, and we show the first $8\times 8$ slice of 256 of these filters that show gradient-like and directional features.
\begin{figure*}[!htbp]
	\centering
	\begin{subfigure}[h]{0.245\textwidth}
		\centering	
		\includegraphics[width=1\textwidth]{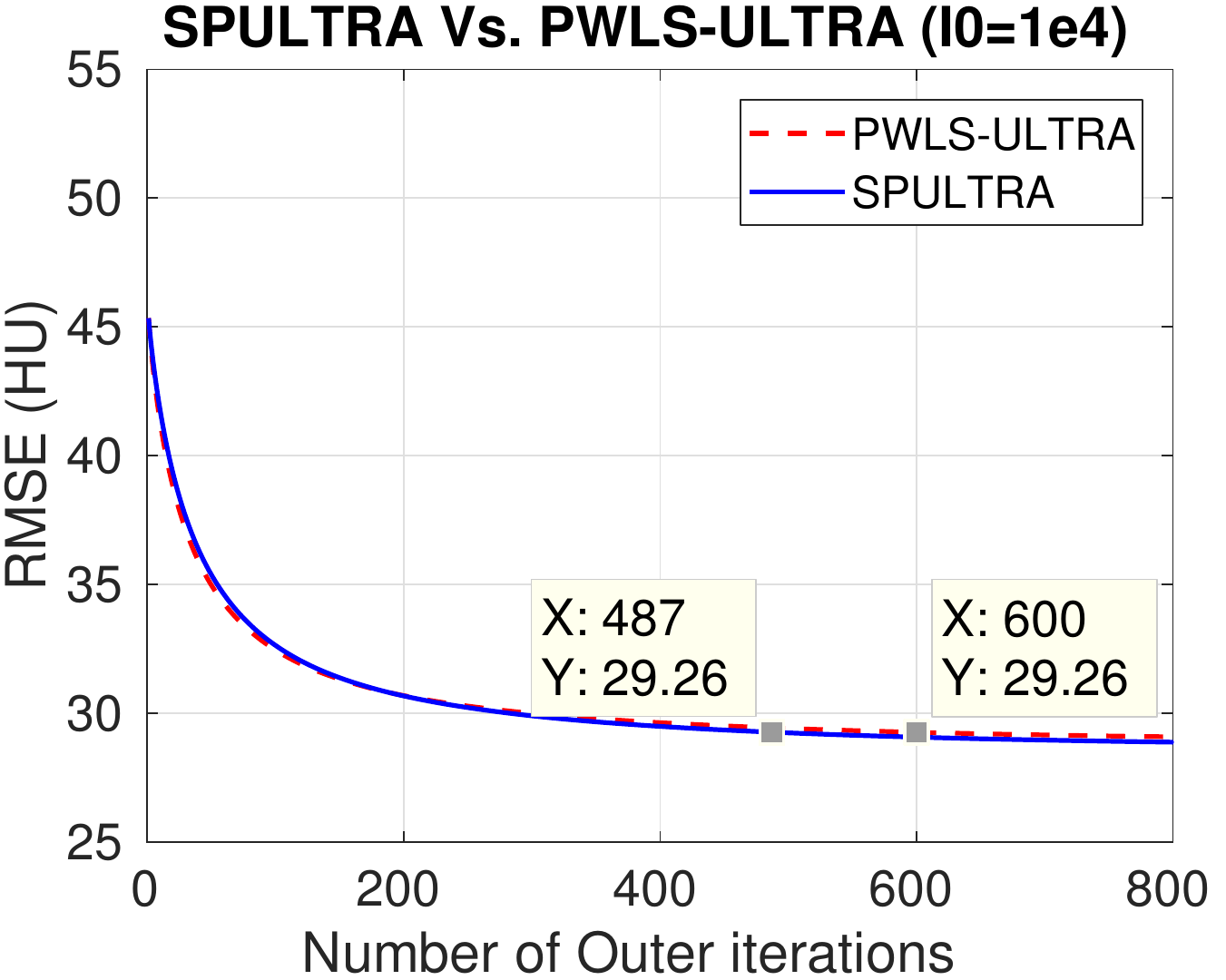}
		\caption{$I_0 = \DIFadd{1\times}10^4$}
		\label{fig:SPULTRA-rmse-1e4}
	\end{subfigure}
	\hfill \hspace{-0.2in}
	\begin{subfigure}[h]{0.245\textwidth}
		\centering	
		\includegraphics[width=1\textwidth]{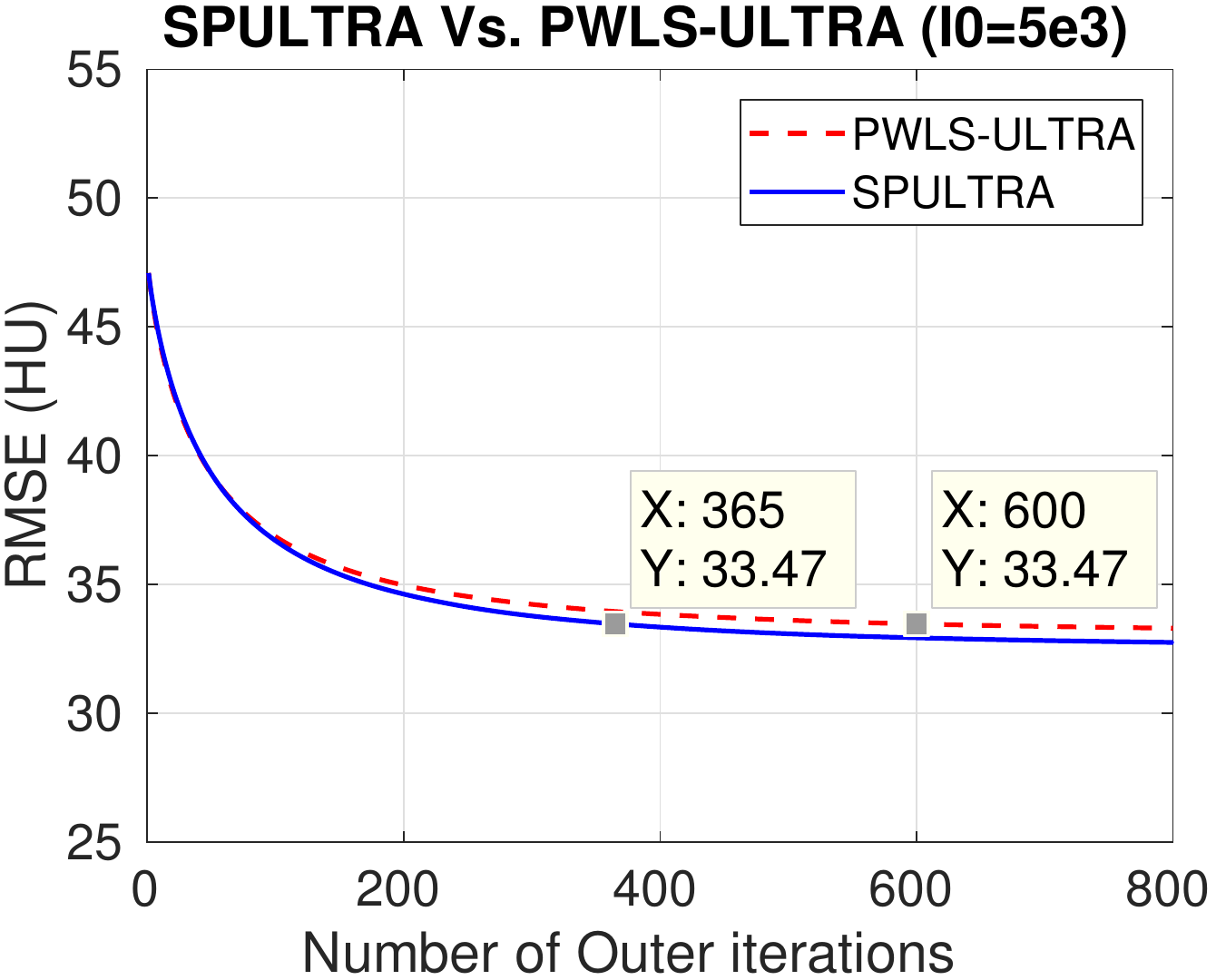}
		\caption{$I_0 = 5\times 10^3$}
		\label{fig:SPULTRA-rmse-5e3}
	\end{subfigure}
	%\hfill
	\hfill \hspace{-0.2in}
	\begin{subfigure}[h]{0.245\textwidth}
		\centering	
		\includegraphics[width=1\textwidth]{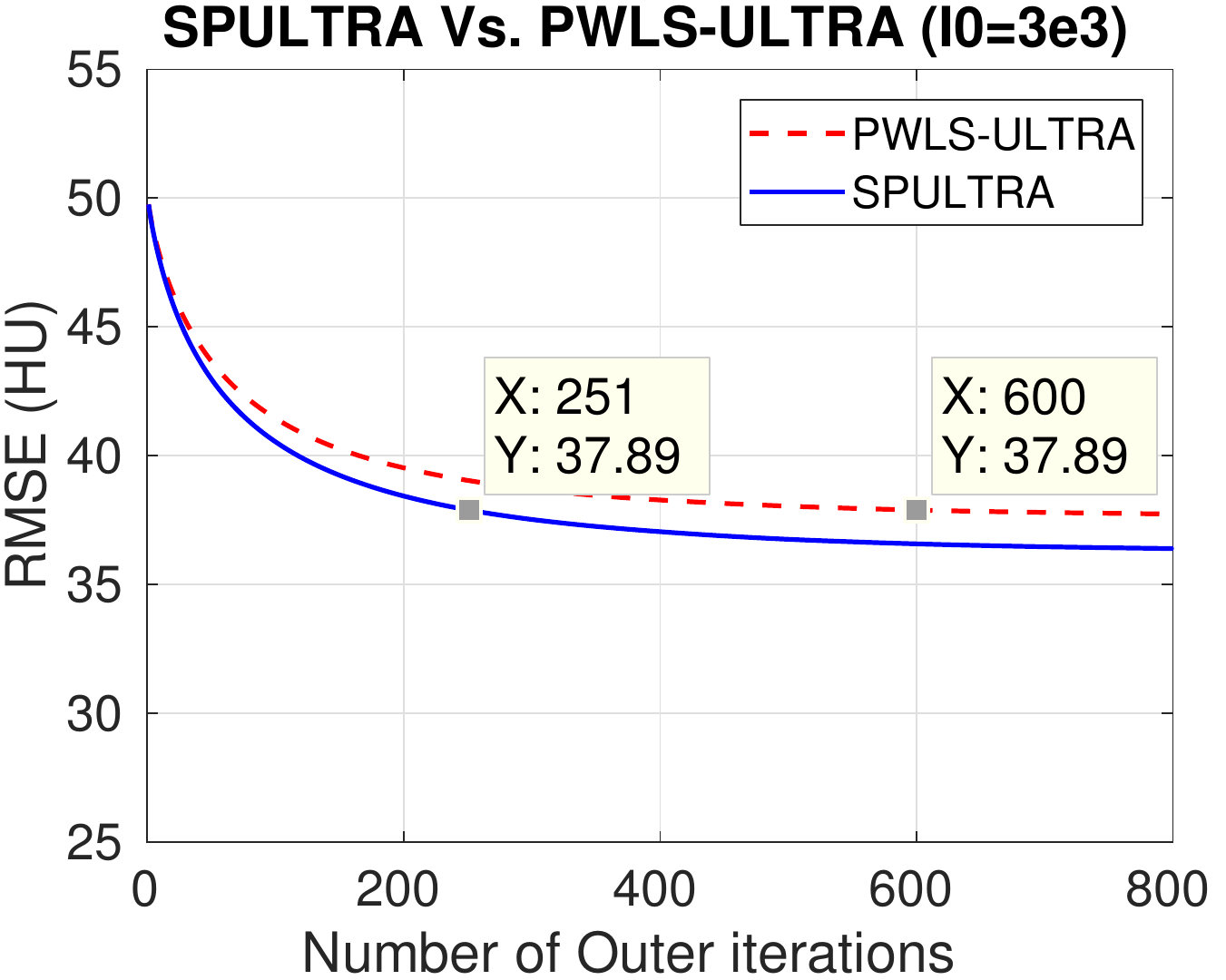}
		\caption{$I_0 = 3\times 10^3$}
		\label{fig:SPULTRA-rmse-3e3}
	\end{subfigure}
	\hfill \hspace{-0.2in}
	\begin{subfigure}[h]{0.245\textwidth}
		\centering	
		\includegraphics[width=1\textwidth]{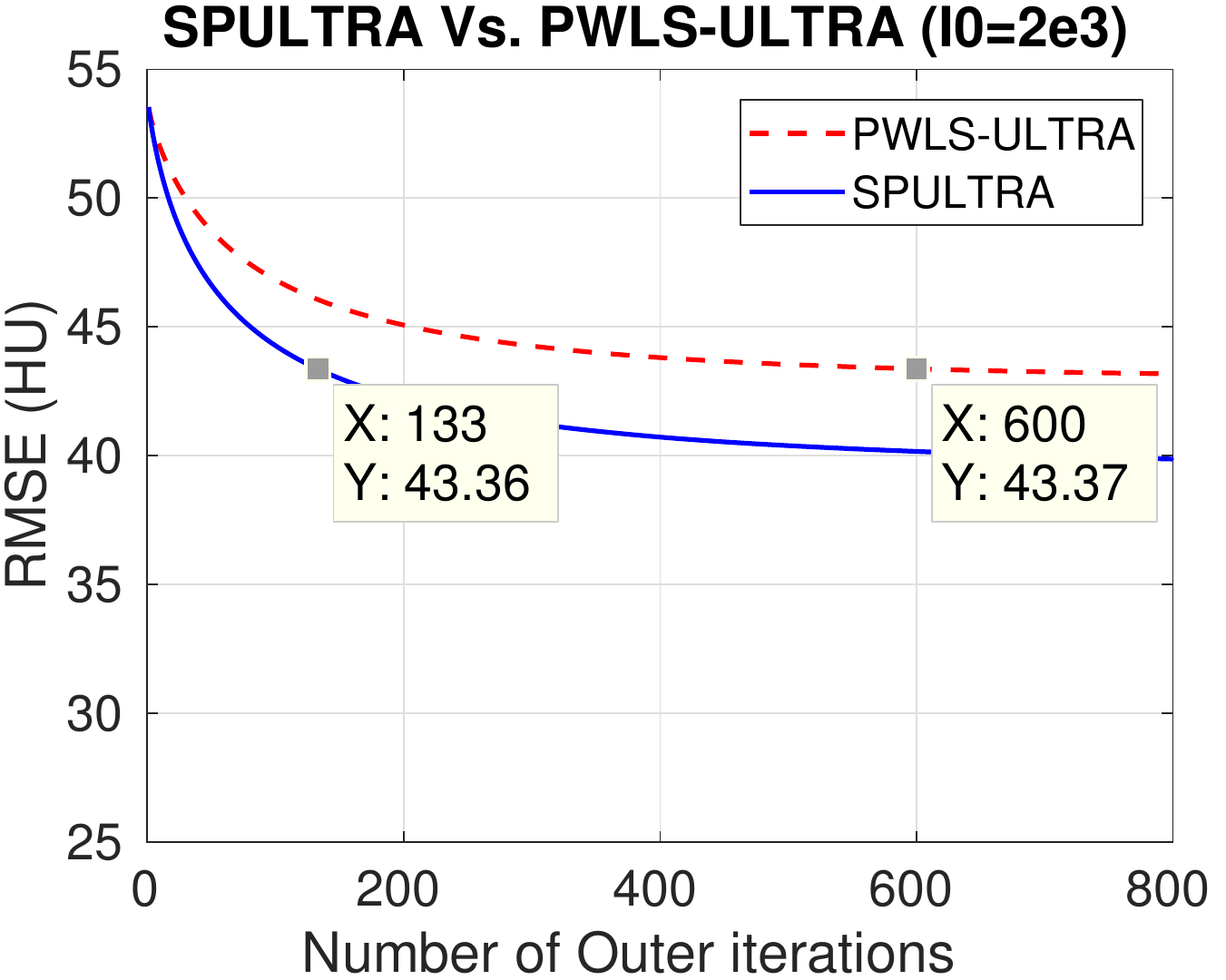}
		\caption{$I_0 = 2\times 10^3$}
		\label{fig:SPULTRA-rmse-2e3}
	\end{subfigure}
	\caption{RMSE comparison of SPULTRA and PWLS-ULTRA. The cursors indicate the RMSEs (Y) at specific number of outer iterations (X).}
	\label{fig:rmse_comp}
	\vspace{-0.16in}
\end{figure*} 
\begin{figure*}[!htp]
	\centering
	\begin{subfigure}[h]{1\textwidth}
		\centering	
		\includegraphics[width=0.9\textwidth]{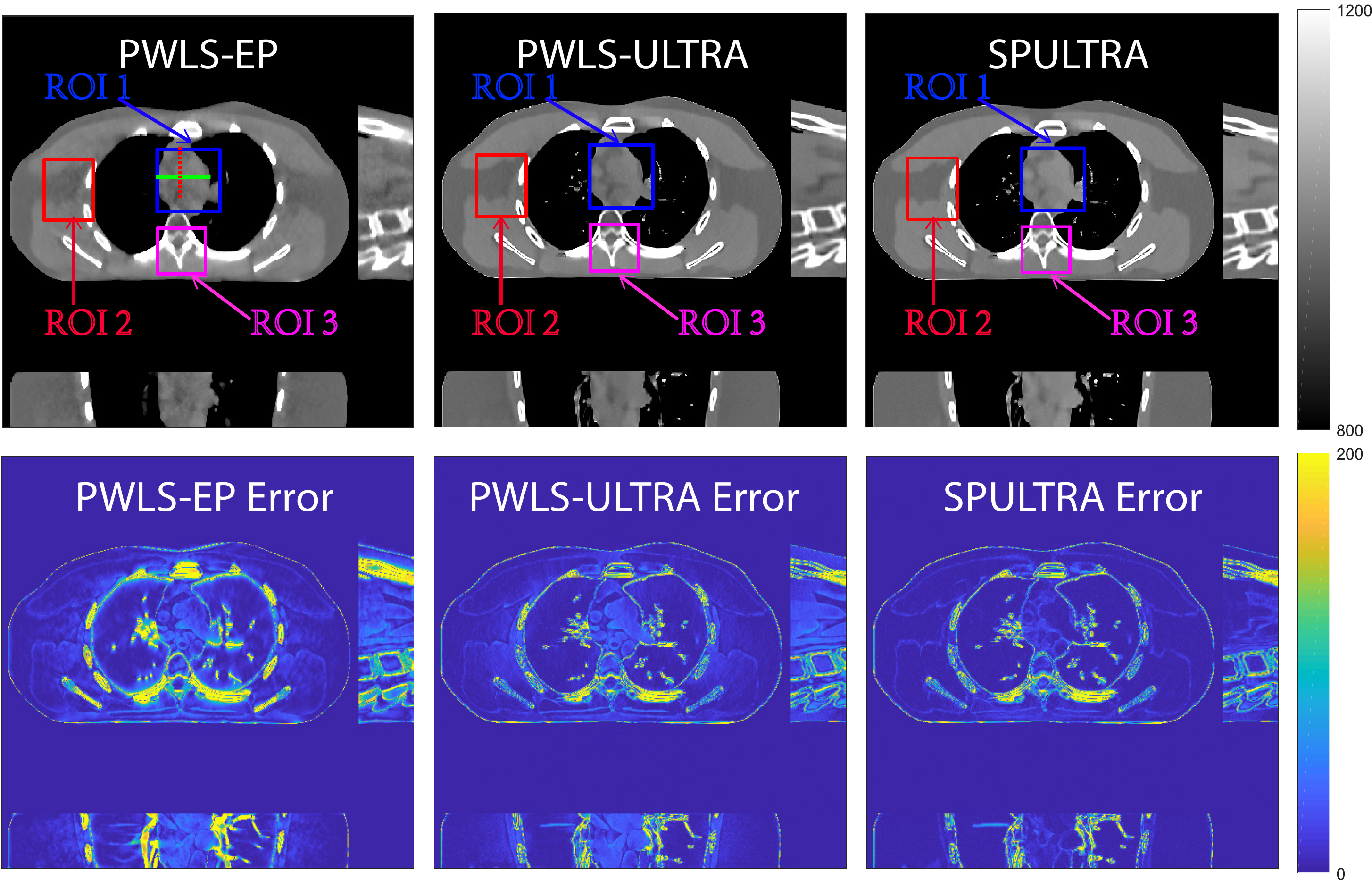}
		\caption{}
		\label{fig:xcat-3e3}
	\end{subfigure}
	\vfill
	\begin{subfigure}[h]{1\textwidth}
		\centering	
		\includegraphics[width=0.9\textwidth]{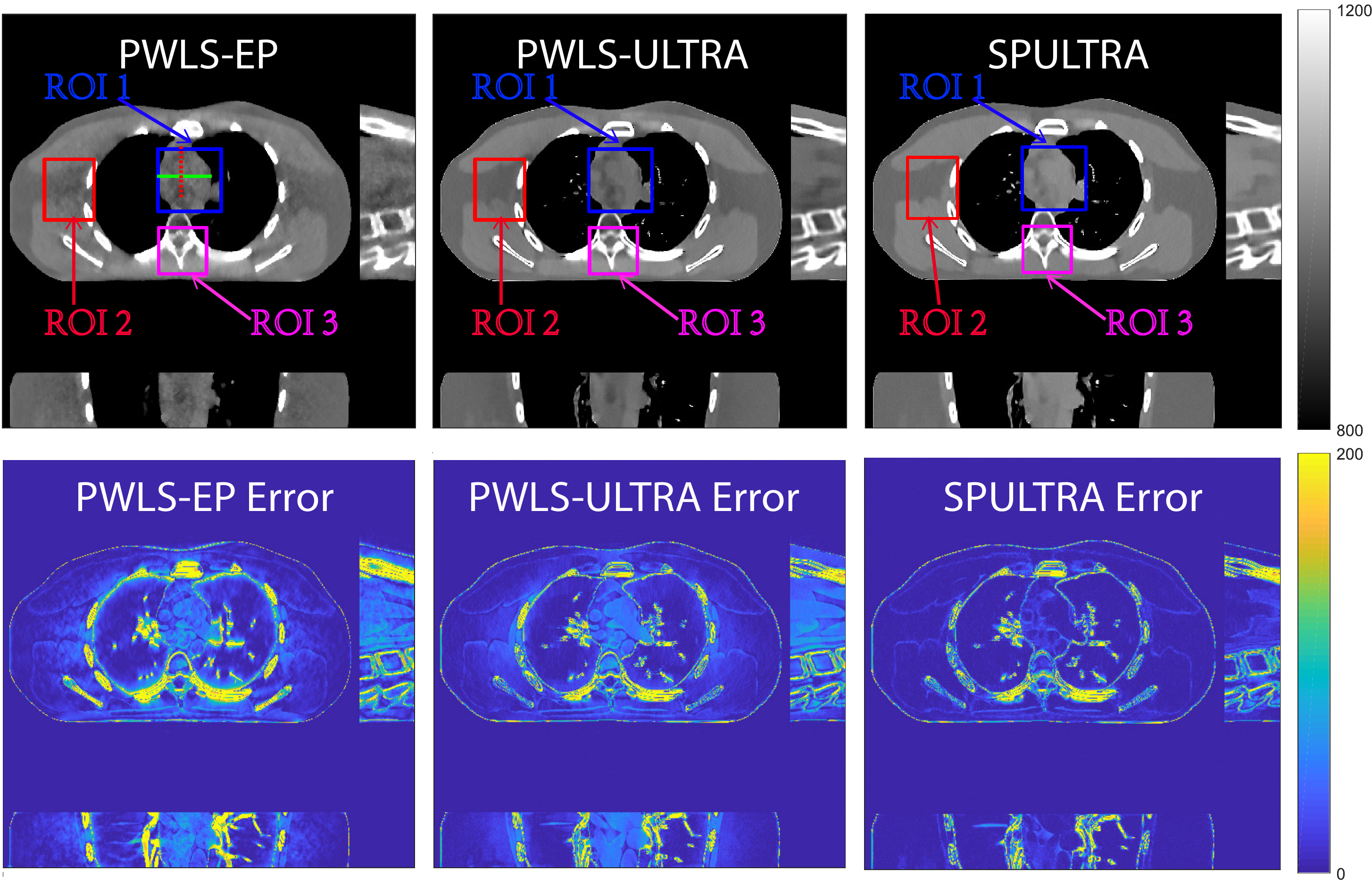}
		\caption{}
		\label{fig:xcat-2e3}
	\end{subfigure}
	\caption{\DIFadd{Comparison of reconstructions and reconstruction errors at (a) $I_0 = 3\times 10^3$ and (b) $I_0 = 2\times 10^3$ dose levels. The 3D images are displayed with the central slices along the axial, sagittal, and coronal directions. The unit of the display windows is HU.}}
	\label{fig:xcat-recon}
	\vspace{-0.1in}
\end{figure*}
Fig.~\ref{fig:sparsecode} also shows the \DIFadd{central axial slice of the }sparse coefficient \DIFadd{maps (volumes)} for different filters of the transforms in the third, fourth and fifth columns. Each \DIFadd{voxel} value in a sparse coefficient \DIFadd{map} is obtained by applying the specific 3D filter to a 3D patch (whose \DIFadd{front} top left corner is at that \DIFadd{voxel}) and hard-thresholding the result. Coefficients for patches \emph{not belonging} to the specific \DIFadd{class }are set to zero (masked out). The sparse code \DIFadd{maps} capture different types of image features (e.g., edges at different orientations or contrasts) depending on the filters and \DIFadd{classes}.

\subsubsection{Numerical Results}
We compare the RMSE and the SSIM for SPULTRA with those for PWLS-EP and PWLS-ULTRA.
Tab.~\ref{tab:numerical} lists the two metrics \DIFadd{for the reconstruction targeted ROI} after sufficient iterations (800 iterations) for convergence of PWLS-EP, PWLS-ULTRA, and SPULTRA, for various dose levels. 
The results show that SPULTRA achieves significant improvements in RMSE and SSIM in low-dose situations. Notably, compared to PWLS-ULTRA, SPULTRA further decreases the RMSE by up to 1.3 HU when $I_0 = 3\times 10^3$, and by around 3.3 HU when $I_0 = 2\times 10^3$. 

The RMSE improvement of SPULTRA over PWLS-ULTRA can be more clearly observed from Fig.~\ref{fig:rmse_comp} that shows the RMSE evolution with the number of outer iterations under different dose levels. At low-doses, SPULTRA decreases the RMSE more quickly (from the same initial value) and to much lower levels than PWLS-ULTRA. 
Fig.~\ref{fig:rmse_comp} shows that to achieve the same RMSE as PWLS-ULTRA at 600 outer iterations, SPULTRA takes 487, 365, 251 and 133 outer iterations under $I_0 = \DIFadd{1\times}10^4, \ 5\times 10^3, \ 3\times 10^3, \text{and } 2\times 10^3$, respectively.

\subsubsection{Computational Costs}
As discussed in Sec. \ref{sec:computation analysis}, SPULTRA has a similar computational cost per iteration as PWLS-ULTRA, except for computing some initializations for image update. Fig.~\ref{fig:rmse_comp} shows that the SPULTRA method requires much fewer number of outer iterations than PWLS-ULTRA to achieve the same RMSE for the reconstruction, especially at low doses. 
\begin{figure}[!htbp]
	\centering
	\begin{subfigure}[h]{0.45\textwidth}
		\centering	
		\includegraphics[width=1\textwidth]{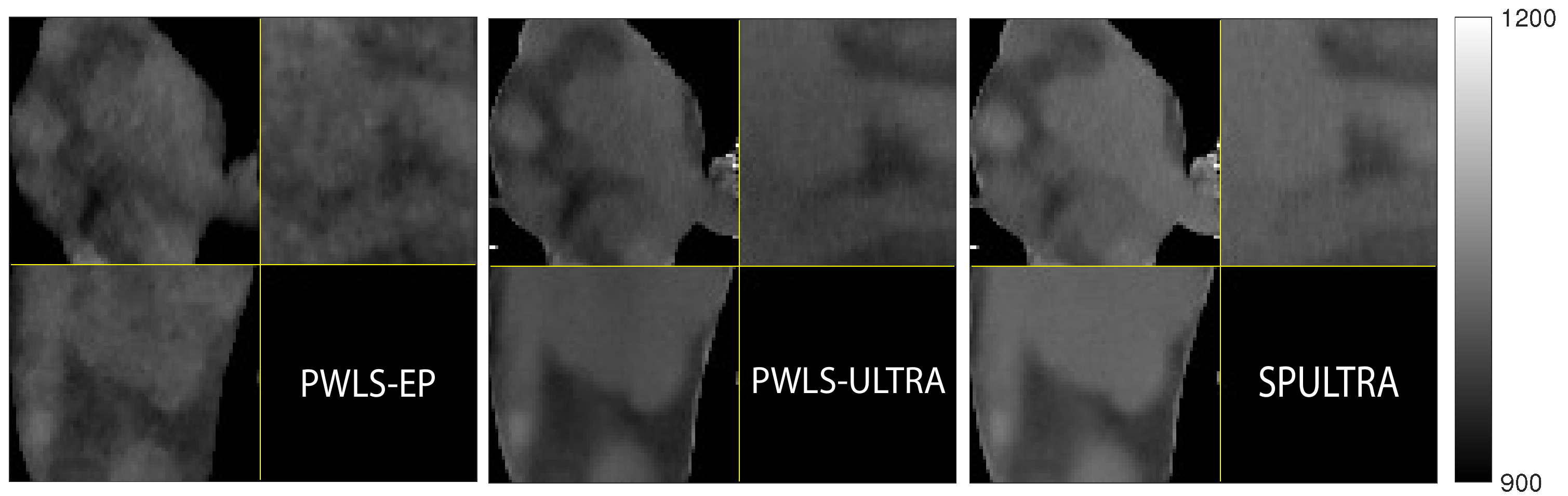}
		\caption{\DIFadd{$I_0 = 3\times 10^3$}}
		\label{fig:xcat-3e3-roi1}
	\end{subfigure}
	\vfill
	\begin{subfigure}[h]{0.45\textwidth}
		\centering	
		\includegraphics[width=1\textwidth]{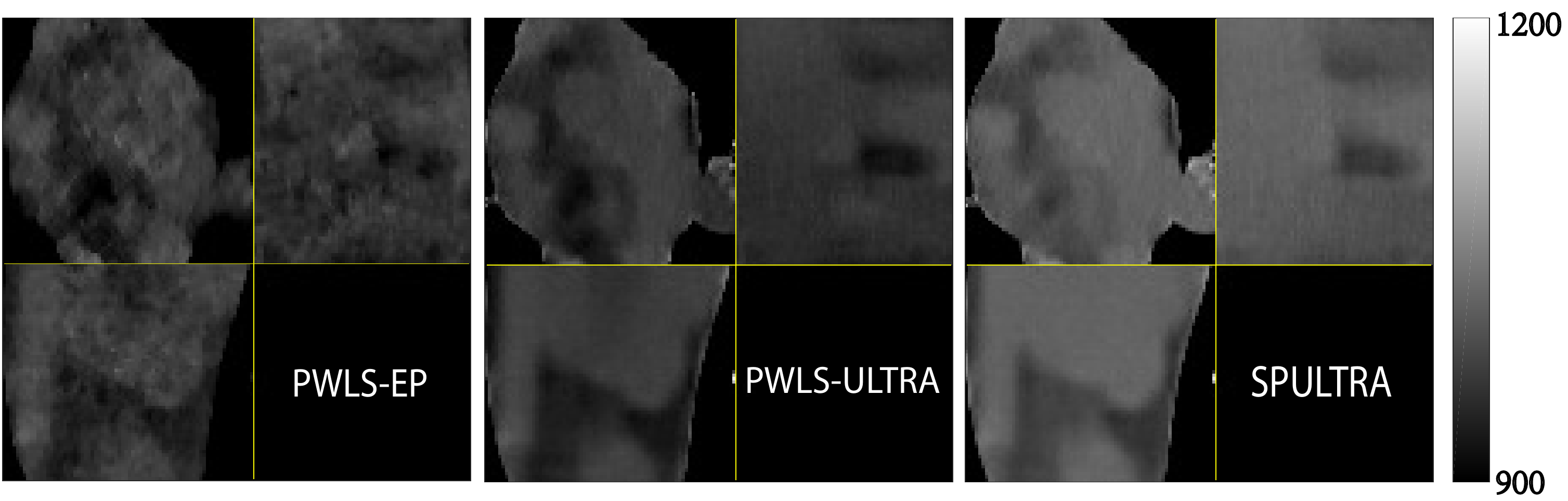}
		\caption{\DIFadd{$I_0 = 2\times 10^3$}}
		\label{fig:xcat-2e3-roi1}
	\end{subfigure}
	\caption{\DIFadd{3D displays of reconstructions of ROI 1 defined in Fig.~\ref{fig:xcat-recon}. The display windows are {[900, 1200]} HU.}}
	\label{fig:xcat-recon-roi1}
	\vspace{-0.06in}
\end{figure}
When the dose is very low, e.g., when $I_0 = 2\times 10^3$, SPULTRA takes only a quarter the number of outer iterations as PWLS-ULTRA to achieve the same RMSE. Thus, the total runtime to achieve a specific reconstruction quality at low doses is typically much lower for SPULTRA than for PWLS-ULTRA.
When the dose is not very low, for example when $I_0 = \DIFadd{1\times}10^4$, the SPULTRA and the PWLS-ULTRA methods have similar computational costs and runtimes. To achieve RMSE of 29.26 HU (see Fig.~\ref{fig:SPULTRA-rmse-1e4}), PWLS-ULTRA requires 600 outer iterations, while SPULTRA requires 487$\times 120\% \approx$ 584 effective outer iterations where the additional $20\%$ runtime is associated with initializations in each SPULTRA outer iteration.

	\subsubsection{Visual Results and Image Profiles}
\begin{figure}[!htbp]
		\centering
		\begin{subfigure}[h]{0.24\textwidth}
			\centering	
			\includegraphics[width=1\textwidth]{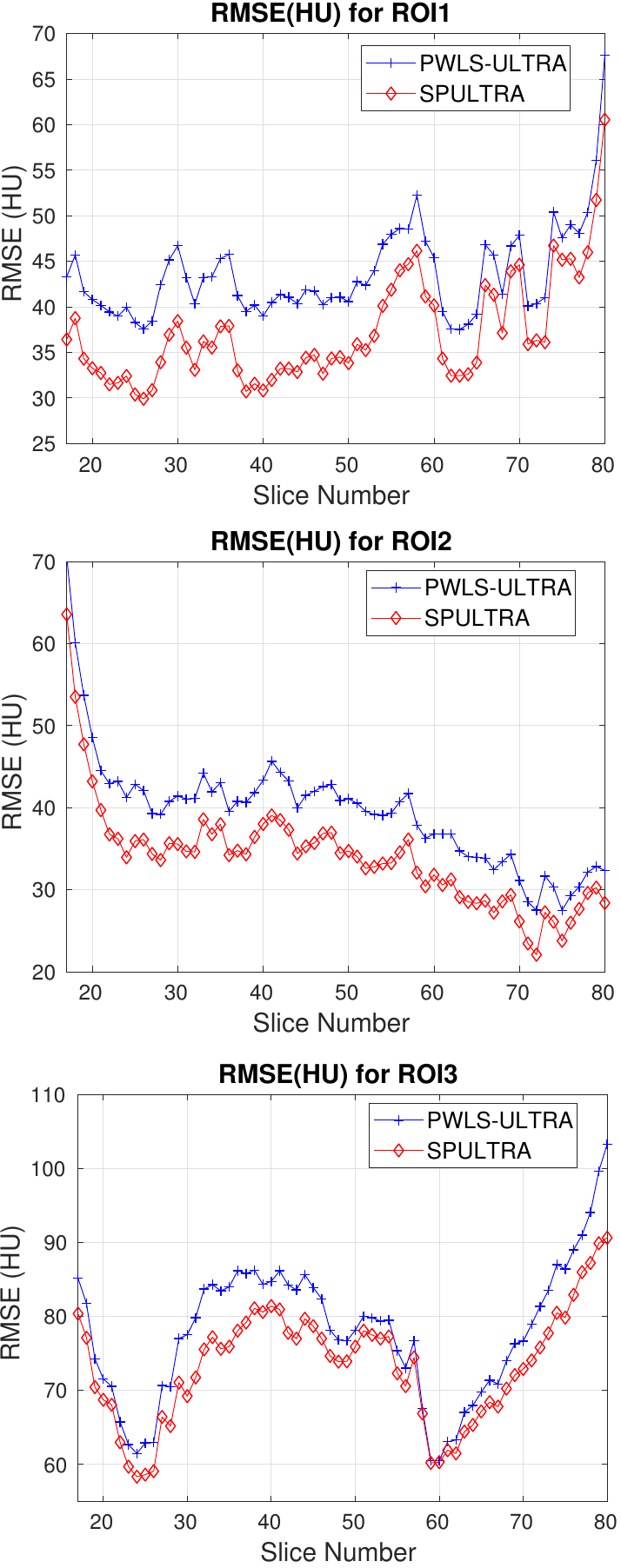}
			\caption{\DIFadd{$I_0 = 3 \times 10^3$}}
			\label{fig:xcat-3e3-roi-metric}
		\end{subfigure}
		\hfill
		\begin{subfigure}[h]{0.24\textwidth}	
			\centering	
			\includegraphics[width=1.03\textwidth]{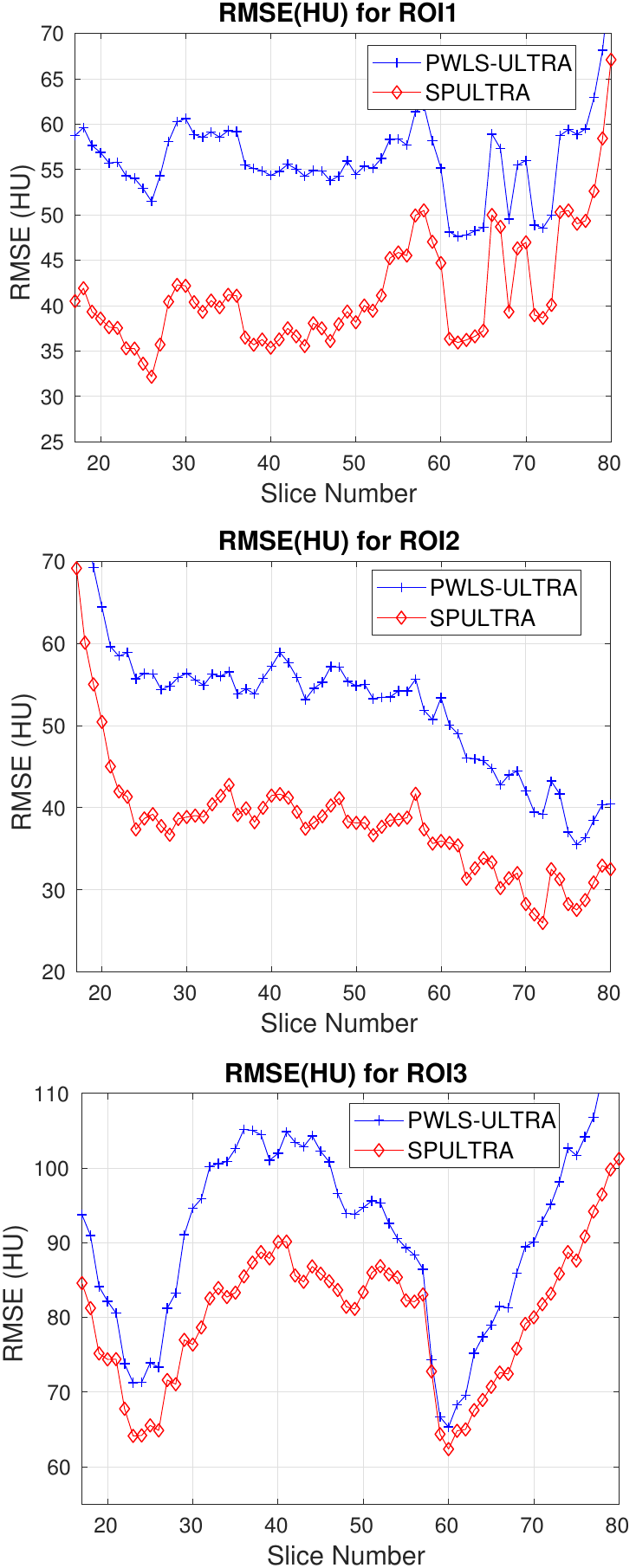}
			\caption{\DIFadd{$I_0 = 2 \times 10^3$}}
			\label{fig:xcat-2e3-roi-metric}
		\end{subfigure}
		\caption{\DIFadd{RMSE (HU) for each axial slice of the 3D ROIs (ROI~1, ROI 2, and ROI 3). The X-axis shows slice indices of the central 64 out of 96 axial slices. Left plot: $I_0 = 3 \times 10^3$. Right plot: $I_0 = 2 \times 10^3$.}}  
		\vspace{-0.1in}
		\label{fig:xcat-metric-roi}
	\end{figure}

	Fig.~\ref{fig:xcat-recon} shows the reconstructed images and the corresponding error images for PWLS-EP, PWLS-ULTRA, and SPULTRA, at $I_0 = 3\times 10^3$ and $I_0 = 2\times 10^3$. Compared to the PWLS-EP result, both PWLS-ULTRA and SPULTRA achieved significant improvements in image quality in terms of sharper reconstructions of anatomical structures such as bones and soft tissues, and suppressing the noise. However, the PWLS-ULTRA method introduces bias in the reconstructions, which leads to larger reconstruction errors compared to the proposed SPULTRA method.
	In Fig.~\ref{fig:xcat-recon}, we \DIFadd{marked three 3D ROIs in the axial plane, i.e., ROI~1, ROI~2, and ROI~3. Fig.~\ref{fig:xcat-recon-roi1} shows the zoom-in images of a 3D plot of ROI~1, and those of ROI~2 and ROI~3 are shown in the supplement. 
		We also plot the evolution of RMSE through the axial slices of the three 3D ROIs in Fig.~\ref{fig:xcat-metric-roi}. The figures demonstrate that SPULTRA clearly outperforms the competing PWLS-EP and PWLS-ULTRA schemes.}

\begin{figure}[!htbp]
	\centering
	\begin{subfigure}[h]{0.24\textwidth}
		\centering	
		\includegraphics[width=1\textwidth]{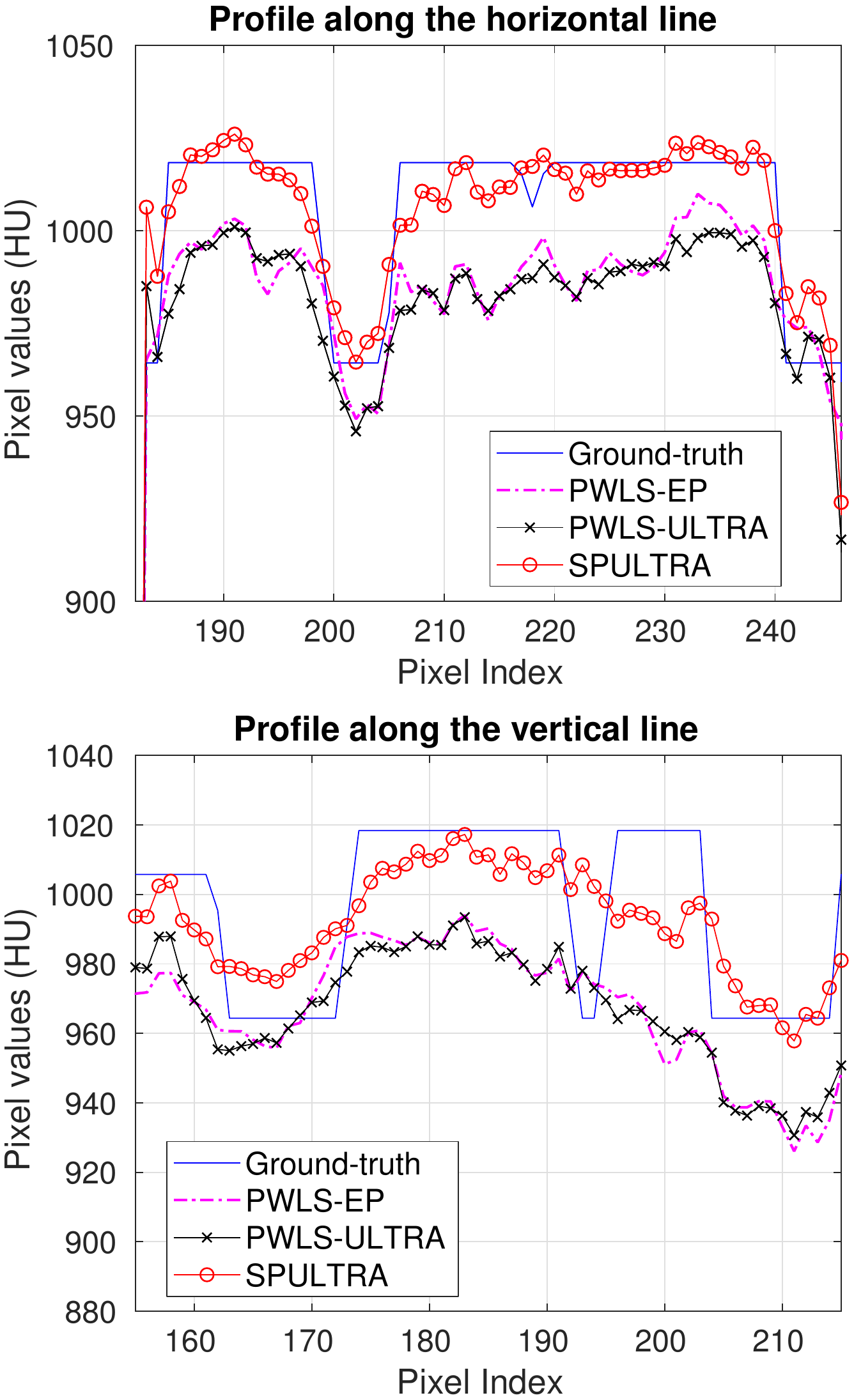}
		\vspace{-0.15in}
		\caption{$I_0 = 3 \times 10^3$}
		\label{fig:xcat-3e3-profile}	
	\end{subfigure}
	\hfill
	\begin{subfigure}[h]{0.24\textwidth}
		\centering	
		\includegraphics[width=1\textwidth]{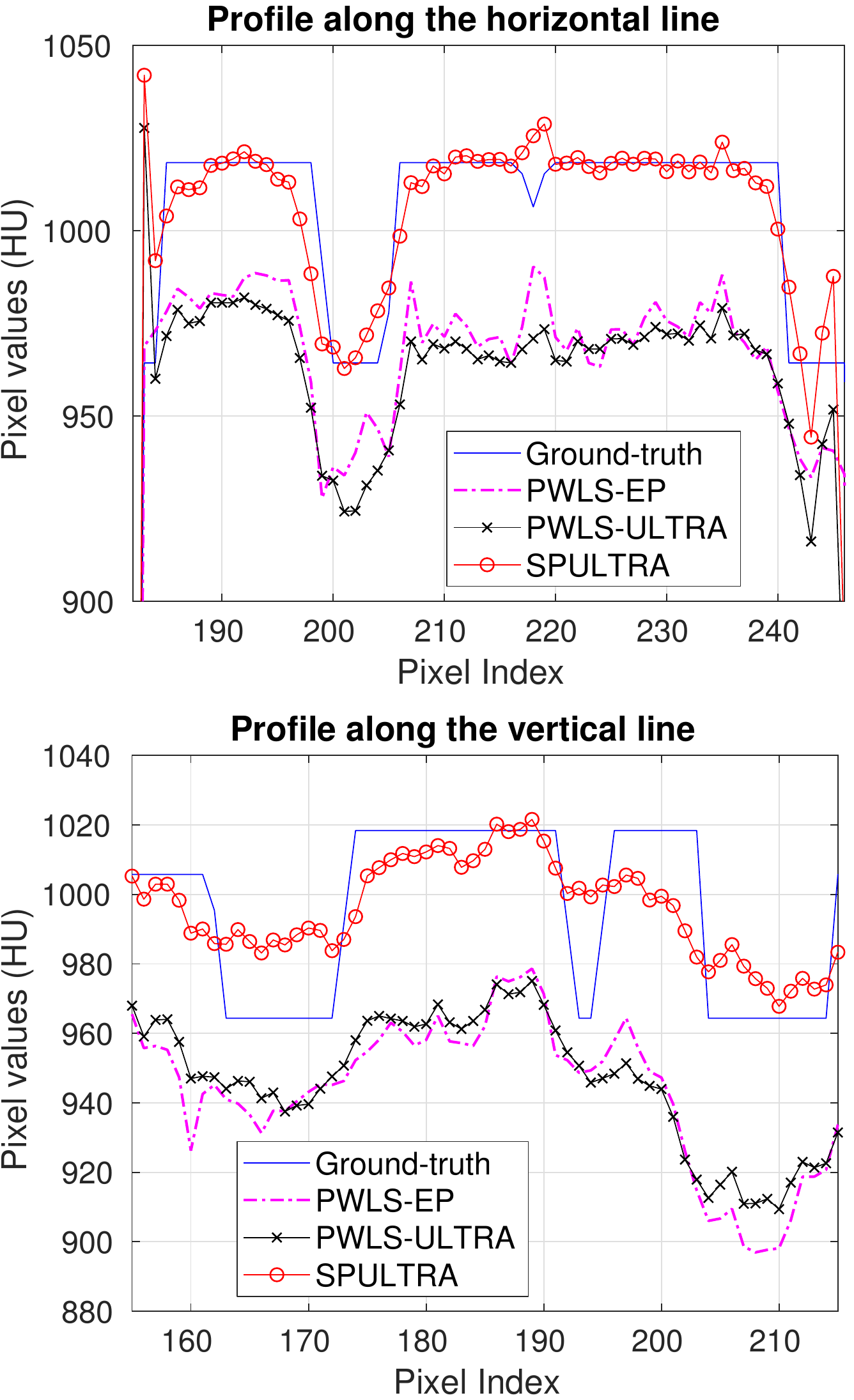}
		\vspace{-0.15in}
		\caption{$I_0 = 2 \times 10^3$}
		\label{fig:xcat-2e3-profile}
	\end{subfigure}
	\caption{Image profiles along the horizontal and vertical lines indicated in Fig.~\ref{fig:xcat-recon}. \DIFadd{Left plot: $I_0 = 3\times 10^3$. Right plot: $I_0 = 2\times 10^3$.}}
	\vspace{-0.2in}
	\label{fig:xcat-recon-profile}
\end{figure}

	The above advantages \DIFadd{of SPULTRA} can be seen more clearly when observing the image profiles. Fig.~\ref{fig:xcat-recon-profile} plots the image profiles \DIFadd{for} the three methods together with that of the ground-truth image. Fig.~\ref{fig:xcat-recon} shows the horizontal green solid line and the vertical red dashed line, whose intensities are plotted in Fig.~\ref{fig:xcat-recon-profile}.
	It is obvious that the \DIFadd{profiles} for SPULTRA \DIFadd{are} closest to the ground-truth among the three compared methods. The gap between the profiles of the PWLS-based methods and the ground-truth shows the bias caused by the compared PWLS methods. 
%\vspace{-0.05in}

\subsection{Synthesized Clinical Data}
\subsubsection{Framework}
\begin{figure}[!htp]\vspace{-0.1in}
	\centering
	\begin{subfigure}[h]{0.24\textwidth}
		\centering	
		\includegraphics[width=1\textwidth]{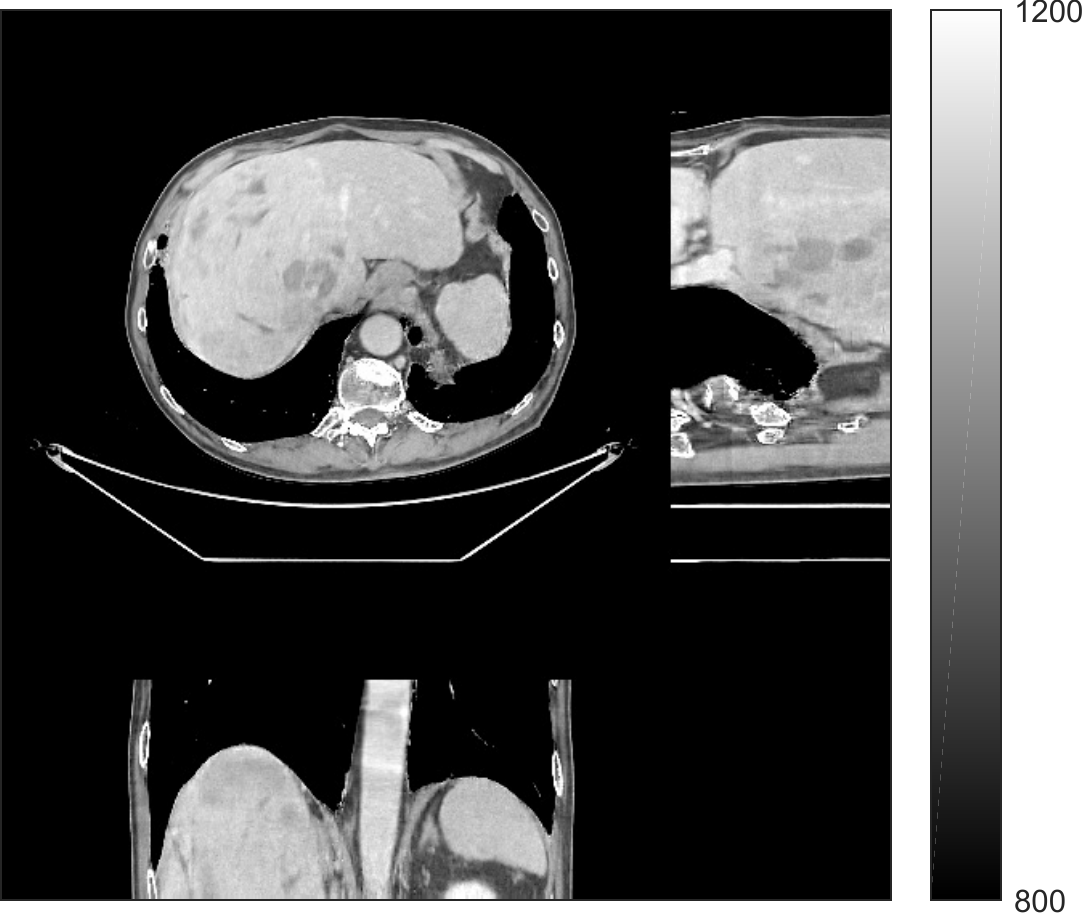}
		\caption{}
		\label{fig:clinic_pwlsultra_zxh}
	\end{subfigure}
	\hfill
	\begin{subfigure}[h]{0.24\textwidth}
		\centering	
	 \includegraphics[width=1\textwidth]{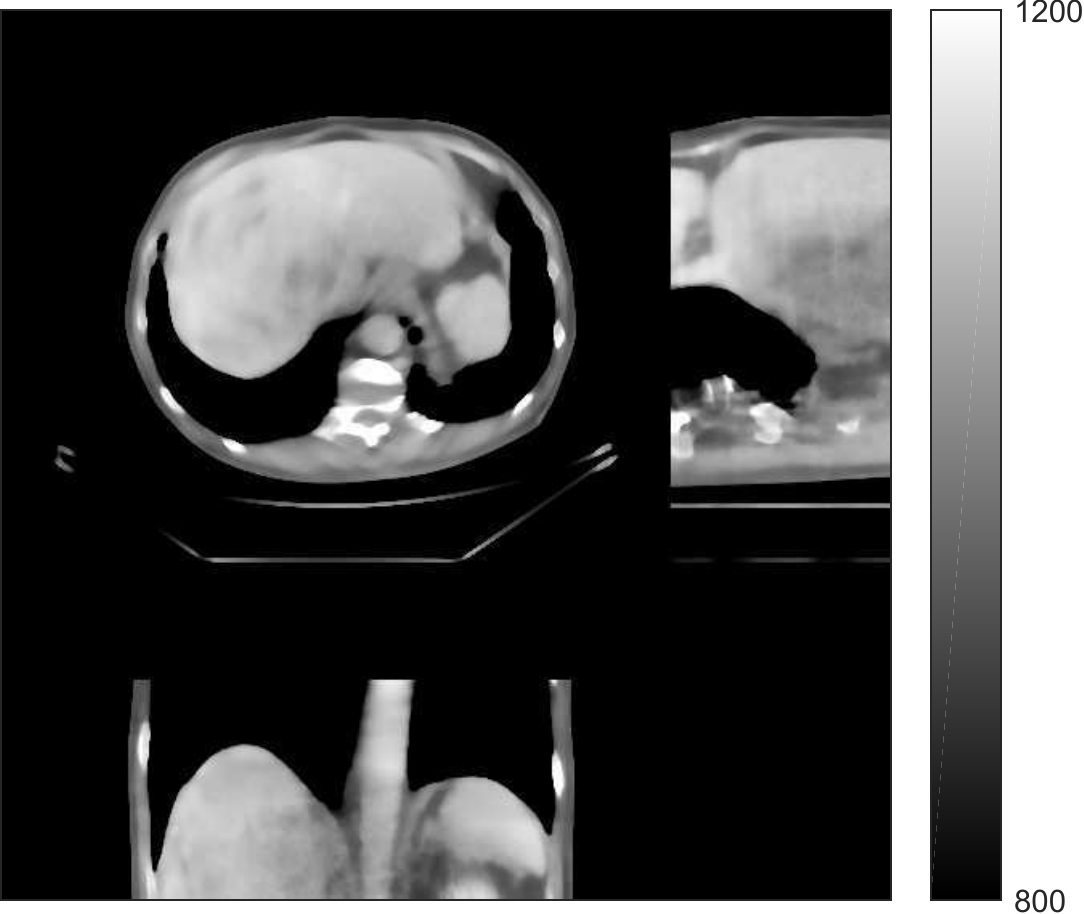}
	\caption{}
		\label{fig:clinic-pwls-ultra-1e4-l2b15}
	\end{subfigure}	
	\vspace{-0.05in}
	\caption{(a) ``true" clinical image \DIFadd{(HU)}, (b) the reconstruction \DIFadd{(HU) }of the synthesized data with PWLS-EP for $I_0 = 1\times 10^4$ with $\beta_{ep} = 2^{15}$. \DIFadd{The central axial, sagittal, and coronal slices of the volume are shown.}}
	\vspace{-0.15in}
	\end{figure}

We used the pre-learned union of 15 square transforms from the XCAT phantom simulations to reconstruct the synthesized helical chest scan volume of size ${420\times 420\times 222}$ with $\Delta_x = \Delta_y = 1.1667$ mm and ${\Delta_z = 0.625}$ mm. The sinograms were of size $888\times 64\times 3611$. 
Since the clinical data is synthesized via the PWLS-ULTRA reconstruction, the noise model for this synthesized data is obscure, making it difficult to determine appropriate low-dose levels for such data. 
We tested the radiation dose of $I_0 =1\times 10^4$ with an electronic noise variance the same as the XCAT phantom simulation, i.e., $\sigma^2 = 25$. The percentage of non-positive pre-log measurements for the synthesized clinical data in this case was around $0.14\%$. Such non-positive values were replaced by $1\times 10^{-5}$ for PWLS-based methods.
Fig.~\ref{fig:clinic_pwlsultra_zxh} shows the ``true'' clinical image that was reconstructed from real clinical regular-dose sinogram using the PWLS-ULTRA method.
\begin{figure*}[!htbp]
	\centering
	\includegraphics[width=1\textwidth]{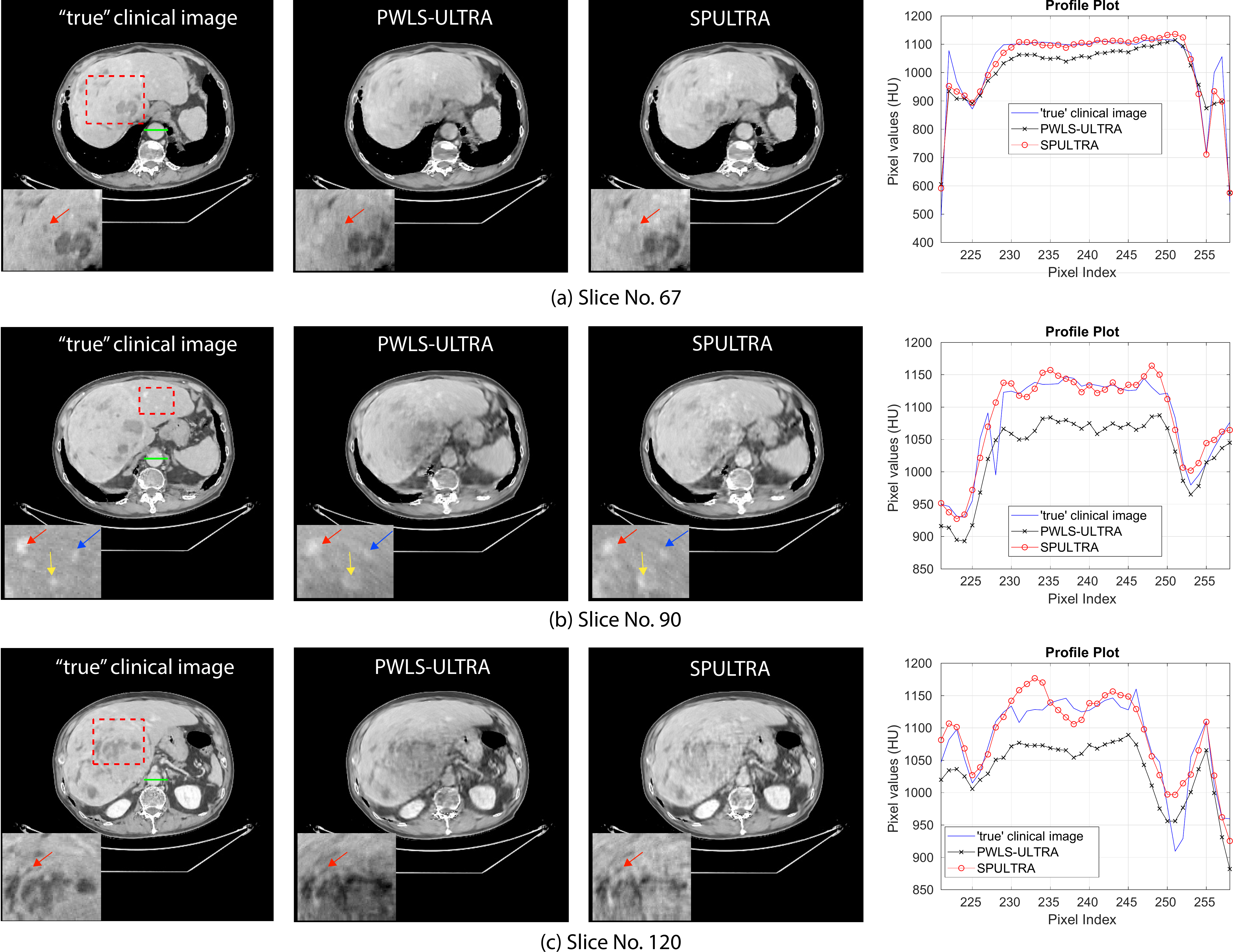}
	\caption{Reconstructed images (columns 1 to 3) and the image profiles (the 4th column) along the green line in the ``true'' clinical image for the synthesized clinical data with $I_0 = \DIFadd{1\times}10^4$ and $\sigma^2 = 25$. (a) Results for axial slice No. 67, (b) results for slice No. 90, and (c) results for slice No. 120. \DIFadd{We selected one ROI for each of these three slices and the arrows point out some small structures in the image. The display windows for reconstructed images are {[800, 1200]} HU, and those for the zoomed-in ROIs are {[950, 1200]} HU.}}
	\label{fig:clinic-recon-profile-1e4}
	\vspace{-0.25in}
\end{figure*} 
Similar to the XCAT phantom simulation, the initial image for both SPULTRA and PWLS-ULTRA was a reconstruction obtained using PWLS-EP. We set the regularizer parameter $\beta_{ep}$ for PWLS-EP \DIFadd{to} $2^{15}$ to generate a smoother (with less noise) initial image, which led to good visual image equality for the SPULTRA and PWLS-ULTRA reconstructions. \DIFadd{Since the optimization problem for PWLS-EP is strictly convex, we simply initialized PWLS-EP with a zero image. Fig.~\ref{fig:clinic-pwls-ultra-1e4-l2b15} shows the PWLS-EP reconstructed image for $I_0 = 1\times 10^4$. We set the regularizer parameters for both PWLS-ULTRA and SPULTRA as $\gamma_c = 5\times 10^{-4}$, and $\beta = 1.5 \times 10^{4}$}.
\subsubsection{Reconstruction results for the synthesized clinical data} 
Fig.~\ref{fig:clinic-recon-profile-1e4} shows three axial slices from the 3D reconstructions with SPULTRA and PWLS-ULTRA at $I_0 = 1\times 10^4$: the middle slice (No. 67) and two slices located farther away from the center (No. 90 and No. 120). The image profiles along a horizontal line \DIFadd{(shown in green)} in the displayed slices are also shown in Fig.~\ref{fig:clinic-recon-profile-1e4}. 
The reconstructed slices using PWLS-ULTRA appear darker around the center compared to the ``true'' clinical image and the reconstructions with SPULTRA. This means PWLS-ULTRA produces a strong bias in the reconstruction. The bias can be observed more clearly in the profile plots: the pixel intensities for the SPULTRA reconstruction better follow those of the ``true" clinical image, while those for the PWLS-ULTRA reconstruction are much \DIFadd{worse} than the ``true" values. Moreover, SPULTRA achieves sharper rising and failing edges compared to PWLS-ULTRA. In other words, SPULTRA also achieves better resolution than PWLS-ULTRA. \DIFadd{Fig.~\ref{fig:clinic-recon-profile-1e4} also shows a zoomed-in ROI for each of the chosen slices, and highlights some small details with arrows. It is clear that in addition to reducing the bias, SPULTRA reconstructs image details better than PWLS-ULTRA.}

\vspace{-0.12in}
 \subsection{\DIFadd{Ultra Low-dose Experiments with Raw Data}}
\subsubsection{\DIFadd{Framework}}
\DIFadd{We obtained from GE a 2D fan-beam raw (pre-log) scan of a shoulder phantom, which included the beam-hardening effect. The provided $200~\text{mA}$ with 1 second scan can be viewed as a standard-dose scan and all the raw measurements are positive. Based on this standard-dose scan, we simulated an ultra low-dose scan as shown in \eqref{eq:shoulder_gen} with $\alpha = 200$, and added Poisson and Gaussian noise ($\sigma = 5$) to the measurements. The simulated measurements have about $0.4\%$ non-positive values. The sinograms were of size ${888\times 984}$, and reconstructed images were of size ${512\times 512}$ with ${\Delta_x = \Delta_y = 0.9766}$ mm.}
		
\DIFadd{For PWLS-ULTRA and SPULTRA, we pre-learned a union of five square transforms using ${8 \times 8}$ overlapping image patches with stride ${1\times 1}$ from five $512 \times 512$ XCAT phantom slices~\cite{pwls-ultra2018}. Here, we also compared SPULTRA with a recent deep-learning based low-dose CT denoising framework ``WavResNet" combined with an RNN architecture\cite{WavResNet18}. The iterative RNN version of WavResNet was pre-trained based on the 2016 Low-Dose CT Grand Challenge data set~\cite{WavResNet18}. During reconstruction, WavResNet, PWLS-ULTRA, and SPULTRA were initialized with the image reconstructed by PWLS-EP with $\beta_{ep}=0.1$. The parameters ${(\beta, \gamma_c)}$ for \DIFadd{both} PWLS-ULTRA and SPULTRA were set as $(0.05, 80)$. \DIFadd{These values worked well in our experiment. In the supplement, we discuss in detail the parameter selection procedure of $(\beta,\ \gamma_c)$ for both PWLS-ULTRA and SPULTRA.}
	\DIFadd{Parameters} for testing WavResNet were set according to \cite{WavResNet18}\DIFadd{, and the pixel values of the input to WavResNet were converted to match the network required scalings}. 
	Since the WavResNet was trained with images reconstructed with the filtered backprojection (FBP) method~\cite{WavResNet18}, we also tested on this shoulder phantom that initialized WavResNet with an FBP reconstructed image. Although initializing WavResNet with an FBP reconstructed image better matches the trained model than the PWLS-EP reconstructed image does, the latter still provided better results. We included in the supplement the denoised image initialized with the FBP reconstruction.
	}
	\begin{figure*}[!htp]
		\centering
		\includegraphics[width=1\textwidth]{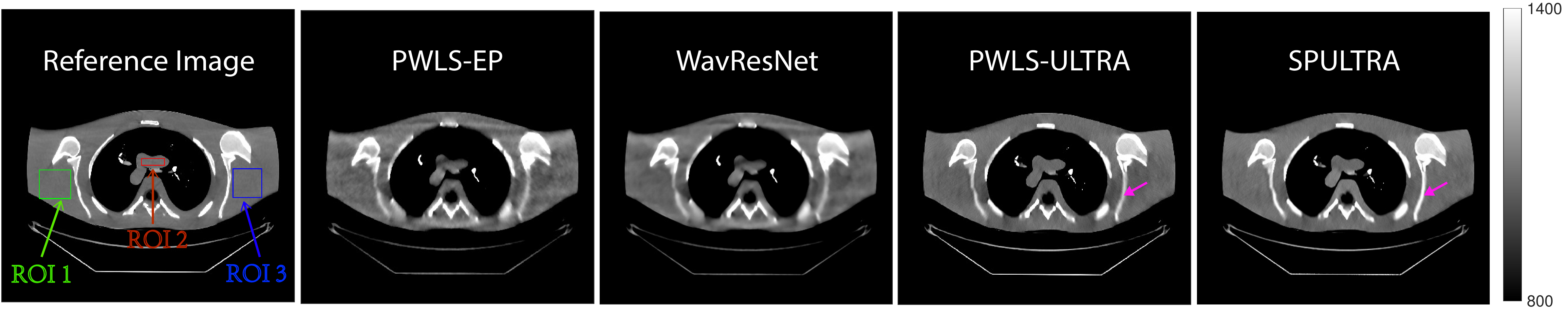}
		\caption{Reconstructions for ultra low-dose 2D scan simulated from raw measurements. The leftmost image is the PWLS-EP reconstructed image for the $200\text{ mA}$ scan. The second image is the PWLS-EP reconstruction for the simulated ultra low-dose scan, and it is the initial image for WavResNet~\cite{WavResNet18}, PWLS-ULTRA~\cite{pwls-ultra2018}, and SPULTRA. The display windows are [800,~1400]~HU.}
		\label{fig:1ma-imgs-recon}
		\vspace{-0.08in}
	\end{figure*}  
	
	\begin{table*}
		\centering
			\caption{\DIFadd{Mean (HU) and standard deviation (STD) (HU) of the ROIs for ultra low-dose shoulder phantom simulations.}} 
		\begin{subtable}{0.49\textwidth}
			\centering
		\caption{\DIFadd{Mean (HU)}} \vspace{-0.02in}
		\begin{tabular}{c r r r}
			\toprule 
			\textit{\DIFadd{Methods}}& \textit{\DIFadd{ROI 1}} & \textit{\DIFadd{ROI 2}}  & \textit{\DIFadd{ROI 3}} \\ 
			\midrule
			\DIFadd{Reference}& \DIFadd{1052.1}&\DIFadd{1060.1}& \DIFadd{1053.4}\\ 
			\midrule
			\DIFadd{PWLS-EP}& \DIFadd{1032.7}&\DIFadd{977.5}& \DIFadd{1026.3}\\ 
			\midrule
			\DIFadd{WavResNet\cite{WavResNet18}}& \DIFadd{1037.6}&  \DIFadd{981.1}&\DIFadd{1031.2}\\
			\midrule
			\DIFadd{PWLS-ULTRA\cite{pwls-ultra2018}}&\DIFadd{1031.1}&\DIFadd{1043.0}& \DIFadd{1024.2}\\
			\midrule
			\DIFadd{SPULTRA}& \textbf{\DIFadd{1054.7}} & \textbf{\DIFadd{1044.0}} & \textbf{\DIFadd{1049.6}}\\
			\bottomrule
		\end{tabular} 
\end{subtable}
\hfill
\begin{subtable}{0.49\textwidth}
		\centering
	\caption{\DIFadd{STD (HU) }} \vspace{-0.02in}
	\begin{tabular}{c r r r}
		\toprule 
		\textit{\DIFadd{Methods}}& \textit{\DIFadd{ROI 1}} & \textit{\DIFadd{ROI 2}}  & \textit{\DIFadd{ROI 3}} \\ 
		\midrule
		\DIFadd{Reference}& \DIFadd{8.12}&\DIFadd{8.81}& \DIFadd{6.98}\\ 
		\midrule
		\DIFadd{PWLS-EP}& \DIFadd{19.45}&\DIFadd{19.45}& \DIFadd{30.46}\\ 
		\midrule
		\DIFadd{WavResNet\cite{WavResNet18}}& \DIFadd{18.91}&  \DIFadd{18.91}&\DIFadd{30.16}\\
		\midrule
		\DIFadd{PWLS-ULTRA\cite{pwls-ultra2018}}&\DIFadd{14.82}&\DIFadd{10.92}& \DIFadd{19.29}\\
		\midrule
		\DIFadd{SPULTRA}& \DIFadd{16.34} & \DIFadd{11.42} & \DIFadd{11.60}\\
		\bottomrule
	\end{tabular} 
\end{subtable}
		\label{tab:shoulder-mean-roi}
		\vspace{-0.16in}
	\end{table*}
\vspace{-0.05in}
	\subsubsection{\DIFadd{Results}}
Fig. \ref{fig:1ma-imgs-recon} shows the reconstructions for the $200\text{ mA}$ scan (reference image) along with the reconstructions for the simulated ultra low-dose scan obtained with PWLS-EP, WavResNet, PWLS-ULTRA, and SPULTRA. Visually, WavResNet fails to reconstruct the image but improves over the initial PWLS-EP reconstruction, while PWLS-ULTRA and SPULTRA provide better image quality. This indicates that the ULTRA-based methods may have a better generalization property than WavResNet, since they learn more fundamental features of CT images (also see~\cite{pwls-ultra2018}).	
We selected three smooth ROIs, where the pixel values are approximately constant. Tab.~\ref{tab:shoulder-mean-roi} shows the mean \DIFadd{and the standard deviation of} pixel values for these ROIs for various methods and the standard-dose reference. Since the iterative RNN version of WavResNet only has small improvements over PWLS-EP, the pixel values do not change much compared with PWLS-EP. PWLS-ULTRA however reduces the bias in the central region of the image (ROI 2), but fails to correct the bias in the regions near the bones (ROI 1 and ROI 3). SPULTRA reduces the bias in the central region of the image, and also significantly corrects the bias near the bone regions. \DIFadd{The standard deviations of the ROIs reconstructed by SPULTRA are comparable to those reconstructed by PWLS-ULTRA, and are close to those of the reference ROIs.} Additionally, SPULTRA reconstructs the bone (indicated by the magenta arrow in the last two subfigures of Fig.~\ref{fig:1ma-imgs-recon}) better than PWLS-ULTRA.

%% file: conclusions_r1.tex
%\vspace{-0.1in}
\section{Conclusions}\label{sec:conclusion} \vspace{-0.01in}
This paper proposes a new LDCT reconstruction method dubbed SPULTRA that combines the shifted-Poisson statistical model with the union of learned transforms or ULTRA regularizer. %To deal with the nonconvexity of the problem, particularly the nonconvex data-fidelity term caused by the shifted-Poisson model, we iteratively designed quadratic surrogate functions for this term in the proposed algorithm.
To deal with the nonconvex data-fidelity term arising from the shifted-Poisson model, we iteratively designed quadratic surrogate functions for this term in the proposed algorithm.
%Within each surrogate function update iteration, the cost function has the similar form of that of PWLS-ULTRA method so that we can optimize it following the algorithm for PWLS-ULTRA.
In each surrogate function update iteration, the overall cost function (i.e., majorizer) has a similar structure as in the very recent PWLS-ULTRA method, and is optimized by performing an image update step with a quadratic cost and a sparse coding and clustering step with an efficient closed-form update.
%In each algorithm iteration, the image is updated by minimizing a quadratic surrogate cost that has a similar form as in the recent PWLS-ULTRA method, and the sparse codes and clusters are updated in closed-form by solving a nonconvex subproblem.  
We evaluated the proposed SPULTRA scheme with numerical experiments on the XCAT phantom, synthesized clinical data, \DIFadd{and beam-hardened ultra low-dose raw measurement simulations.
	SPULTRA outperformed }prior methods in terms of eliminating bias and noise in the reconstructed image while maintaining the resolution of the reconstruction under very low X-ray doses.
SPULTRA was also much faster than PWLS-ULTRA in achieving a desired reconstruction quality for low-doses\DIFadd{, and it had better generalization property than the WavResNet based denoising scheme. Moreover, we investigated the }convergence guarantees of the proposed surrogate function based alternating minimization scheme. \DIFadd{We will investigate }variations or generalizations of the SPULTRA model such as exploiting unions of overcomplete or tall transforms, or rotationally invariant transforms in future work.

%% file: multimedia.tex
Here, we present additional proofs and experimental results to accompany our manuscript~\cite{TMI-SPULTRA-as-submit}. First, we present a sketch of the proof for the convergence theorem in Section~IV of~\cite{TMI-SPULTRA-as-submit}. Then, we include some additional experimental results.
\setcounter{section}{6}	
\setcounter{equation}{15}
\setcounter{figure}{10}
\setcounter{table}{3}

\section{Proof Sketch for Convergence Theorem}
As stated in Section IV, the objective function is written as follows:
\begin{equation}\label{eq:P0}
\begin{aligned}
G(\x,\Z,\Gamma)	= \mathsf{L}(\x)+ \R(\x,\Z,\Gamma) + \mathfrak{X}(\x), 
\end{aligned}
\tag{P0}
\end{equation}
where $\mathfrak{X}(\x)$ is a barrier function that takes the value 0 when the constraint on $\x$ is satisfied and is $+\infty$ otherwise, and $\mathsf{L}(\x)$ is the data fidelity function of the form ${\mathsf{L}(\x) = \sum_{i = 1}^{N_d}h_i([\A\x]_i)}$ in which $\A \in \mathbb{R}^{N_d \times N_p}$ is the CT system matrix. $\Z$ is the sparse code matrix concatenated by column vectors $\z_{j}$, and ${\Gamma \in \mathbb{R}^{\tilde{N}}}$ is a vector whose elements represent the classes indices for the patches, i.e., $\Gamma_j \in \{1, \cdots, K\}$. 
With ${l_i \triangleq [\A\x]_i}$, $h_i(l_i)$ was defined as
\begin{equation}
h_i(l_i) \triangleq (I_0e^{-f_i(l_i)} + \sigma^2) - Y_i\log(I_0e^{-f_i(l_i)} + \sigma^2).
\tag{2}
\end{equation}
The regularizer $\R(\x,\Z,\Gamma)$ was defined as 
%\begin{equation}\label{eq:Rx}
%\begin{aligned}
%\R(\x,\Z,\Gamma) \triangleq \beta \sum_{k=1}^{K}  \bigg\{  \sum_{j\in \Gamma} \|\omg_k \P_j \x - \z_{j}\|^2_2 + \gamma_c^2\|\z_{j}\|_0 \bigg\} ,
%\end{aligned}
%%\vspace{-0.05in}
%\end{equation}
\begin{equation}\label{eq:Rx}
\begin{aligned}
\R(\x,\Z,\Gamma) \triangleq \beta \sum_{j=1}^{\tilde{N}}  \bigg\{ \|\omg_{\Gamma{j}} \P_j \x - \z_{j}\|^2_2 + \gamma_c^2\|\z_{j}\|_0 \bigg\} ,
\end{aligned}
%\vspace{-0.05in}
\end{equation}
where $\beta >0$ is a parameter for balancing the data-fidelity and regularizer penalties, and $\tilde{N}$ is the number of patches.

\begin{theorem}\label{them1}
\DIFadd{Assume the image update step is solved exactly.} For an initial $(\x^0, \Z^0, \Gamma^0)$, iterative sequence $\{\x^n, \Z^n, \Gamma^n\}$ generated by the SPULTRA algorithm is bounded, and the corresponding objective sequence $\{G(\x^n, \Z^n, \Gamma^n)\}$ decreases monotonically and converges to ${G^* \triangleq G^*(\x^0, \Z^0, \Gamma^0)}$. Moreover, all the accumulation points of the iterate sequence are equivalent and achieve the same value $G^*$ of the objective. Each accumulation point $(\x^*,\Z^*,\Gamma^*)$ also satisfies the following partial optimality conditions:
\begin{equation}\label{eq:critical-point}
	\begin{aligned}
	&\mathbf{0} \in \partial_{\x}G(\x,\Z^*,\Gamma^*)|_{\x = \x^*},\\
	&(\Z^*,\Gamma^*) \in \arg\min_{\Z,\Gamma} G(\x^*,\Z,\Gamma),
	\end{aligned}
	\tag{14}
\end{equation}
where $\partial_{\x}$ denotes the sub-differential operator for the function $G$ with respect to $\x$ \cite{rockafellar2009variational,mordukhovich2006variational,2015BCS-ST}.
Finally,  ${\|\x^{n+1} - \x^n\|_2 \to 0}$ as $n\to \infty$.
\end{theorem}

\subsection{Preliminaries}\label{sec:preliminaries}
\subsubsection{Surrogate Function design}
To optimize the non-convex function $G(\cdot)$, we design a series of quadratic majorizers for each $h_i(\l_i)$:
		\begin{equation}\label{eq:surrogate_i}
	\begin{aligned}
	q(l_i;l_i^n) = h_i(\l_i^n) +\dot{h_i}(l_i^n)(l_i - l_i^n)+\frac{1}{2}c_i(l_i^n)(l_i - l_i^n)^2.		
	\end{aligned}
	\end{equation}
	Here, $c_i(l_i)$ is the curvature defined in (5) of \cite{TMI-SPULTRA-as-submit}. According to \cite{TMI99}, such a choice of $c_i(l_i)$ is an optimum curvature that ensures 
majorizer conditions:
\begin{subequations}
		\begin{equation}\label{eq:srr_cond1}
	h_i(l_i) \leq q(l_i;l_i^n), \ \forall l_i \geq 0,
%	\tag{5.1}	
%	\vspace{-0.2in}
	\end{equation}
	\begin{equation}\label{eq:srr_cond2}
	h_i(l_i^n) = q(l_i^n;l_i^n).
%	\tag{5.2}	
	\end{equation}
\end{subequations}

In general, when minimizing a majorizing function or updating $l_i$, let 
\begin{equation}\label{eq:sol_q}
l_i^{n+1} = \arg\min q(l_i;l_i^n).
\end{equation}
Then, using \eqref{eq:srr_cond1} and \eqref{eq:srr_cond2} yields
\begin{equation}\label{eq:srr_inequality}
h_i(l_i^{n+1})\leq q_i(l_i^{n+1};l_i^n)\leq q(l_i^n;l_i^n) = h_i(l_i^n).
\end{equation}
Thus, in general, minimizing a majorizer monotonically decreases the original cost.

Clearly, \eqref{eq:surrogate_i} can be rewritten as follows: 
\begin{equation}
\begin{aligned}
q(l_i;l_i^n) &=\frac{1}{2}\big[\big(c_i(l_i^n)^{\frac{1}{2}}(l_i - l_i^n)\big)^2 + 2\dot{h_i}(l_i^n)(l_i - l_i^n) \\ &\quad+  \big(c_i(l_i^n)^{-\frac{1}{2}}\dot{h_i}(l_i^n)\big)^2  \big] \\
& \quad  + h_i(l_i^n) - \frac{1}{2}c_i(l_i^n)^{-1}\dot{h_i}(l_i^n)^2\\
& = \frac{c_i(l_i^n)}{2}\big[ (l_i - l_i^n) + c_i(l_i^n)^{-1}\dot{h_i}(l_i^n) \big]^2 + q_c^n.
\end{aligned}
\end{equation}
When optimizing $q(l_i;l_i^n)$, $q_c^n$ is a constant that can be ignored, and we can optimize 
\begin{equation}\label{eq:srr_drop}
\varphi_n (l_i) \triangleq \frac{c_i(l_i^n)}{2}\big[ (l_i - l_i^n) + c_i(l_i^n)^{-1}\dot{h_i}(l_i^n) \big]^2.
\end{equation}
The minimizer of \eqref{eq:srr_drop} also solves \eqref{eq:sol_q}, which makes \eqref{eq:srr_inequality} hold for every iteration.\\

Since in SPULTRA, we majorize the entire function $\mathsf{L}(\x)$, its majorizer is therefore
\begin{equation}\label{eq:L_major}
\mathbf{Q}(\x;\x^n) = \phi(\x;\x^n) + \underbrace{ \mathsf{L}(\x^n) - \frac{1}{2}||\bm{d}_h(l^n)||_{(\W^n)^{-1}}^2}_{Q_c^n},
\end{equation}
where
\begin{equation}\label{eq:surrogate}
\begin{aligned}
\phi(\x;\x^n) \triangleq \frac{1}{2}||\tilde{\y}^n - \A\x||_{\W^n}^2,
\end{aligned}
\end{equation}
and $\bm{d}_h(l^n) \in \mathbb{R}^{N_d}$ is the row vector whose entries are $\dot{h_i}(l_i^n)$, ${\W^n \triangleq \diag\{c_i(l_i^n)\}}$, $\tilde{\y}^n \triangleq \A\x^n -\big(\W^n\big)^{-1}[\bm{d}_h(l^n)]^T$.

Adding the regularizer $\R(\x,\Z,\Gamma)$ to $\mathbf{Q}(\x;\x^n) $, we obtain the following majorizer for $G(\x,\Z,\Gamma)$:
\begin{equation}\label{eq:major_all}
\begin{aligned}
F(\x,\Z,\Gamma;\x^n) &\triangleq \phi(\x;\x^n) + Q_c^n+  \R(\x,\Z,\Gamma) + \mathfrak{X}(\x).
\end{aligned}
\end{equation}
Dropping the constant term $Q_c^n$, the overall surrogate function for $G(\x,\Z,\Gamma)$ in the $n$th iteration becomes
	\begin{equation}\label{eq:surr_all}
	\Phi(\x,\Z,\Gamma;\x^n) = \phi(\x;\x^n) + \R(\x,\Z,\Gamma) + \mathfrak{X}(\x).
	\end{equation}
\subsection{Proof of Theorem~\ref{them1} - Part 1}
Here, we show that for an initial $(\x^0, \Z^0, \Gamma^0)$, iterative sequence $\{\x^n, \Z^n, \Gamma^n\}$ generated by the SPULTRA algorithm is bounded, and the corresponding objective sequence $\{G(\x^n, \Z^n, \Gamma^n)\}$ decreases monotonically and converges to ${G^* \triangleq G^*(\x^0, \Z^0, \Gamma^0)}$.
\subsubsection{Boundedness of the sequence $\{\x^n, \Z^n, \Gamma^n\}$}
It is obvious that the sequences $\{\x^n\}$ and $\{\Gamma^n\}$ are bounded, because of the constraints in \eqref{eq:P0}. Since $\z_j^{n}= \mathit{H}_{\gamma_c}(\omg_{\Gamma_j^n}\P_j\x^n)$ is obtained by hard-thresholding a bounded input, the sequence $\{\Z^n\}$ is also bounded.

\subsubsection{Monotone decrease of the objective function $G(\x,\Z,\Gamma)$}\label{sec:decrease}
First, we discuss the objective behavior in each step of the algorithm.
\paragraph{Image update step}\label{sec:img_upd}With $\Z$ and cluster assignments $\Gamma$ fixed, the cost function for the image update step is ${\Phi(\x,\Z^n,\Gamma^n;\x^n)}$. $\Phi(\cdot)$ as in \eqref{eq:surr_all} is a sum of quadratic functions and the simple barrier function $\mathfrak{X}(\x)$, and many approaches can be used to minimize it. Assuming it is solved exactly, we have 
\begin{equation}\label{eq:img_converge1}
\x^{n+1}\in \arg \min_{\x}  \Phi(\x,\Z^n,\Gamma^n;\x^n),
\end{equation}
or equivalently, ${\x^{n+1}\in \arg \min_{\x}  F(\x,\Z^n,\Gamma^n;\x^n)}.$

Since $F(\x,\Z,\Gamma;\x^n)$ is the majorizer of $G(\x,\Z,\Gamma)$, we have
\begin{equation}\label{eq:mono_img}
\begin{aligned}
&G(\x^{n+1},\Z^n,\Gamma^n) \leq F(\x^{n+1},\Z^n,\Gamma^n;\x^n) \\
&\leq F(\x^{n},\Z^n,\Gamma^n;\x^n)  = G(\x^{n},\Z^n,\Gamma^n)
\end{aligned}
\end{equation}

\paragraph{Sparse coding and clustering step}
    With $\x$ fixed, the relevant part of the cost function for the sparse coding and clustering step is ${\mathsf{R}(\x^{n+1}, \Z, \Gamma)}$. Since the solution with respect to ${(\Z, \Gamma)}$ is computed exactly as described in {Section~III.~C} in \cite{TMI-SPULTRA-as-submit}, we have
\begin{equation}
(\Z^{n+1},\ \Gamma^{n+1}) \in \arg \min_{\Z, ~\Gamma}  \mathsf{R}(\x^{n+1}, \Z, \Gamma).
\end{equation}
This then implies
\begin{equation}\label{eq:spa_clu_convergence_G}
(\Z^{n+1},\ \Gamma^{n+1}) \in \arg \min_{\Z, ~\Gamma}  G(\x^{n+1},\Z, \Gamma).
\end{equation}\\
	
Therefore, $G(\x^{n+1},\Z^{n+1}, \Gamma^{n+1}) \leq G(\x^{n+1},\Z, \Gamma)$. Combining this with \eqref{eq:mono_img} implies that the objective decreases in each outer iteration. In other words, the objective sequence $\{G^n \triangleq G(\x^n, \Z^n, \Gamma^n)\}$ is monotonically decreasing. Moreover, the objective $G$ is readily lower bounded by ${N_d \sigma^2 - (\sum_{i = 1}^{N_d}Y_i)\log(I_0 + \sigma^2)}$. Therefore, it converges to some limit ${G^* \triangleq G^*(\x^0, \Z^0, \Gamma^0)}$.

\subsection{Proof of Theorem~\ref{them1} - Part 2}\label{sec:property2}
Here, we show that all the accumulation points of the iterate sequence are equivalent and achieve the same value $G^*$ of the objective function.

%\subsubsection{Equivalence of the accumulation points}
 Since the sequence $\{\x^n, \Z^n, \Gamma^n\}$ is bounded, it follows from the Bolzano-Weierstrass Theorem that there exists a convergent subsequence and a corresponding accumulation point. In order to show that all the accumulation points of $\{\x^n, \Z^n, \Gamma^n\}$ achieve the same value of $G^*$, we consider an arbitrary convergent subsequence $\{\x^{q_m}, \Z^{q_m}, \Gamma^{q_m}\}$, and show that $G(\x^*, \Z^*, \Gamma^*) = G^*$ for the accumulation point $(\x^*, \Z^*, \Gamma^*)$.
 
 First, the objective satisfies
\begin{equation}\label{eq:acc_srr}
G^{q_m} \triangleq G(\x^{q_m}, \Z^{q_m}, \Gamma^{q_m}) = \mathsf{L}(\x^{q_m})+ \R(\x^{q_m}, \Z^{q_m}, \Gamma^{q_m}).
\end{equation}
Clearly, $\{G^{q_m}\}$ converges to $G^*$. Since ${\x^{q_m} \to \x^*}$ and ${\Z^{q_m} \to \Z^*}$ as $m\to \infty$, and $\mathsf{L}(\x)$ is a continuous function, therefore, $\mathsf{L}(\x^{q_m} )\to \mathsf{L}(\x^{*})$. Since $\Z^{q_m}$ does not contain any \DIFadd{non-zero} entries with magnitude less than $\gamma_c$ and $\Z^{q_m} \to \Z^*$, clearly, the support (i.e., locations of non-zeros) of $\Z^{q_m}$ must coincide with the support of $\Z^*$ after finitely many iterations. Similarly, because $\{\Gamma^{q_m}\}$is an integer-vector sequence, $\Gamma^{q_m}$ converges to $\Gamma^{*}$ in a finite number of iterations. Therefore, taking the limit $m\to \infty$ term by term in $G(\x^{q_m}, \Z^{q_m}, \Gamma^{q_m})$ yields
\begin{equation}\label{eq:acc_G}
\lim_{m\to \infty} G(\x^{q_m}, \Z^{q_m}, \Gamma^{q_m}) = G(\x^*, \Z^*, \Gamma^*).
\end{equation}
Combining \eqref{eq:acc_G} with the fact that $G^{q_m} \to G^*$, we obtain 
\begin{equation}\label{eq:propert2}
G(\x^*, \Z^*, \Gamma^*) = G^*.
\end{equation}
Thus, any accumulation point of $\{\x^n, \Z^n, \Gamma^n\}$ achieves the value $G^*$ for the cost.

\subsection{Proof of Theorem~\ref{them1} - Part 3}\label{sec:property3}
Here, we show that each accumulation point $(\x^*,\Z^*,\Gamma^*)$ satisfies the partial optimality conditions in \eqref{eq:critical-point}. The proof uses the following Lemma~1.
\begin{lemma}\label{lemma1}
Consider the subsequence $\{\x^{q_m}, \Z^{q_m-1}, \Gamma^{q_m-1}\}$  that converges to the accumulation point $(\x^*, \Z^{**}, \Gamma^{**})$, then the subsequence $\{\x^{q_m-1}\}$ also converges to $\x^*$, with $\x^*$ being the unique minimizer of $F(\x, \Z^{**}, \Gamma^{**}; \x^{*})$ with respect to $\x$.
\end{lemma}
\textit{Proof of Lemma~\ref{lemma1}}:\\
Since $\{\x^{q_m-1}\}$ is bounded, there exists a convergent subsequence $\{\x^{q_{m_t}-1}\}$ which converges to $\x^{**}$.

The following inequalities follow from \eqref{eq:mono_img} and \eqref{eq:spa_clu_convergence_G}:
\begin{equation}\label{eq:mono_subseq}
\begin{aligned}
G^{q_{m_t}} &= G(\x^{q_{m_t}}, \Z^{q_{m_t}},\Gamma^{q_{m_t}}) \leq G(\x^{q_{m_t}}, \Z^{q_{m_t}-1},\Gamma^{q_{m_t}-1})\\
&\leq F(\x^{q_{m_t}}, \Z^{q_{m_t}-1},\Gamma^{q_{m_t}-1};\x^{q_{m_t}-1}) \\
&\leq F(\x^{q_{m_t}-1}, \Z^{q_{m_t}-1},\Gamma^{q_{m_t}-1};\x^{q_{m_t}-1})  \\
& = G(\x^{q_{m_t}-1}, \Z^{q_{m_t}-1},\Gamma^{q_{m_t}-1}) =G^{q_{m_t}-1} .
\end{aligned}
\end{equation}
Since $G^{q_{m_t}}$ and $G^{q_{m_t}-1}$ are successive elements from the sequence $\{G^n\}$, and $\{G^n\}$ converges to $G^*$, then taking the limit $t\to \infty$ throughout \eqref{eq:mono_subseq} yields
\begin{subequations}
\begin{equation}
G^* \leq F(\x^{*}, \Z^{**}, \Gamma^{**};\x^{**}) \leq F(\x^{**}, \Z^{**}, \Gamma^{**};\x^{**})\leq G^*.
\end{equation}
Thus, 
\begin{equation}\label{eq:37b}
F(\x^{*}, \Z^{**}, \Gamma^{**};\x^{**}) = F(\x^{**}, \Z^{**}, \Gamma^{**};\x^{**}) = G^*.
\tag{35(b)}
\end{equation}
\end{subequations}

Since \eqref{eq:major_all} is a quadratic cost with simple box constraints on $\x$, the Hessian of the quadratic terms with respect to $\x$ is
\begin{equation}\label{eq:H_F}
\mathbf{H}(\x) = \A^T\W^n\A + 2\beta \sum_{j=1}^{N}   \P_j^T\omg_{\Gamma{j}}^T\omg_{\Gamma{j}}\P_j .
\end{equation}

Clearly, $\A^T\W^n\A$ is non-negative definite, and $\sum_{j=1}^{N}   \P_j^T\omg_{\Gamma{j}}^T\omg_{\Gamma{j}}\P_j $ is positive definite\DIFadd{\cite{STlearning15,2015BCS-ST}}. Since $\beta$ is a positive scalar, the Hessian in \eqref{eq:H_F} is positive definite. This implies that the minimization of $F(\cdot)$ (quadratic with a box constraint) has a unique solution~\cite{bjorck1996numerical, mead2010least,rojas2002interior}. Moreover, since the following inequality holds for all $\x$ satisfying the problem constraints
\begin{equation}
\begin{aligned}
   & F(\x^{q_{m_t}}, \Z^{q_{m_t-1}},\Gamma^{q_{m_t-1}};\x^{q_{m_t}-1}) \\
&\leq F(\x, \Z^{q_{m_t-1}},\Gamma^{q_{m_t-1}};\x^{q_{m_t}-1}),
\end{aligned}
\end{equation}
taking the limit $t\to \infty$ above and using similar arguments as for \eqref{eq:acc_G} yields
\begin{equation}
\begin{aligned}
  F(\x^{*}, \Z^{**},\Gamma^{**};\x^{**}) 
\leq F(\x, \Z^{**},\Gamma^{**};\x^{**}),
\end{aligned}
\end{equation}
implying that $\x^*$ is a minimizer of ${F(\x, \Z^{**},\Gamma^{**};\x^{**})}$. Since the minimizer of ${F(\x, \Z^{**},\Gamma^{**};\x^{**})}$ with respect to $\x$ is unique, and using \eqref{eq:37b} immediately implies ${\x^{**} = \x^{*}}$.

Since $\{\x^{q_{m_t}-1}\}$  is an arbitrary subsequence of $\{\x^{q_{m}-1}\}$, $\x^*$ is the limit of any convergent subsequence of $\{\x^{q_{m}-1}\}$. In other words, $\x^*$ is the unique accumulation point of the bounded sequence, i.e., $\{\x^{q_{m}-1}\}$ itself converges to $\x^*$.

This completes the proof of the Lemma.\\

%\subsubsection{Proof of theorem part 2}
We have shown in the proof of Lemma~\ref{lemma1} that $\x^{**}$ is a unique minimizer of the quadratic function ${F(\x, \Z^{**},\Gamma^{**};\x^{**})}$. This means that ${{\mathbf{0} \in \partial_{\x} F(\x, \Z^{**},\Gamma^{**};\x^{**})|_{\x = \x^{**}}}}$.
It is easy to show that we can equivalently consider the sequence $\{\x^{q_m}, \Z^{q_m}, \Gamma^{q_m}\}$ converging to $(\x^*,\Z^*,\Gamma^*)$ for which
\begin{equation}\label{eq:partial_F_0}
\mathbf{0} \in \partial_{\x} F(\x,\Z^*,\Gamma^*; \x^*)|_{\x = \x^{*}}.
\end{equation}
Based on the definition of the majorizer of L(x), we have
\begin{equation}\label{eq:dphi*}
\nabla\phi(\x;\x^*)|_{\x = \x^{*}}= \nabla\mathsf{L}(\x)|_{\x = \x^{*}},
\end{equation}
where $\nabla$ is the gradient operator on continuous functions.
Since the quadratic surrogate and regularizer components of $F(\cdot)$ have exact gradients, combining \eqref{eq:dphi*} with \eqref{eq:partial_F_0} yields
\begin{equation}
\mathbf{0} \in \partial_{\x}G(\x,\Z^*,\Gamma^*)|_{\x = \x^{*}}.
\end{equation}
In other words, $\x^*$ is a critical point of $G(\x,\Z^*,\Gamma^*)$.

To show the partial optimality condition for $(\Z^*, \Gamma^*)$ as in \eqref{eq:critical-point}, we first use \eqref{eq:spa_clu_convergence_G} for the subsequence $\{\x^{q_m}, \Z^{q_m}, \Gamma^{q_m}\}$ yielding
\begin{equation}
G(\x^{q_m}, \Z^{q_m}, \Gamma^{q_m}) \leq G(\x^{q_{m-1}}, \Z, \Gamma),  \  \forall (\Z,\Gamma).
\end{equation}
Then, taking the limit ${m \to \infty}$ above and using \eqref{eq:acc_G} and Lemma~\ref{lemma1}, we get
\begin{equation}
G(\x^*, \Z^*, \Gamma^*) \leq G(\x^*, \Z, \Gamma), \  \forall (\Z,\Gamma),
\end{equation}
which can be equivalently written as
	\begin{equation}\label{eq:z_Ck_PGO*}
	(\Z^{*},\Gamma^{*}) \in \arg \min_{\Z,\Gamma}  G(\x^{*},\Z,\Gamma).
	\end{equation}
	
\subsection{Proof of Theorem~\ref{them1} - Part 4}
Here, we show that $\|\x^{n+1} - \x^n\|_2 \to 0$ as $n\to \infty$. Since $\{\x^n\}$ is bounded, $\|\x^n\|_2 \leq C$ for some $C>0$ and all $n$. Therefore, the sequence $\{e^n\}$ is also bounded, with $e^n \triangleq \|\x^{n+1} - \x^n\|_2 \leq 2C$, $\forall$ $n$. Hence, there exists a convergent subsequence $\{e^{q_m}\}$ of $\{e^n\}$. For the bounded sequence 
$\{\x^{q_{m}+1}, \Z^{q_{m}},\Gamma^{q_{m}}\}$, 
there exists a convergent subsequence $\{\x^{q_{m_t}+1},\Z^{q_{m_t}},\Gamma^{q_{m_t}}\}$ 
converging to $(\x^*,\Z^*, \Gamma^*)$. Moreover, by Lemma~\ref{lemma1}, the sequence $\{\x^{q_{m_t}}\}$ also converges to $\x^*$. Therefore, clearly the subsequence $\{e^{q_{m_t}}\}$ with $e^{q_{m_t}}\triangleq ||\x^{q_{m_t}+1}- \x^{q_{m_t}}||_2$ converges to 0. Since $\{e^{q_{m_t}}\}$ is a subsequence of the convergent $\{e^{q_{m}}\}$, then $\{e^{q_{m}}\}$ has the same limit, i.e., 0. As the convergent subsequence $\{e^{q_{m}}\}$ is chosen arbitrarily from $\{e^n\}$, we conclude that $0$ is the only accumulation point of $\{e^n\}$. Thus, $||\x^{n+1} - \x^n||_2 \to 0$ as $n\to \infty$.

\section{Additional Experimental Results}
\subsection{Behavior of the Learned ULTRA Models}
Here, we further illustrate the sparse coefficient maps generated by SPULTRA.
\begin{figure}[!htbp]
	\centering
	\includegraphics[width=0.4\textwidth]{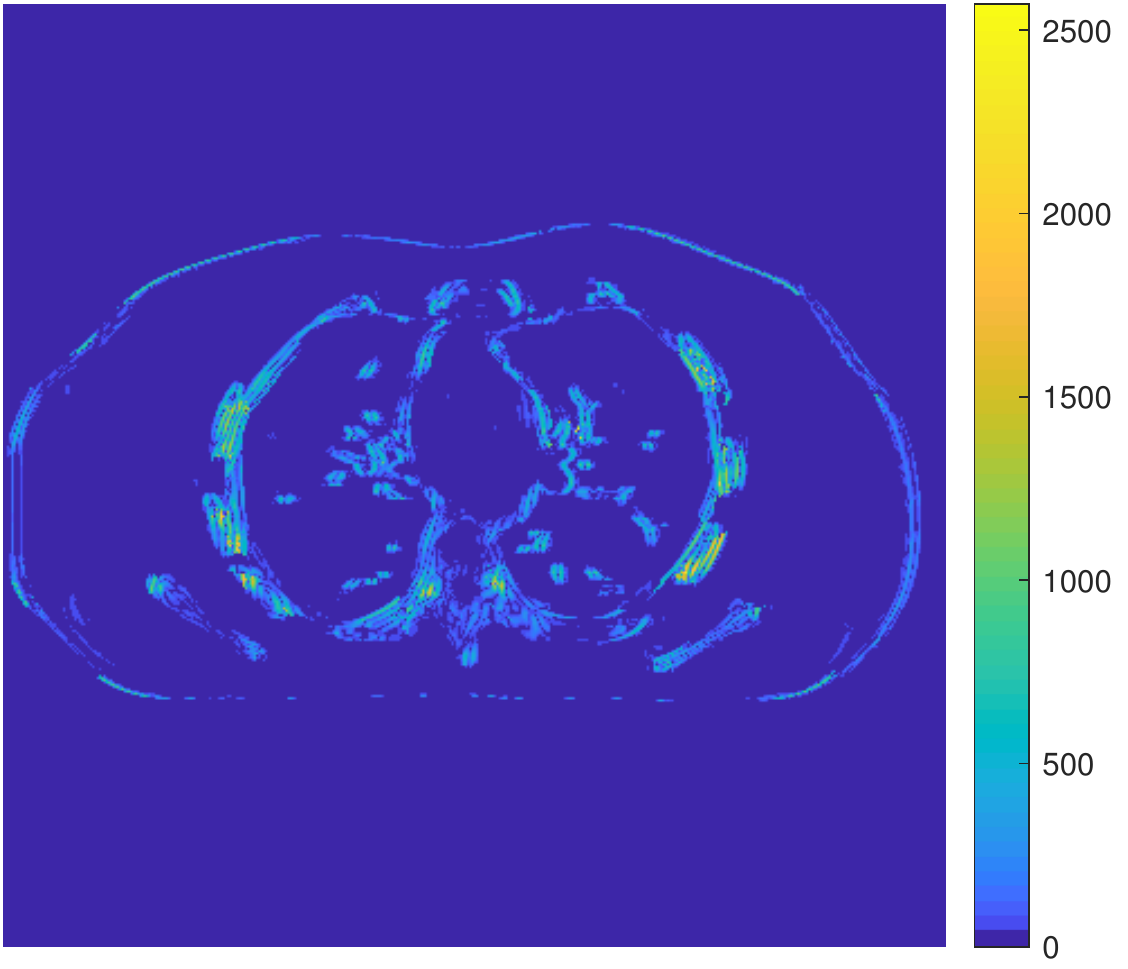}
	\caption{Sum of the disjoint sparse coefficient maps generated by the 81st filter from all classes.}
	\label{fig:spacod_row81_all}
	\vspace{-0.1in}
\end{figure}
The sparse code vectors $\z_j$ in (3) can be concatenated as columns of a sparse code matrix $\Z$. Fig.~2 in~\cite{TMI-SPULTRA-as-submit} displays the axial slice of the sparse coefficient volume obtained from the 81st row of $\Z$.
This represents the effective map for the 81st filter of all classes (composed as the sum of the 81st filter's map from each class).
Fig.~\ref{fig:spacod_row81_classes} shows the underlying maps for the 81st filter for all classes obtained by masking out (or setting to zero) pixels in Fig.~\ref{fig:spacod_row81_all} that correspond to patches not in the class. The filters are shown at the top left corner of the sparse coefficient images. Thus, in the ULTRA model, several filters with different properties and different features or edges collaboratively help form the ``effective" sparse coefficient maps.
\begin{figure}[!htp]
	\centering
	\includegraphics[width=0.5\textwidth]{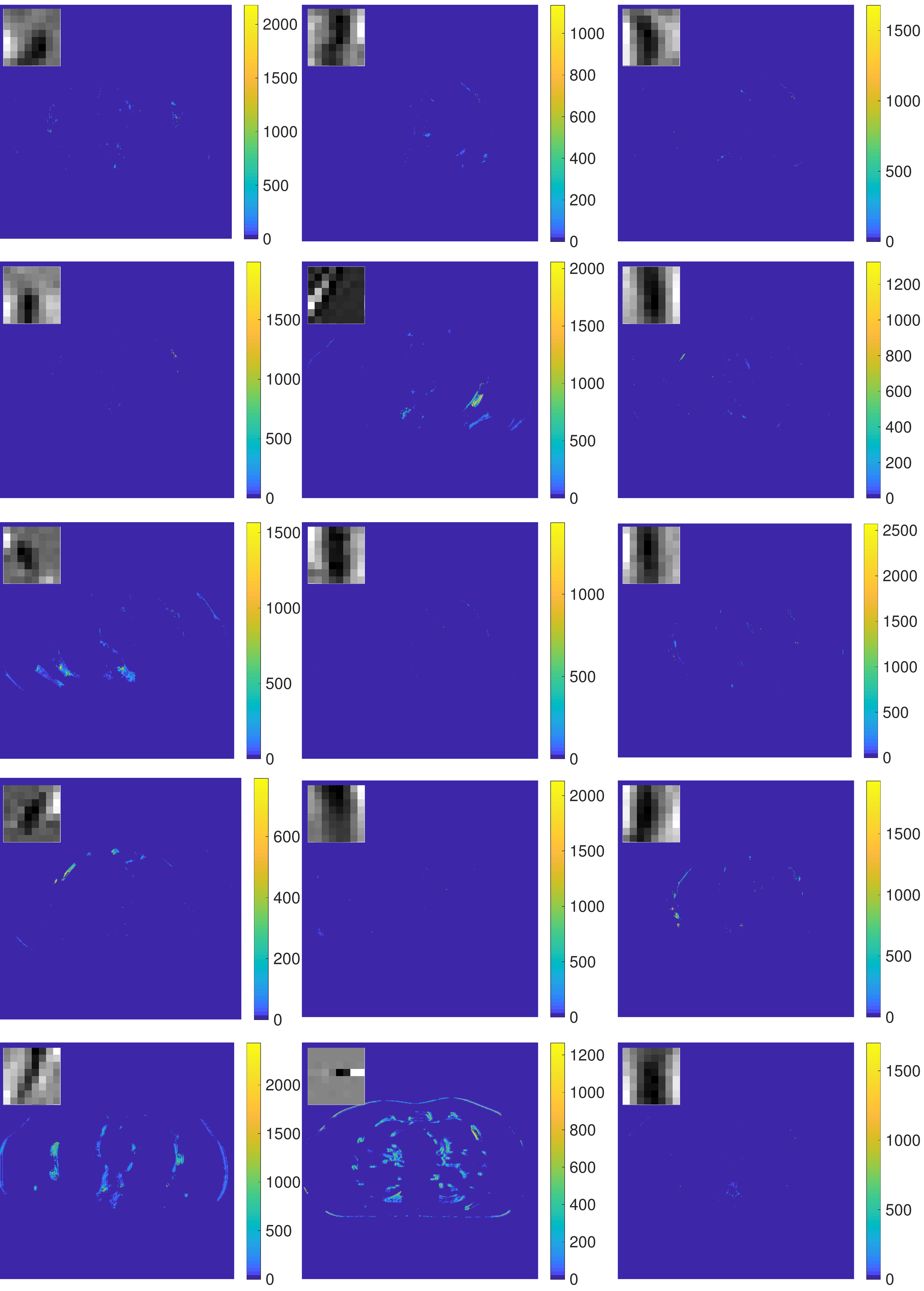}
	\caption{Sparse coefficient map (axial slice) for the $81$st filter of each class.}
	\label{fig:spacod_row81_classes}
	\vspace{-0.2in}
\end{figure}
\subsection{\DIFadd{Clustering results in low-dose situations}}
\begin{figure}[!htp]
	\centering
	\includegraphics[width=0.45\textwidth]{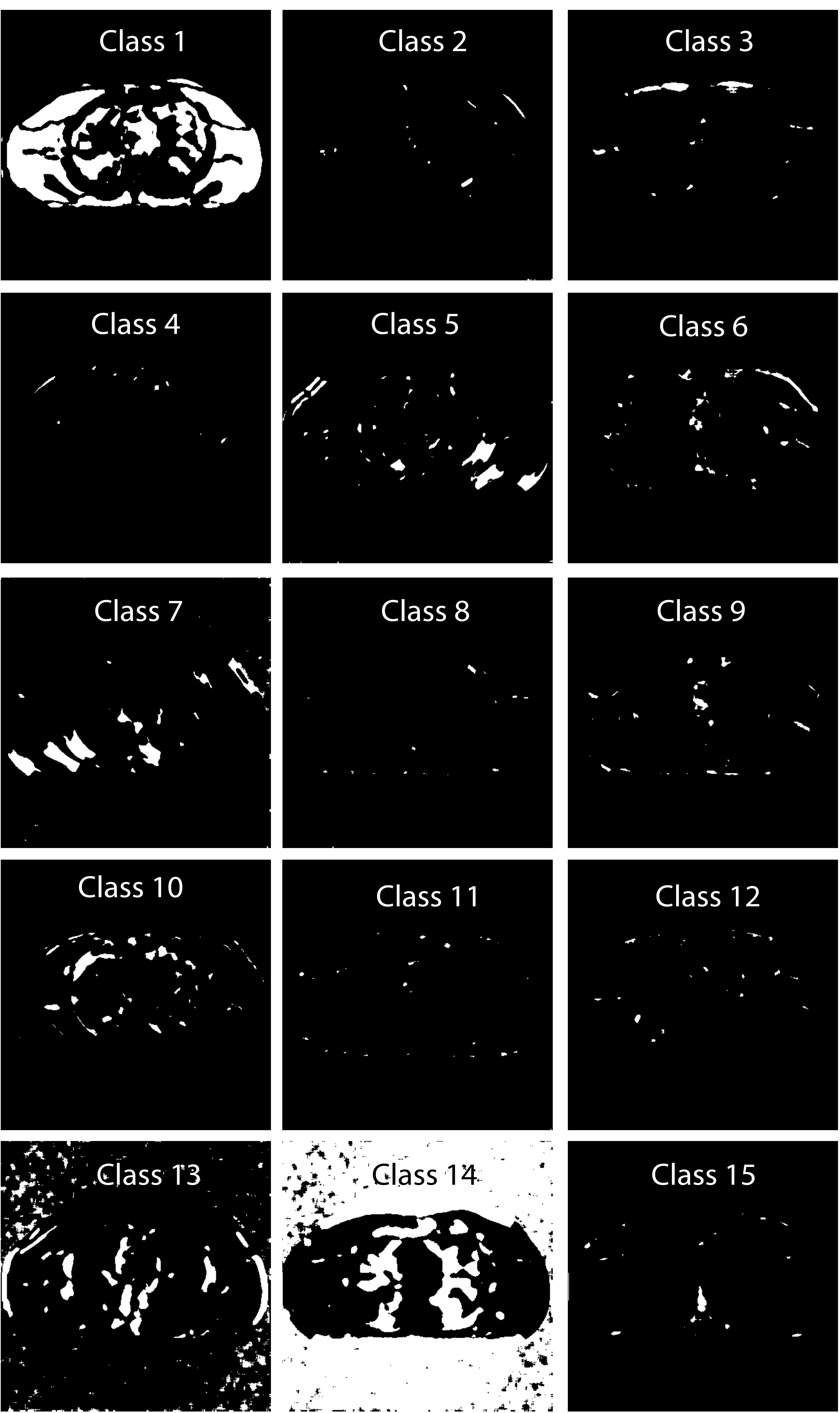}
	\caption{\DIFadd{Binary images showing the clustering memberships of pixels in the central axial slice of the XCAT phantom reconstructed at $I_0 = 2\times 10^3$.}}
	\label{fig:cluster_2e3_all}
	\vspace{-0.05in}
\end{figure}

\DIFadd{Fig.~2 in the manuscript showed 3 out of 15 voxel-level clustering results of the reconstructed image at $I_0 = 1\times 10^4$. Here, Fig.~\ref{fig:cluster_2e3_all} is a binary image showing clustering memberships of all the classes for the reconstruction at $I_0 = 2\times 10^3$. The white regions indicate pixels assigned to the corresponding class. The voxel-level clustering results (that display the pixels using their reconstructed intensities) at $I_0 = 2\times 10^3$ are actually similar to the ones shown in Fig.~2 (first column) in the manuscript. Specifically, Tab.~\ref{tab:cluster_perc} shows the percentages of pixels assigned to Class~1, 13 and 14 respectively. Although $I_0 = 2\times 10^3$ is a much lower dose compared with $I_0 = 1\times 10^4$, the clustering results only have slight changes. This illustrates that the voxel clustering based on majority vote of overlapping patches is robust in low-dose situations.}

\begin{table}[!htp]
	\centering
		\caption{\DIFadd{Percentages of pixels belonging to Class~1, Class~13, and Class~14.}}
		\begin{tabular}{l c c c}
		\toprule
	\DIFadd{$\quad I_0$}	&\DIFadd{Class 1} & \DIFadd{Class 13} &\DIFadd{Class 14}\\ \midrule
	\DIFadd	{$1\times 10^4$ } & \DIFadd{18.7 $\%$}	& \DIFadd{6.6 $\%$}&\DIFadd{65.5 $\%$}  \\   \midrule
	\DIFadd	{$2\times 10^3$} &\DIFadd{18.0 $\%$} & \DIFadd{6.9 $\%$}& \DIFadd{ 64.4 $\%$}\\   \bottomrule
	\end{tabular}
\vspace{-0.21in}
\label{tab:cluster_perc}
\end{table}
\subsection{Zoom-ins of ROI 2  and ROI 3 in the XCAT phantom simulations}
%\vspace{-0.03in}
\begin{figure}[!htbp]
	\centering
	\begin{subfigure}[h]{0.48\textwidth}
		\centering	
		\includegraphics[width=1\textwidth]{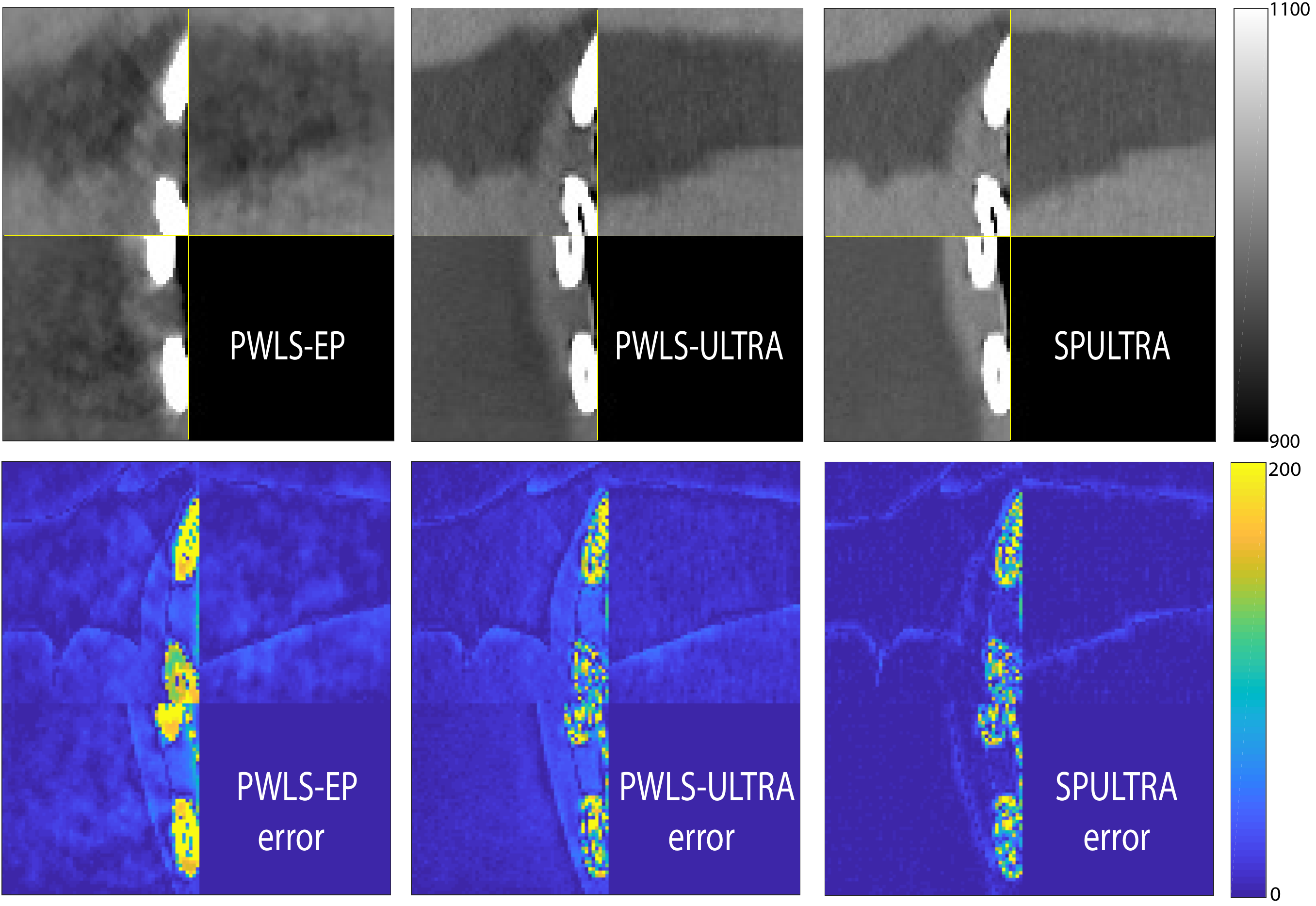}
		\caption{$I_0 = 3\times 10^3$}
		\label{fig:xcat-3e3-roi2}
	\end{subfigure}
	\vfill
	\begin{subfigure}[h]{0.48\textwidth}
		\centering	
		\includegraphics[width=1\textwidth]{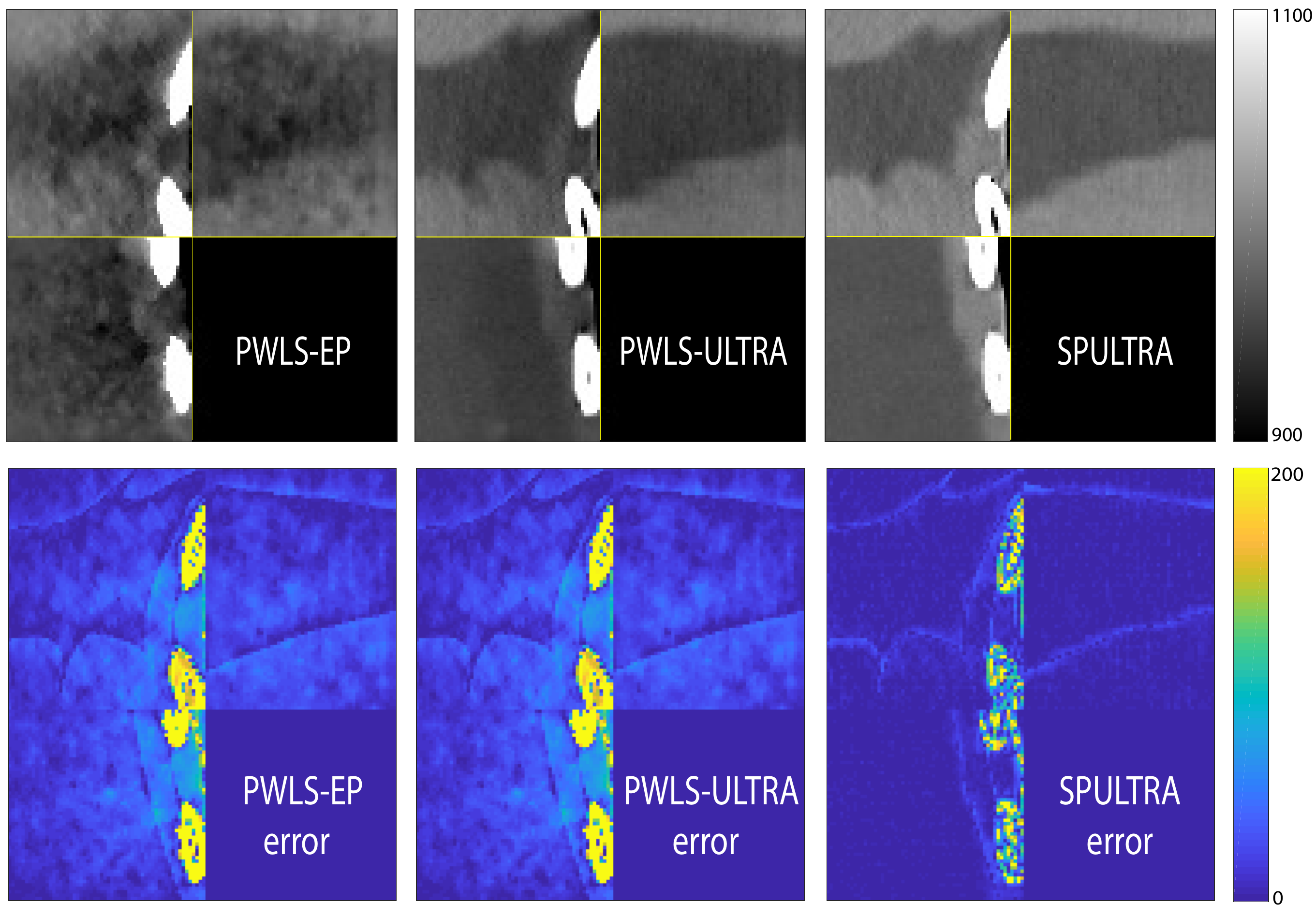}
		\caption{$I_0 = 2\times 10^3$}
		\label{fig:xcat-2e3-roi2}
	\end{subfigure}
	\caption{\DIFadd{Plots of the ROI 2 (central axial, sagittal, and coronal slices of the 3D volume). The display windows for the reconstructed ROI and the corresponding error image are {[900,~1100]} HU and [0,~200]~HU, respectively.}}
	\label{fig:xcat-recon-roi2}
	\vspace{-0.1in}
\end{figure}

\begin{figure}[!htbp]
	\centering
	\begin{subfigure}[h]{0.45\textwidth}
		\centering	
		\includegraphics[width=1\textwidth]{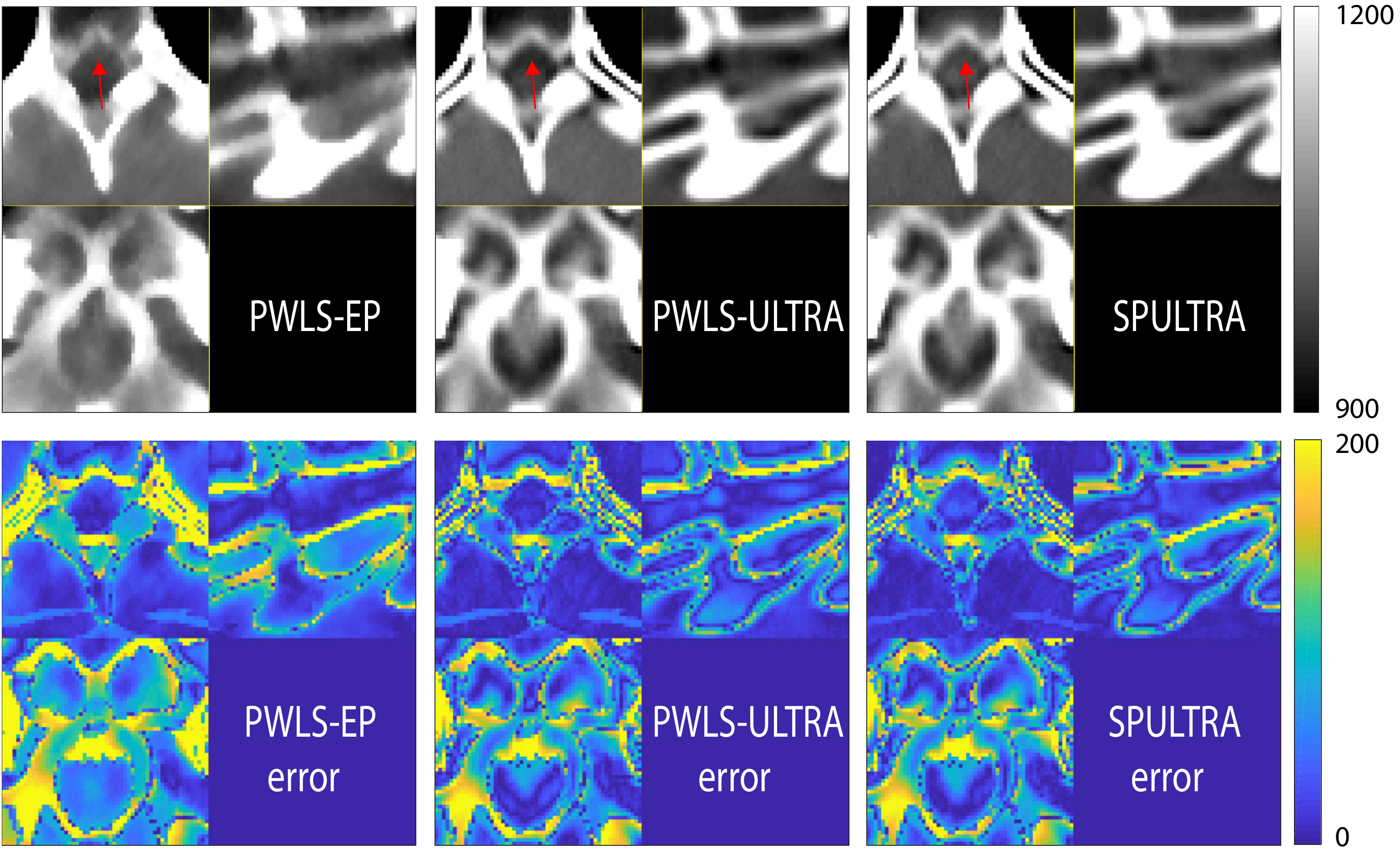}
		\caption{$I_0 = 3\times 10^3$}
		\label{fig:xcat-3e3-roi3}
	\end{subfigure}
	\vfil
	\begin{subfigure}[h]{0.45\textwidth}
		\centering	
		\includegraphics[width=1\textwidth]{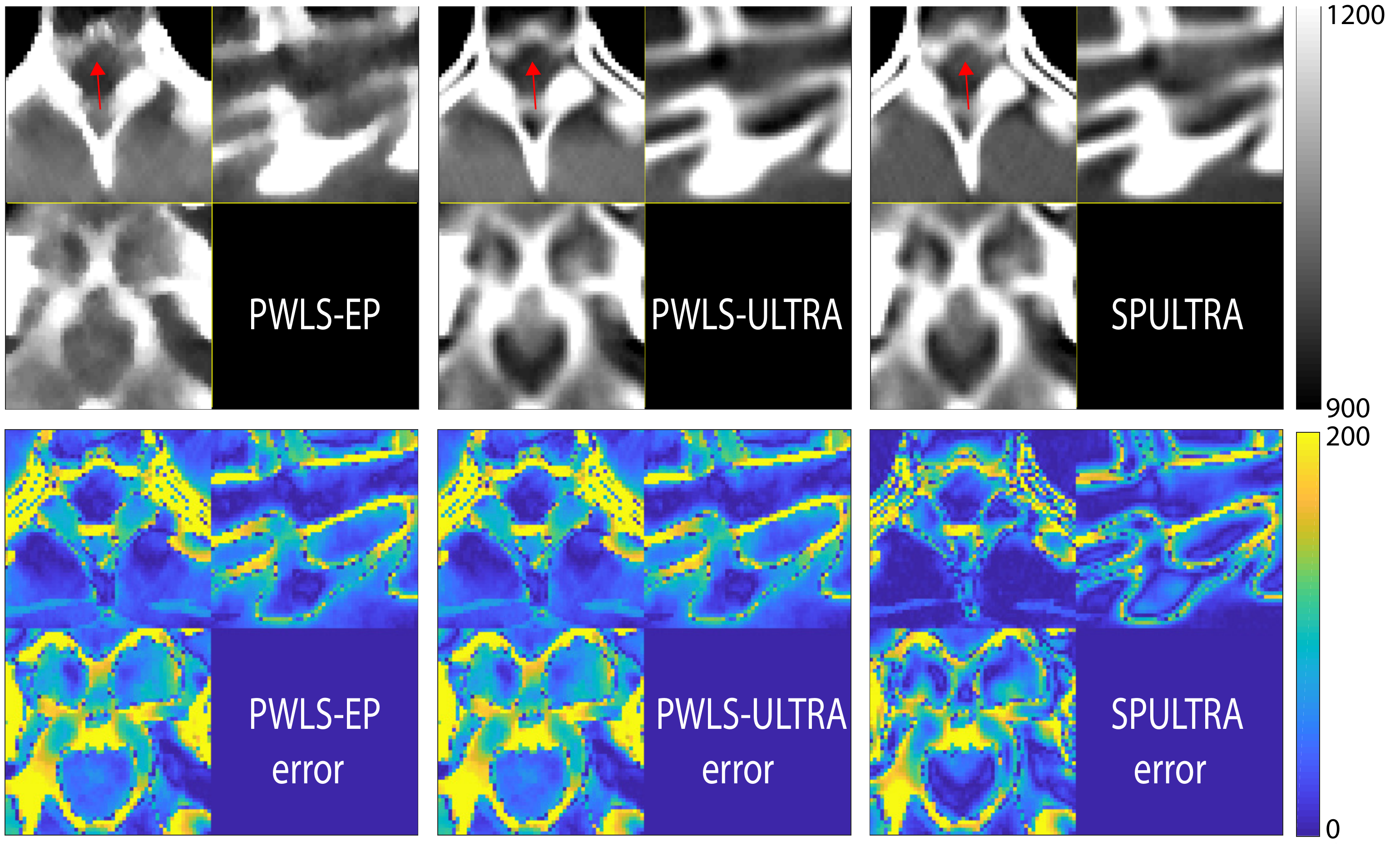}
		\caption{$I_0 = 2\times 10^3$}
		\label{fig:xcat-2e3-roi3}
	\end{subfigure}
	\caption{\DIFadd{Plots of the ROI 3 (central axial, sagittal, and coronal slices of the 3D volume). The display windows for the reconstructed ROI and the corresponding error image are {[900,~1100]} HU and [0,~200]~HU, respectively.}}
	\label{fig:xcat-recon-roi3}
	\vspace{-0.1in}
\end{figure}
	\begin{figure}[!htbp]
	\centering
	\includegraphics[width=0.47\textwidth]{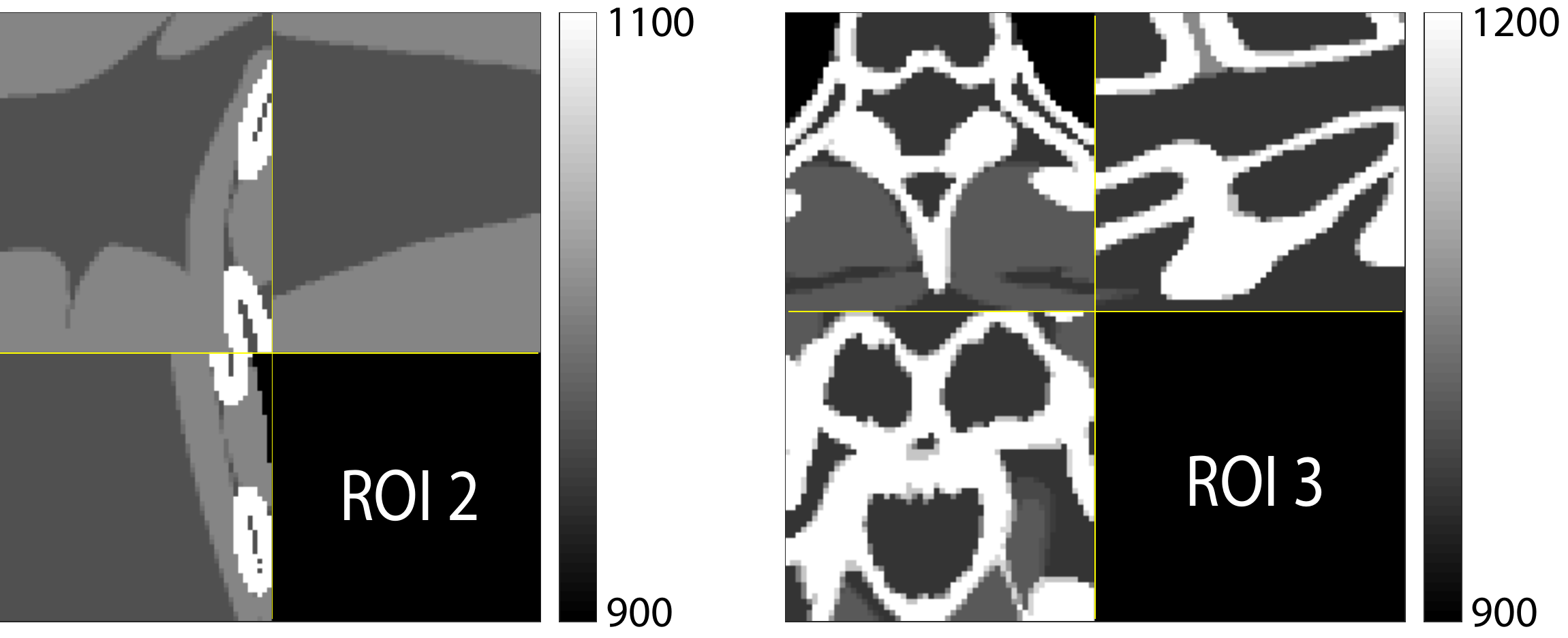}
	\caption{\DIFadd{3D plots of the ground-truth ROI 2 and ROI 3. The display windows for ROI~2 and ROI~3 are [900,~1100]~HU and [900,~1200]~HU, respectively.}}
	\vspace{-0.2in}
	\label{fig:xtrue-roi2-3}
\end{figure}
\DIFadd{Fig.~\ref{fig:xcat-recon-roi2} and Fig.~\ref{fig:xcat-recon-roi3}} plot the zoom-ins \DIFadd{and the corresponding error images} of ROI 2 and ROI 3 for the XCAT phantom simulations in Section~V.A, with $I_0 = 3 \times 10^3$ and $I_0 = 2 \times 10^3$, respectively. In Fig.~\ref{fig:xcat-recon-roi3}, we highlighted a region in the axial slice with small red arrows. \DIFadd{We show the zoom-ins of the ground-truth ROI 2 and ROI 3 of the XCAT phantom in Fig.~\ref{fig:xtrue-roi2-3}. }The results show that SPULTRA improves image quality over PWLS-EP and PWLS-ULTRA by reducing bias and improving image edges.
\subsection{\DIFadd{FBP images of XCAT phantom simulations}}
\begin{figure}[!htbp]
	\centering
	\begin{subfigure}[h]{0.22\textwidth}
		\centering	
		\includegraphics[width=1\textwidth]{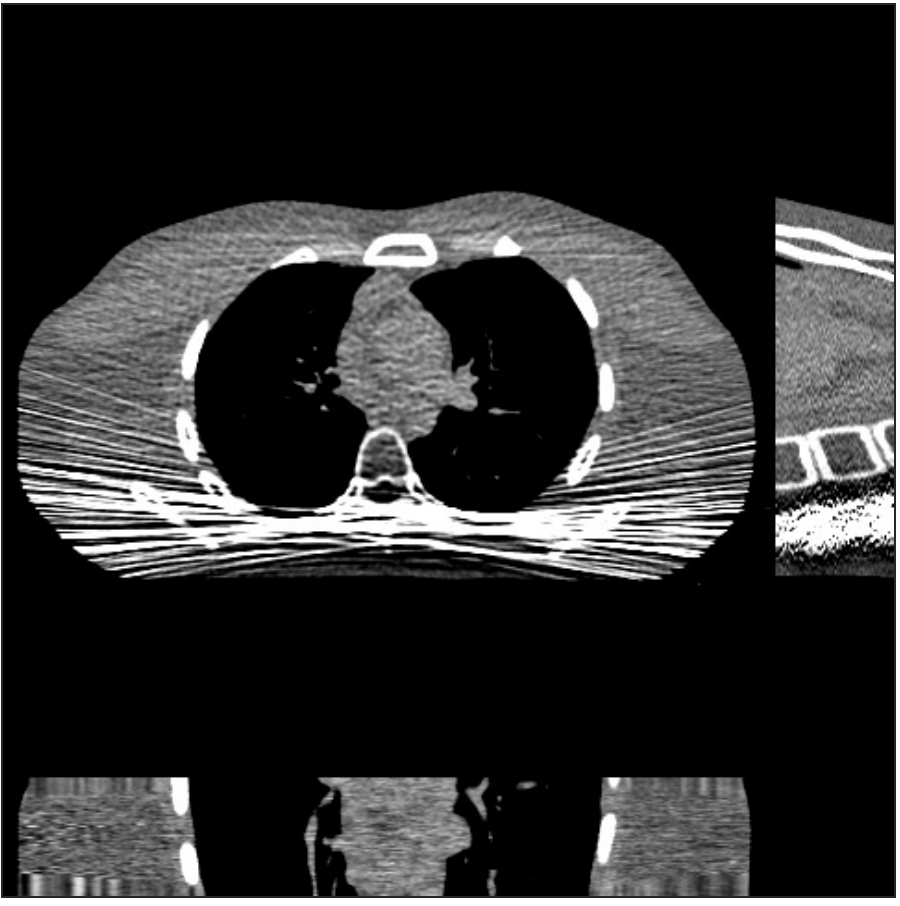}
		\caption{$I_0 = 1\times 10^4$}
		\label{fig:1e4_xfdk}
	\end{subfigure}
	\hfil
	\begin{subfigure}[h]{0.22\textwidth}
		\centering	
		\includegraphics[width=1\textwidth]{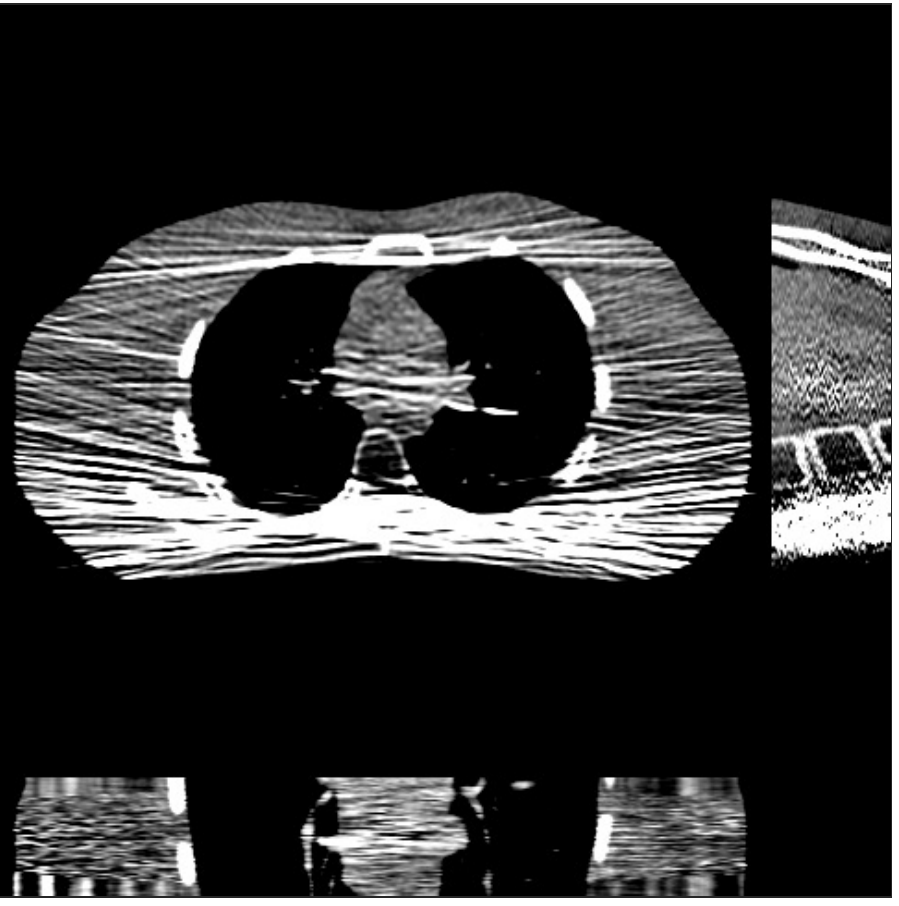}
		\caption{$I_0 = 5\times 10^3$}
		\label{fig:5e3_xfdk}
	\end{subfigure}
	\vfil
	\begin{subfigure}[h]{0.22\textwidth}
		\centering	
		\includegraphics[width=1\textwidth]{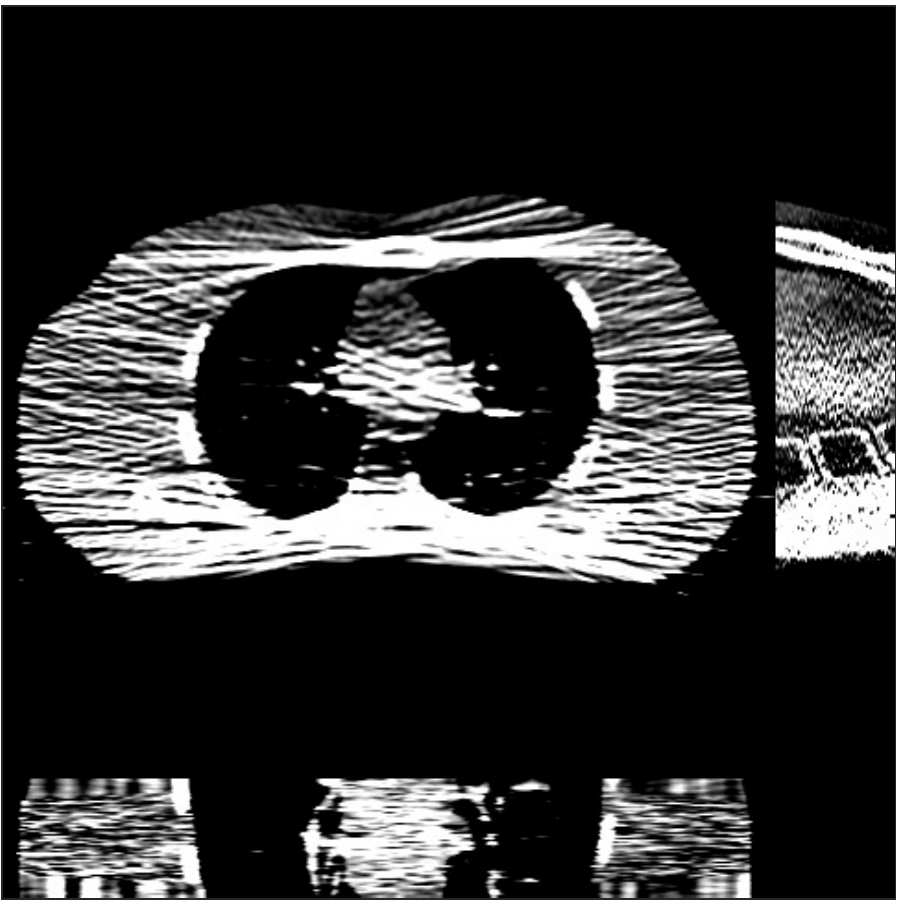}
		\caption{$I_0 = 3\times 10^3$}
		\label{fig:3e3_xfdk}
	\end{subfigure}
	\hfil
	\begin{subfigure}[h]{0.22\textwidth}
		\centering	
		\includegraphics[width=1\textwidth]{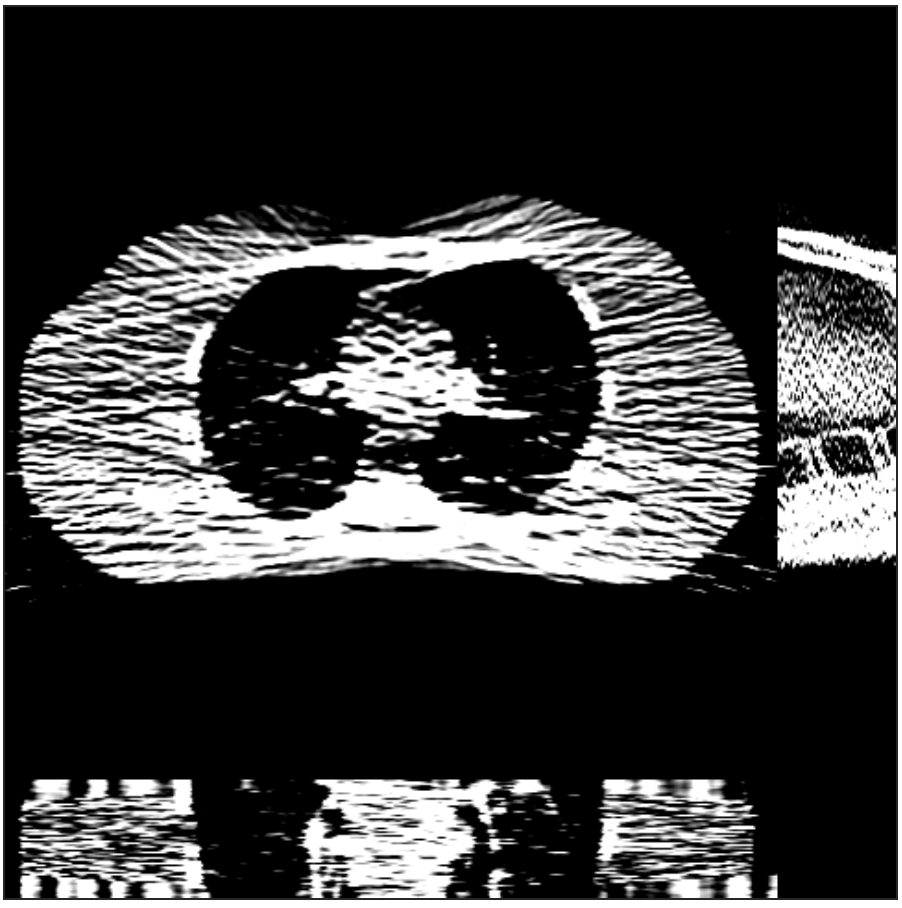}
		\caption{$I_0 = 2\times 10^3$}
		\label{fig:2e3_xfdk}
	\end{subfigure}
	\caption{\DIFadd{FDK reconstructions for XCAT phantom simulations at different doses. The display window is [800,~1200]~HU.}}
	\label{fig:fdk_xcat}
	\vspace{-0.1in}
\end{figure}

\vspace{-0.05in}
\DIFadd{In XCAT phantom simulations, the PWLS-EP algorithm was initialized with an image reconstructed by the FDK \cite{feldkamp1984practical} method. Fig.~\ref{fig:fdk_xcat} shows the FDK reconstructed images for all the tested doses in Section V.A. These images have substantial streak artifacts and noise.
}
\subsection{Ultra Low-dose 2D Shoulder Data Simulations}
\subsubsection{\DIFadd{Initialize WavResNet with the FBP image}}
\begin{figure}[!htp]
	\centering
	\begin{subfigure}[h]{0.22\textwidth}
		\centering	
		\includegraphics[width=1\textwidth]{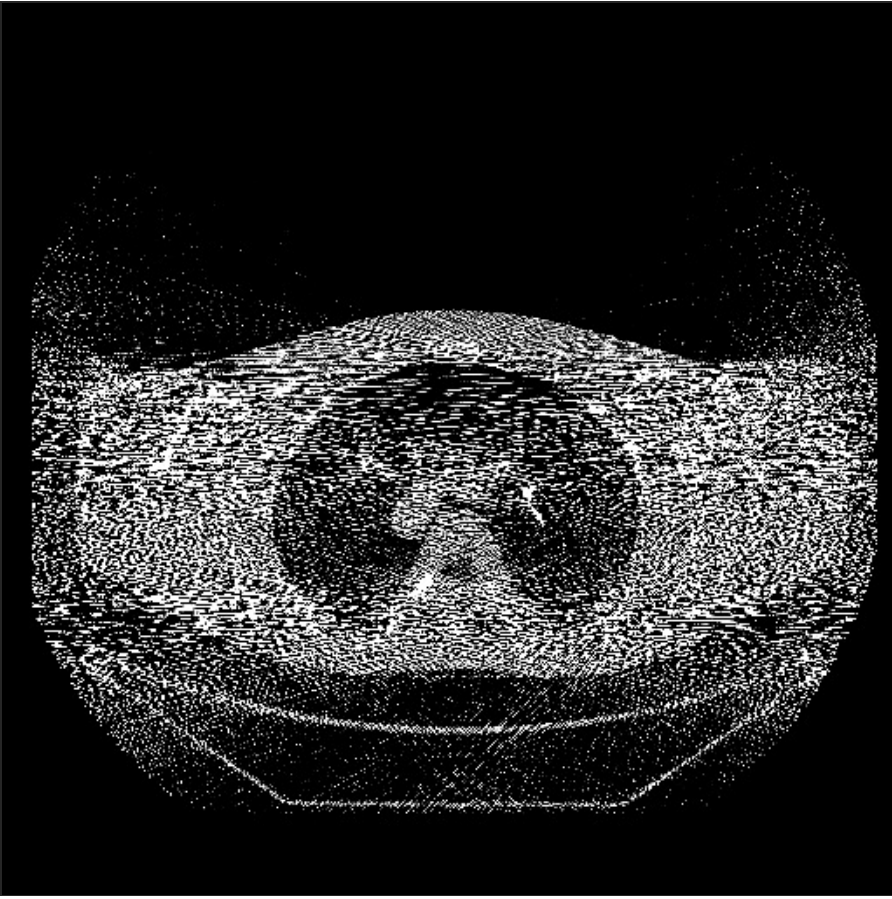}
		\caption{FBP input}
		\label{fig:shoulder_fbp}
	\end{subfigure}
	\hfil
	\begin{subfigure}[h]{0.22\textwidth}
		\centering	
		\includegraphics[width=1\textwidth]{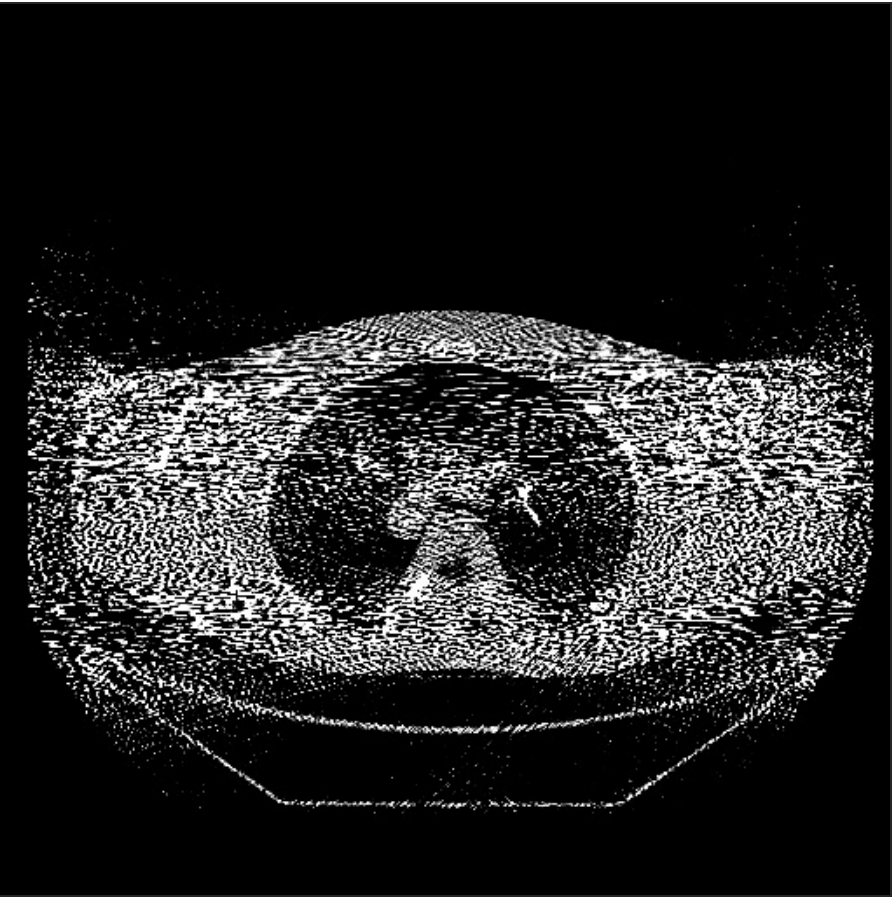}
		\caption{RNN versioned WavResNet}
		\label{fig:wavresnet-6iter}
	\end{subfigure}
	\caption{Iterative RNN versioned WavResNet result with an FBP image input. The display window is [800, 1400] HU.}
	\label{fig:wavresnet-fbpinit}
	\vspace{-0.1in}
\end{figure}

In~\cite{TMI-SPULTRA-as-submit}, we presented the denoised image obtained using the iterative RNN version of WavResNet with the PWLS-EP reconstructed image as input. 
Since we used the optimal parameters reported in~\cite{WavResNet18} for WavResNet, wherein the inputs are reconstructed images using the filtered backprojection (FBP) method, here we also show the result obtained by using the FBP reconstructed shoulder phantom as input to WavResNet. Fig.~\ref{fig:wavresnet-fbpinit} shows the initial FBP image and the denoised image using the RNN versioned WavResNet with 6 iterations (as reported in~\cite{WavResNet18}, and more iterations did not provide much improvements in this case). As we see from Fig.~\ref{fig:wavresnet-fbpinit}, the denoised image is still quite noisy, and the image quality is clearly worse than the result with the PWLS-EP input shown in~\cite{TMI-SPULTRA-as-submit}. Hence, we used the PWLS-EP reconstruction as the input to WavResNet in the comparisons.

\subsubsection{\DIFadd{Regularizer Parameters Selection Procedure}}
\begin{figure}[!htbp]
	\centering
	\begin{subfigure}[h]{0.22\textwidth}
		\centering	
		\includegraphics[width=1\textwidth]{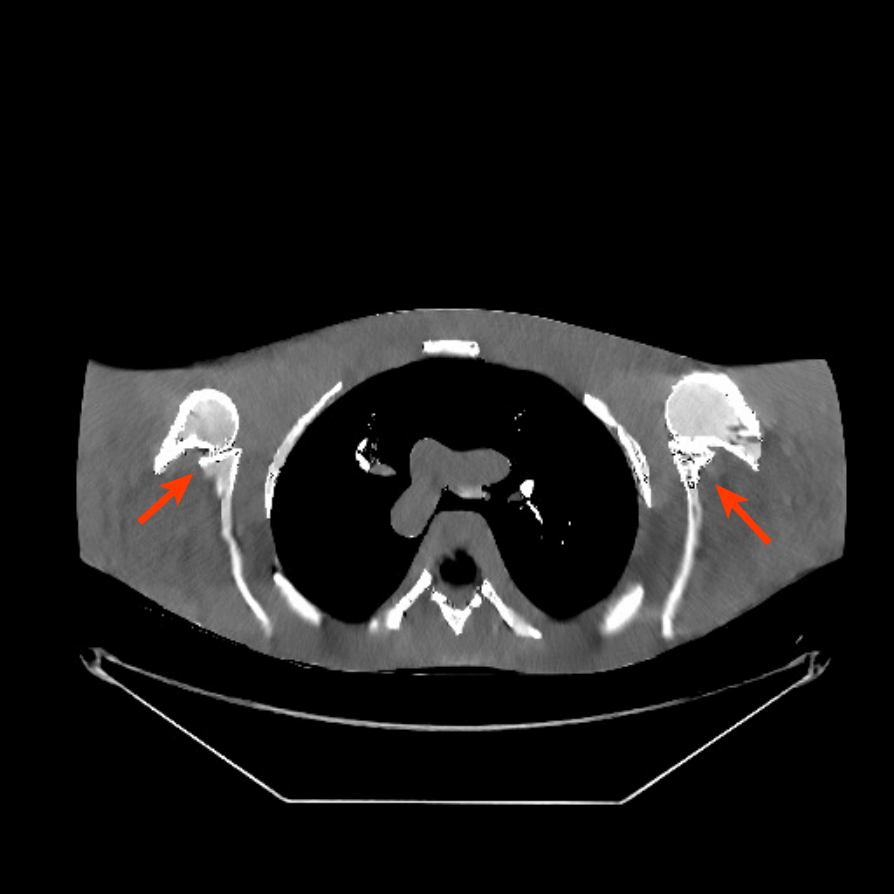}
		\caption{\DIFadd{$(0.05,\ 40)$}}
		\label{fig:shoulder_pwls_0540}
	\end{subfigure}
	\hfil
	\begin{subfigure}[h]{0.22\textwidth}
		\centering	
		\includegraphics[width=1\textwidth]{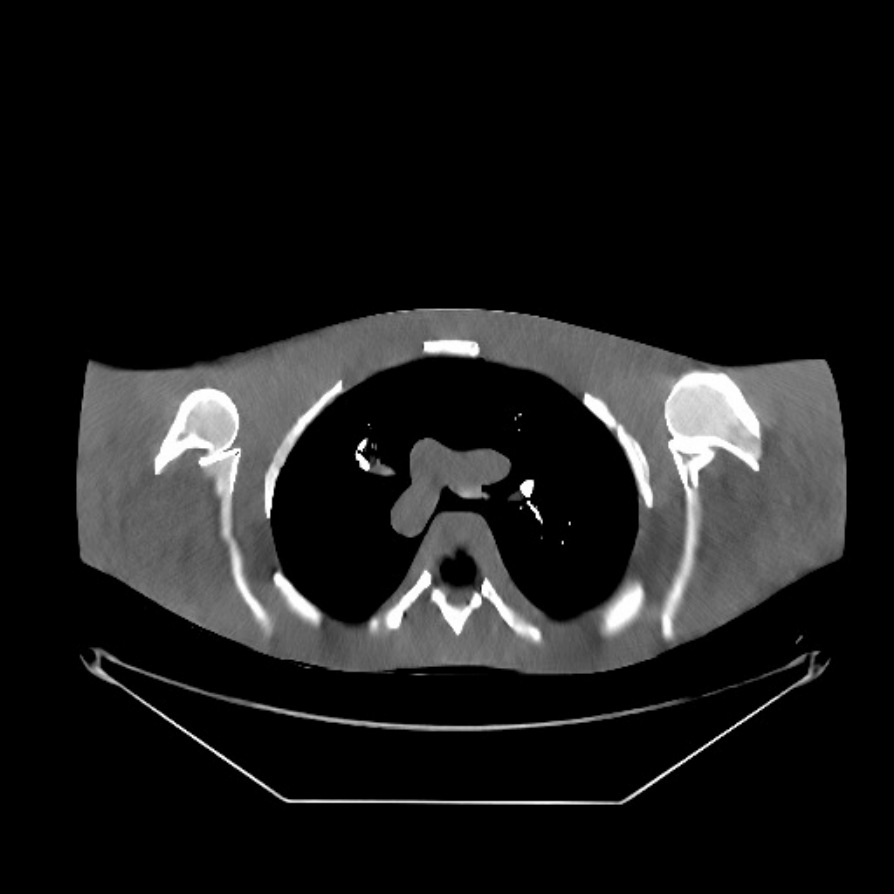}
		\caption{\DIFadd{$(0.05,\ 60)$}}
		\label{fig:shoulder_pwls_bt0560}
	\end{subfigure}
	\vfil
	\begin{subfigure}[h]{0.22\textwidth}
		\centering	
		\includegraphics[width=1\textwidth]{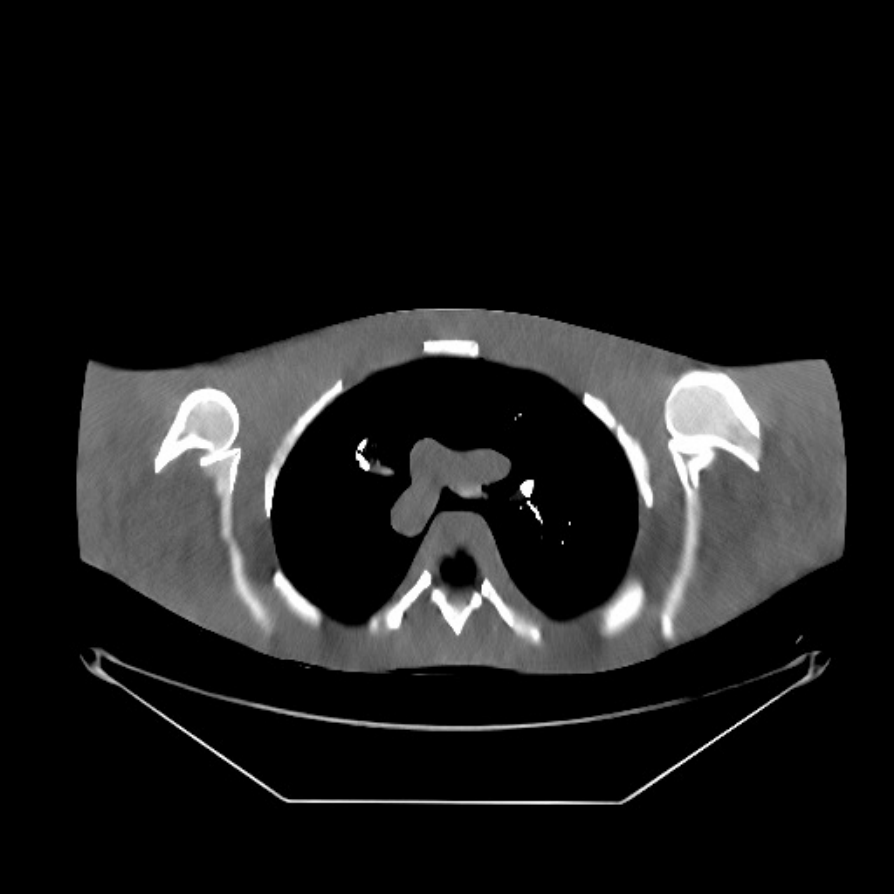}
		\caption{\DIFadd{$(0.05,\ 80)$}}
		\label{fig:shoulder_pwls_0580}
	\end{subfigure}
	\hfil
	\begin{subfigure}[h]{0.22\textwidth}
		\centering	
		\includegraphics[width=1\textwidth]{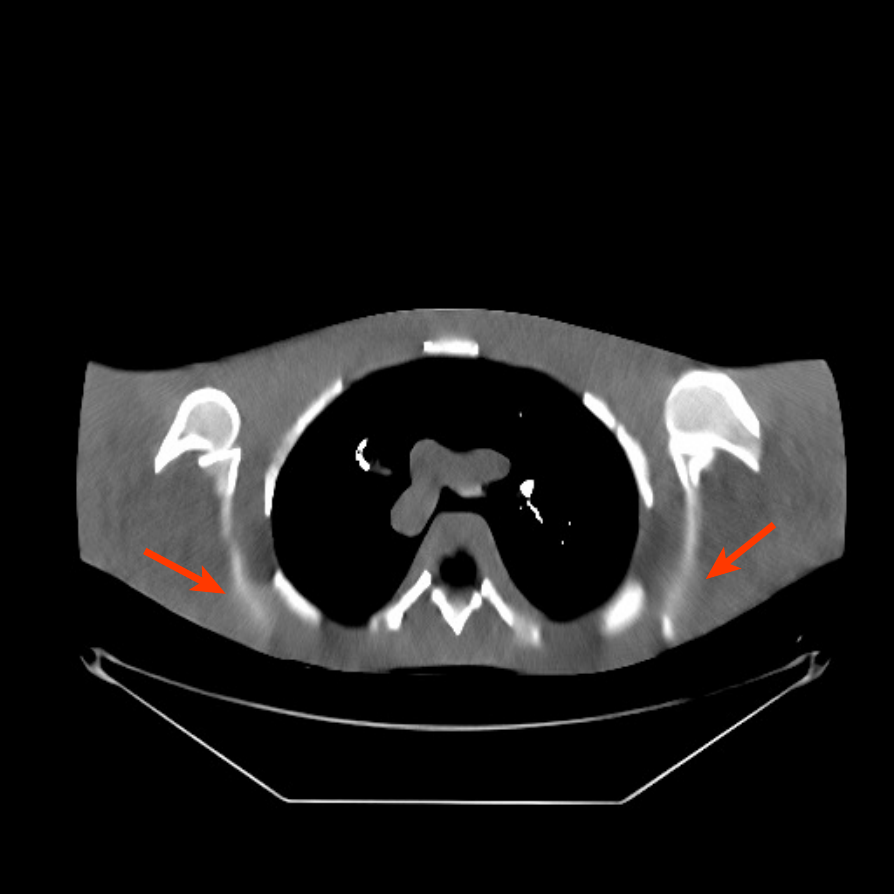}
		\caption{\DIFadd{$(0.05,\ 120)$}}
		\label{fig:shoulder_pwls_05120}
	\end{subfigure}
	\vfil
	\begin{subfigure}[h]{0.22\textwidth}
		\centering	
		\includegraphics[width=1\textwidth]{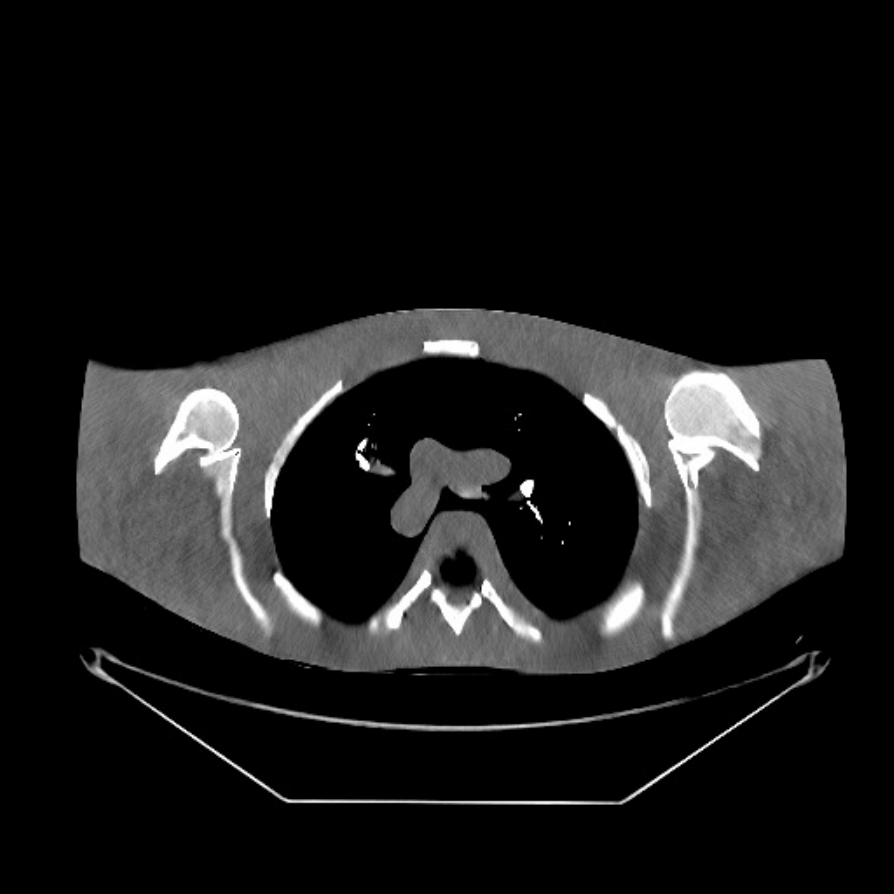}
		\caption{\DIFadd{$(0.03,\ 80)$}}
		\label{fig:shoulder_pwls_0380}
	\end{subfigure}
	\hfil
	\begin{subfigure}[h]{0.22\textwidth}
		\centering	
		\includegraphics[width=1\textwidth]{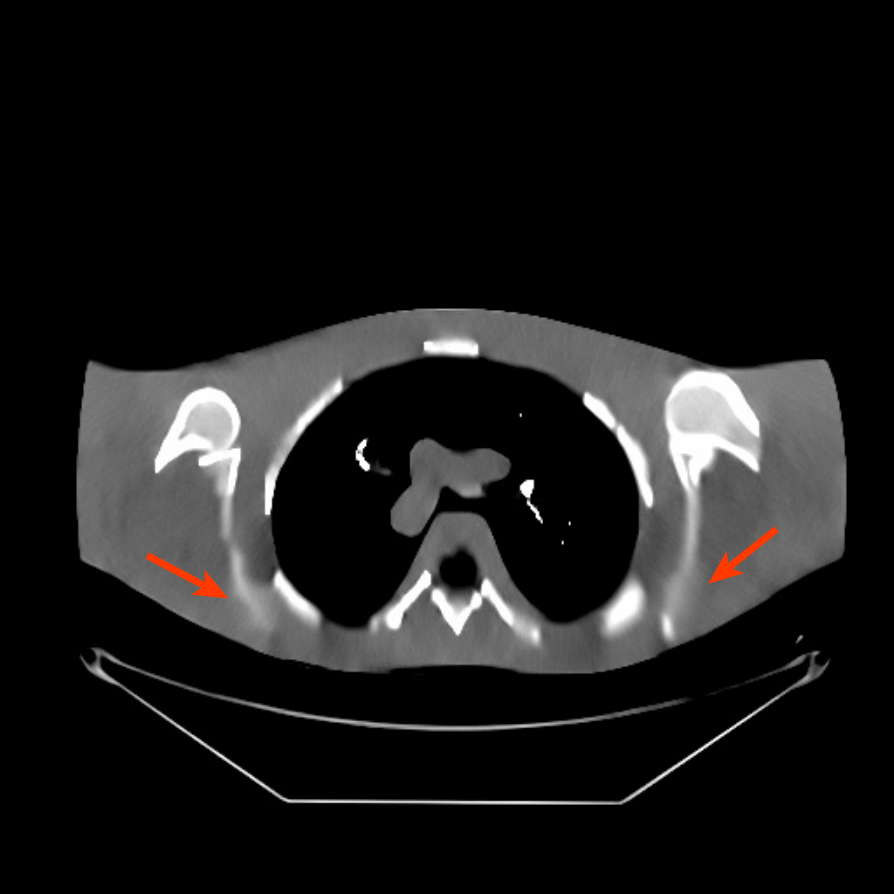}
		\caption{\DIFadd{$(0.1,\ 80)$}}
		\label{fig:shoulder_pwls_180}
	\end{subfigure}
\caption{\DIFadd{PWLS-ULTRA reconstructions with different $(\beta,\ \gamma_c)$ values. The red arrows point to some blurry areas or artifacts.}}
\label{fig:shoulder_pwls_para}
\end{figure}

\begin{figure*}[!ht]
	\centering
	\begin{subfigure}[h]{0.22\textwidth}
	\centering	
	\includegraphics[width=1\textwidth]{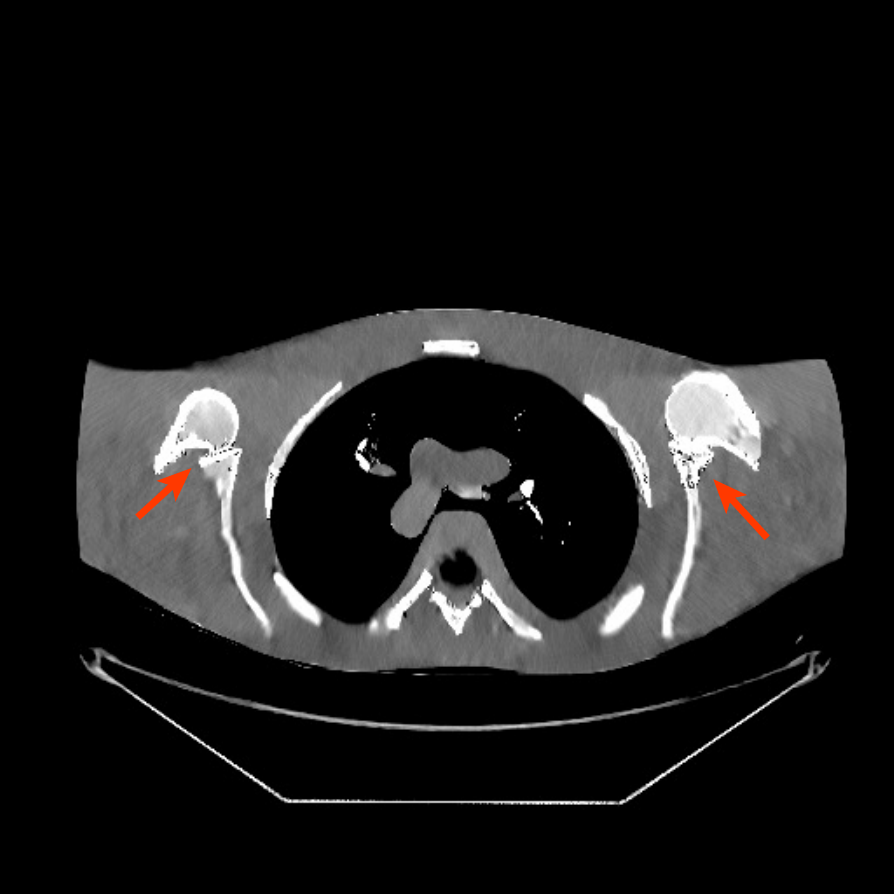}
	\caption{\DIFadd{$(0.05,\ 40)$}}
	\label{fig:shoulder_sp_0540}
\end{subfigure}
\hfil
\begin{subfigure}[h]{0.22\textwidth}
	\centering	
	\includegraphics[width=1\textwidth]{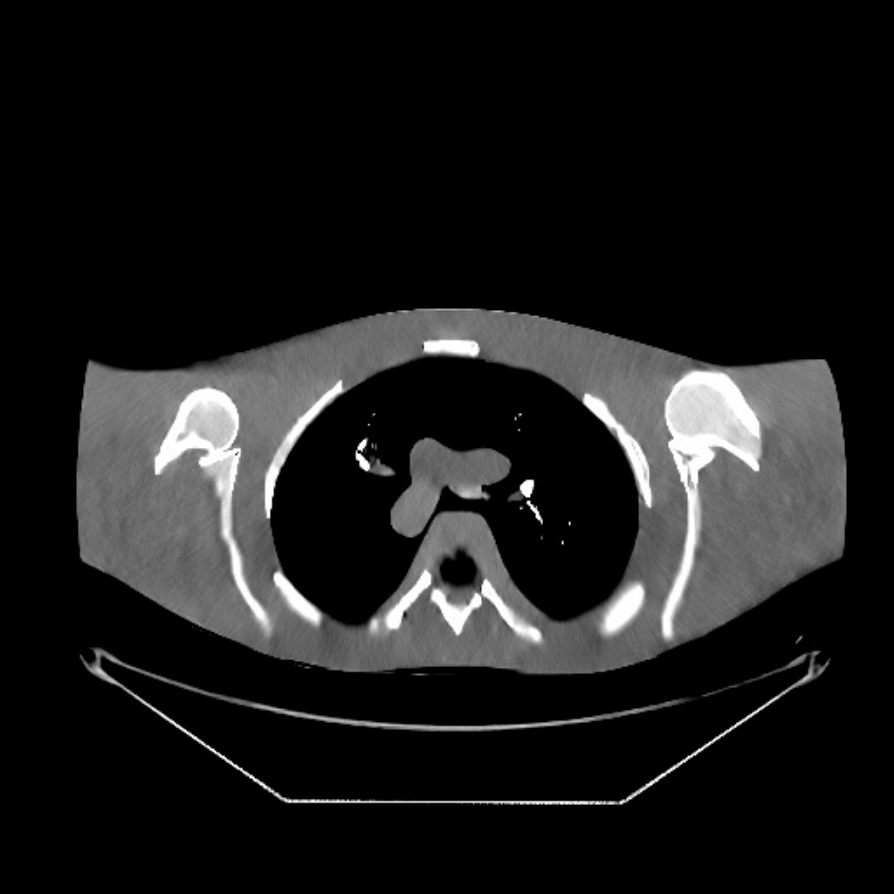}
	\caption{\DIFadd{$(0.05,\ 60)$}}
	\label{fig:shoulder_sp_0560}
\end{subfigure}
\hfil
\begin{subfigure}[h]{0.22\textwidth}
	\centering	
	\includegraphics[width=1\textwidth]{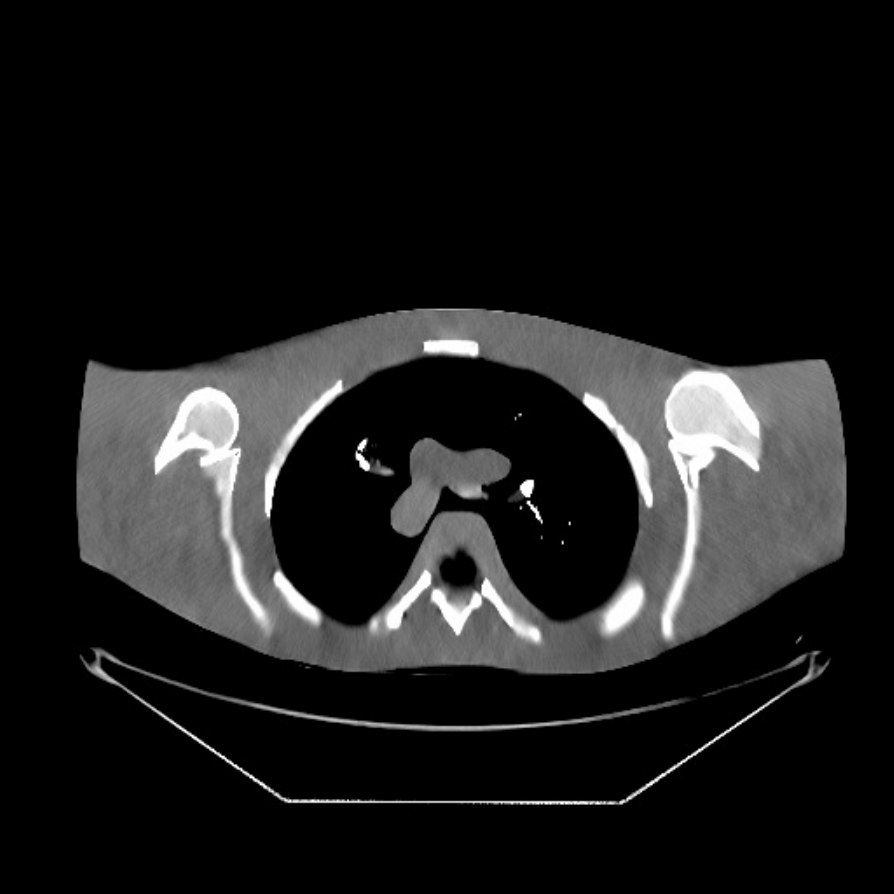}
	\caption{\DIFadd{$(0.05,\ 80)$}}
	\label{fig:shoulder_sp_0580}
\end{subfigure}
\vfil
\begin{subfigure}[h]{0.22\textwidth}
	\centering	
	\includegraphics[width=1\textwidth]{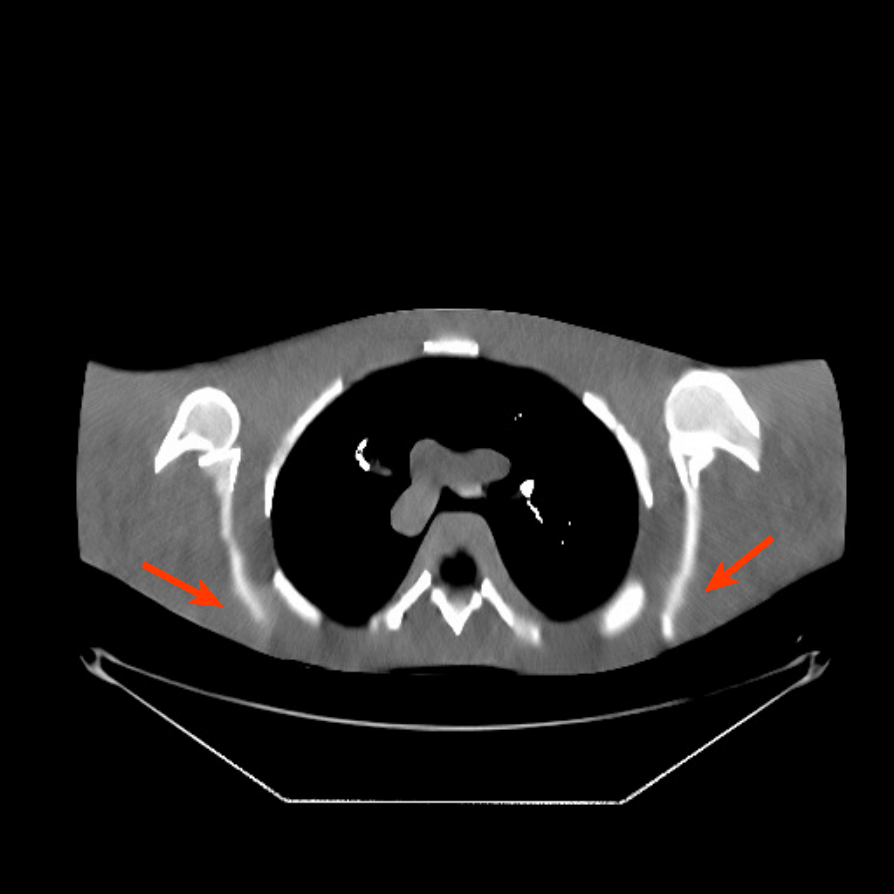}
	\caption{\DIFadd{$(0.05,\ 120)$}}
	\label{fig:shoulder_sp_05120}
\end{subfigure}
\hfil
\begin{subfigure}[h]{0.22\textwidth}
	\centering	
	\includegraphics[width=1\textwidth]{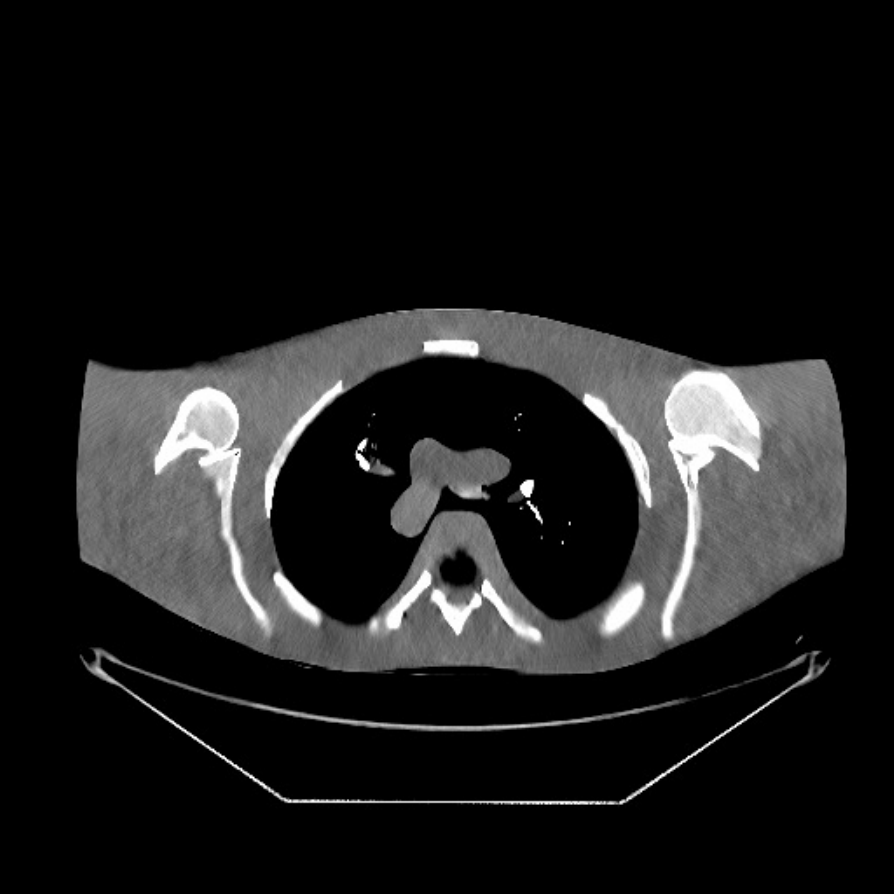}
	\caption{\DIFadd{$(0.03,\ 80)$}}
	\label{fig:shoulder_sp_0380}
\end{subfigure}
\hfil
\begin{subfigure}[h]{0.22\textwidth}
	\centering	
	\includegraphics[width=1\textwidth]{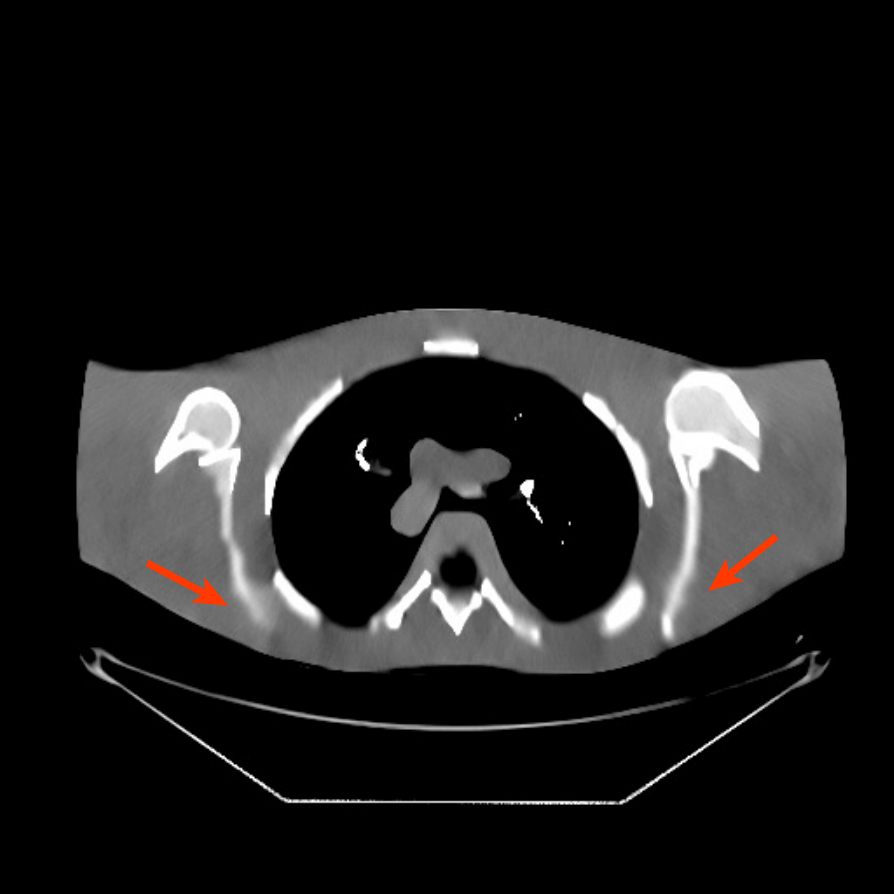}
	\caption{\DIFadd{$(0.1,\ 80)$}}
	\label{fig:shoulder_sp_180}
\end{subfigure}
\caption{\DIFadd{SPULTRA reconstructions with different $(\beta,\ \gamma_c)$ values. The red arrows point to some blurry areas or artifacts.}}
	\label{fig:shoulder_sp_para}
%	\vspace{-0.1in}
\end{figure*}

\DIFadd{In tuning the regularizer parameters for 2D shoulder data simulations where the beam-hardening model is involved, we considered the sparsity level, i.e., the percentage of non-zero entries in the sparse coefficients $\Z$ corresponding to the reconstructed image, and the trade-off among the bias, image resolution, and noise. Based on our heuristic parameters tuning in the XCAT and synthesized clinical data experiments, well reconstructed images usually have sparsity levels around $3\%$ or $4\%$. Therefore, we first roughly chose $\beta=0.05$ that reconstructed a reasonable image, and swept over several $\gamma_c$ values, which controls the sparsity level for both PWLS-ULTRA and SPULTRA, e.g. $\gamma_c = 40,\ 60,\ 80$, and $120$. Tab.~\ref{tab:shoulder-mean-roi} (the second column) reports the sparsity levels of reconstructions with different $(\beta,\gamma_c)$ values. The reconstructed images corresponding to sparsity levels larger than $5\%$ are shown in Fig.~\ref{fig:shoulder_pwls_0540} (PWLS-ULTRA) and Fig.~\ref{fig:shoulder_sp_0540} (SPULTRA). These figures clearly have some artifacts (pointed by red arrows), which verifies the rationale for picking $\gamma_c$ based on the sparsity level. Among $\gamma_c = 60,\ 80$, and $120$, we compared the mean values and standard deviations of the selected ROIs (marked in Fig.~10 in \cite{TMI-SPULTRA-as-submit}), and observed that $\gamma_c=120$ made the reconstructions blurry (see Fig.~\ref{fig:shoulder_pwls_05120} and Fig.~\ref{fig:shoulder_sp_05120}), while $\gamma_c = 60$ and $\gamma_c=80$ can provide good resolution-noise trade-off for reconstructed images. 
	Hereafter, we fixed $\gamma_c = 80$ and swept over several $\beta$ values. Taking $\beta=0.05$ as a baseline, we selected $\beta = 0.03$ and $\beta=0.1$, which are (approximately) $0.5\times$ and $2\times$ of the baseline value. From both numerical results (Mean and STD in Tab.~\ref{tab:shoulder-mean-roi}) and visual results (Fig.~\ref{fig:shoulder_pwls_para} and Fig.~\ref{fig:shoulder_sp_para}), we found that $\beta=0.05$ gave the good bias-resolution-noise trade-off. 
	In the manuscript \cite{TMI-SPULTRA-as-submit}, we showed the results with $\beta=0.05$ and $\gamma_c = 80$.
}
%\begin{figure}[!htp]
%	\centering
%	\begin{subfigure}[h]{0.48\textwidth}
%		\centering	
%		\includegraphics[width=1\textwidth]{figures/shoulder2d/para_pwls-ultra_v2}
%		\caption{\DIFadd{Reconstructed images using PWLS-ULTRA.}}
%		\label{fig:shoulder_pwls_bt}
%	\end{subfigure}
%	\vfill
%	\begin{subfigure}[h]{0.48\textwidth}
%		\centering	
%		\includegraphics[width=1\textwidth]{figures/shoulder2d/para_spultra_v2}
%		\caption{\DIFadd{Reconstructed images using SPULTRA.}}
%		\label{fig:shoulder_spultra_bt}
%	\end{subfigure}
%	\caption{\DIFadd{Reconstructions with $\beta = 0.05,\ 0.07,\ 0.1$ from left to right. The display window is [800, 1400] HU.}}
%	\label{fig:shoulder_bt}
%	\vspace{-0.1in}
%\end{figure}

\begin{table*}[!htbp]
	\centering
	\caption{\DIFadd{Metrics used to tune parameters $(\beta,\gamma_c)$ for PWLS-ULTRA and SPULTRA in ultra low-dose shoulder phantom simulations. The \textit{sparsity (\%)} is the percentage of non-zero entries in the sparse coefficient matrix $\Z$. The \textit{Mean} and the standard deviation (\textit{STD}) are computed for ROIs marked in Fig.~10 of \cite{TMI-SPULTRA-as-submit}, and the unit is HU.} }
	\begin{subtable}{1\textwidth}
		\centering
		\caption{\DIFadd{PWLS-ULTRA}}
	\begin{tabular}{cccc}
		\toprule 
		\textit{\DIFadd{$(\beta,\gamma_c)$}}&\textit{\DIFadd{sparsity (\%)}} & \textit{\DIFadd{Mean (ROI 1 / ROI 2 / ROI3)}}  & \textit{\DIFadd{STD (ROI 1 / ROI 2 / ROI3)}} \\ 
		\midrule
		\DIFadd{\textit{Reference}}& \DIFadd{-}&\DIFadd{\textit{1052.1 / 1060.1 / 1053.4}}& \DIFadd{\textit{8.12 / 8.81 / 6.98}}\\  
		\midrule 
		\DIFadd{(0.05, 40)}& \DIFadd{5.9}&\DIFadd{1031.0 / 1043.2 / 1023.6}& \DIFadd{14.70 / 19.65 / 19.93}\\ 
		\midrule
		\DIFadd{\textbf{(0.05, 60)}}& \DIFadd{\textbf{4.1}}&\DIFadd{\textbf{1031.1 / 1045.4 / 1023.9}}& \DIFadd{\textbf{14.03 / 11.35 / 19.38}}\\ 
		\midrule
		\DIFadd{\textbf{(0.05, 80)}}& \DIFadd{\textbf{3.3}}&\DIFadd{\textbf{1031.1 / 1043.0 / 1024.2}}& \DIFadd{\textbf{14.82 / 10.92 / 19.29}}\\ 
		\midrule
		\DIFadd{(0.05, 120)}&\DIFadd{2.5}&\DIFadd{1032.2 / 1026.7 / 1025.7}& \DIFadd{15.15 / 13.46 / 19.74}\\
		\midrule
		\DIFadd{(0.03, 80)}& \DIFadd{3.6} & \DIFadd{1031.0 / 1043.2 / 1023.5} & \DIFadd{19.08 / 14.96 / 23.23}\\
		\midrule
		\DIFadd{(0.1, 80)}& \DIFadd{3.0} & \DIFadd{1031.8 / 1027.0 / 1025.7} & \DIFadd{12.17 / 13.12 / 16.51}\\			
		\bottomrule
		\end{tabular} 	
	\end{subtable}
	\vfil
	\vspace{0.1in}
	\begin{subtable}{1\textwidth}
		\centering
		\caption{\DIFadd{SPULTRA}}
		\begin{tabular}{ccc  c}
			\toprule 
			\textit{\DIFadd{$(\beta,\gamma_c)$}}&\textit{\DIFadd{sparsity (\%)}} & \textit{\DIFadd{Mean (ROI 1 / ROI 2 / ROI3)}}  & \textit{\DIFadd{STD (ROI 1 / ROI 2 / ROI3)}} \\ 
			\midrule
			\DIFadd{\textit{Reference}}& \DIFadd{-}&\DIFadd{\textit{1052.1 / 1060.1 / 1053.4}}& \DIFadd{\textit{8.12 / 8.81 / 6.98}}\\  
			\midrule 
			\DIFadd{(0.05, 40)}& \DIFadd{7.4}&\DIFadd{1054.7 / 1043.2 / 1049.2}& \DIFadd{16.95 / 12.14 / 13.06}\\ 
			\midrule
			\DIFadd{(0.05, 60)}& \DIFadd{5.0}&\DIFadd{1054.7 / 1047.6 / 1049.1}& \DIFadd{15.96 / 12.26 / 11.93}\\ 
			\midrule
			\DIFadd{\textbf{(0.05, 80)}}& \DIFadd{\textbf{3.9}}&  \DIFadd{\textbf{1054.7 / 1044.0 / 1049.6}}&\DIFadd{\textbf{16.34 / 11.42 / 11.60}}\\
			\midrule
			\DIFadd{(0.05, 120)}&\DIFadd{2.8}&\DIFadd{1055.3 / 1036.9 / 1050.8}& \DIFadd{16.13 / 14.11 / 11.55}\\
			\midrule
			\DIFadd{(0.03, 80)}& \DIFadd{4.2} & \DIFadd{1054.5 / 1042.7 / 1049.2} & \DIFadd{20.65 / 15.81 / 17.36}\\
			\midrule
			\DIFadd{(0.1, 80)}& \DIFadd{3.6} & \DIFadd{1054.7 / 1037.3 / 1050.6} & \DIFadd{12.73 / 12.31 /  6.59}\\			
			\bottomrule
		\end{tabular} 	
	\end{subtable}
	\label{tab:shoulder-mean-roi}
	\vspace{-0.1in}
\end{table*}